\documentclass[11pt]{article}
\usepackage{amsmath}
\usepackage{graphicx,psfrag,epsf}
\usepackage{enumerate}
\usepackage{natbib}
\usepackage{url} 

\usepackage{amsthm}
\usepackage{bigints}
\usepackage{commath}
\usepackage{upgreek}
\usepackage{amsfonts}
\usepackage{mathrsfs}
\usepackage{dsfont}
\usepackage{mathtools}
\usepackage{makeidx}
\usepackage{tabularx}
\usepackage{hyperref}
\usepackage{float}
\usepackage{setspace}
\usepackage{booktabs}
\usepackage{colortbl}
\usepackage{xcolor}
\usepackage{color}
\usepackage{dcolumn}
\usepackage{eurosym}
\usepackage{subcaption}
\usepackage{comment}

\usepackage{xr}
\newcommand{\blind}{0}

\newcolumntype{d}[1]{D{.}{.}{#1}}  

\DeclareMathOperator{\tr}{tr}
\DeclareMathOperator{\vect}{vec}
\DeclareMathOperator{\vecth}{vech}

\newtheorem{theorem}{Theorem}[section]

\newtheorem{lemma}{Lemma}

{
\newtheorem{assumption}{Assumption}
  }

\newcommand\tab[1][1cm]{\hspace*{#1}}

\begin{document}

\def\spacingset#1{\renewcommand{\baselinestretch}%
{#1}\small\normalsize} \spacingset{1}



\if0\blind
{\title{\bf A robust score-driven filter for multivariate time series}
  \author{
  Enzo D'Innocenzo\thanks{Corresponding author.}\\
  Department of Econometrics and Data Science, Vrije Universiteit Amsterdam\\
   De Boelelaan 1105, 1081 HV Amsterdam, Netherlands\\
   e-mail: e.dinnocenzo@vu.nl\\
   \\
   Alessandra Luati\hspace{.2cm}\\
Department of Statistical Sciences, University of Bologna,\\
 Via delle Belle Arti 41, 40126 Bologna, Italy,\\
  e-mail: alessandra.luati@unibo.it\\
  and\\
   \\
   Mario Mazzocchi\hspace{.2cm}\\
Department of Statistical Sciences, University of Bologna,\\
 Via delle Belle Arti 41, 40126 Bologna, Italy,\\
  e-mail: mario.mazzocchi@unibo.it}
  \maketitle
} \fi

\if1\blind
{
  \bigskip
  \bigskip
  \bigskip
  \begin{center}
    {\LARGE\bf A robust score-driven filter for multivariate time series}
\end{center}
  \medskip
} \fi

\newpage
\begin{abstract}
A  multivariate score-driven filter is developed to extract signals from noisy vector processes. By assuming that the conditional location vector from a multivariate Student's \emph{t} distribution changes over time, we construct a robust filter which is able to overcome several issues that naturally arise when modeling heavy-tailed phenomena and, more in general, vectors of dependent non-Gaussian time series. We derive conditions for stationarity and invertibility and estimate the unknown parameters by maximum likelihood (ML). Strong consistency and asymptotic normality of the estimator are proved and the finite sample properties are illustrated by a Monte-Carlo study. From a computational point of view, analytical formulae are derived, which consent to develop estimation procedures based on the Fisher scoring method. The theory is supported by a novel empirical  illustration that shows how the model can be effectively applied to estimate consumer prices from home scanner data.
\end{abstract}

\noindent%

\vfill

\newpage
\spacingset{1.2} 
\section{Introduction}
The analysis of multivariate time series has a long history, due to the empirical evidence, from most research fields, that time series resulting from complex phenomena do not only depend on their own past, but also on the history of other variables. For this reason, from \cite{Hannan1970}, the literature on multivariate time series has grown very fast. The leading example is the dynamic representation of the conditional mean of a vector process which gives rise to vector autoregressive processes, see \cite{Hamilton1994} and \cite{Lutkepohl2007}.

Following the taxonomy proposed in \cite{Cox1981}, two main classes of models can be considered when analysing dynamic phenomena: parameter-driven and observation-driven models. The class of parameter driven model is a broad class, which involves  unobserved component models and state space models (\citealp{Harvey1989, West_Harrison1997}). Within this framework, parameters are allowed to vary over time as dynamic processes driven by idiosyncratic innovations. Hence, likelihood functions are analytically tractable only in specific cases, notably linear Gaussian models, where inference can be handled by the Kalman filter. On the other hand, parameter-driven models are very sensitive to small deviations from the distributional assumptions. In addition, the Gaussian assumption often turns out to be  restrictive, and flexible specifications may be more appropriate. Thus, a fast growing field of research is dealing with nonlinear or non-Gaussian state-space models, resting on computer intensive simulation methods like the particle filter discussed in \cite{durbkoopman2012}. Although these methods provide extremely powerful instruments for estimating nonlinear and/or non-Gaussian models, they can be computationally demanding. Furthermore, it may be difficult to derive the statistical properties of the implied estimators, due to the complexity of the joint likelihood function. 

In contrast, in observation-driven models, the dynamics of time varying parameters depend on deterministic functions of lagged variables. This enables a stochastic evolution of the parameters which become predictable given the past observations. \cite{Koopman_Lucas_Scharth2016} assess the performances and optimality properties of the two classes of models, in terms of their predictive likelihood. The main advantage of observation-driven models is that the likelihood function is available in closed form, even in nonlinear and/or non-Gaussian cases. Thus, the asymptotic analysis of the estimators becomes feasible and computational costs are reduced drastically. 

Within the class of observation-driven models, score-driven models are a valid option for modeling time series that do not fall in the category of linear Gaussian processes. Examples have been proposed in the context of volatility estimation and originally referred to as generalised autoregressive score (GAS) models, \cite{CrealKoopLucas2013}, and as dynamic conditional score (DCS) models, \cite{Harvey2013}. The key feature of these models is that the dynamics of time-varying parameters are driven by the score of the conditional likelihood, which needs not necessarily to be Gaussian but can be heavy tailed. For example, it may follow a Student's \emph{t} distribution as in \cite{HarveyLuati2014} and \cite{Linton2020}, an exponential generalized beta distribution, as in \cite{Caivano_Harvey_Luati2016}, a binomial distribution as in the vaccine example by \cite{hansen.2019}, or represented by a mixture, see \cite{Lucas_Sch_Sch_2019}. 
The optimality of the score as a driving force for time varying parameters in observation-driven models is discussed in \cite{Blasques_Koopman_Lucas2015}. According to which conditional distribution is adopted, specific situations may be conveniently handled due to the properties of the score. As an example for the univariate case, if a heavy-tailed distribution is specified, namely Student's \emph{t}, the resulting score-driven model yields a simple and natural model-based  signal extraction filter which is robust to extreme observations, without any external interventions or diagnostics, like dummy variables or outlier detection, see \cite{HarveyLuati2014}. 

In score-driven models, as well as in all observation-driven models, the time varying parameters are updated by filtering procedures, i.e. weighted sums of functions of past observations, given some initial conditions that can be fixed or estimated along with the static parameters. A robust filtering procedure should assign less weight to extreme observations in order to prevent biased inference of the signal and the parameters. In particular, the work of \cite{Calvet_Czellar_Ronchetti2015} provides a remarkable application of robust methods when dealing with contaminated observations. The authors show that a substantial efficiency gain can be achieved by huberizing the derivative of the $\log$-observation density. As we show in the present study, the same holds if one considers an alternative robustification method, based on the specification of a conditional multivariate Student's \emph{t} distribution. A similar approach can be found in \cite{Prucha_Kelejian1984} and \cite{Fiorentini_Sentana_Calzolari2003}, where the multivariate Student's \emph{t} distribution provides a valid alternative to relax the normality assumption. In the context of score-driven models, \cite{creal2014} mention the relevance of modeling high-frequency data with outliers and heavy tails by means of the multivariate Student's $t$ distribution.

In this paper, we develop a score-driven filter for the time-varying location of a multivariate Student's \emph{t} distribution. The specification is similar to the multivariate model for the location addressed in \cite{Harvey2013} and has some traits in common with the  quasi-vector autoregressive model by \cite{Blazsekwp}, though our perspective is more focused on the aspects of the filter and its stochastic properties. A spatial extension of the model developed in this paper is considered by \cite{Gasperoni2021}.  We envisage three main contributions to the existing literature.
 
The first contribution is the derivation of the probabilistic theory behind the multivariate dynamic score-driven filter for conditional Student's \emph{t} distributions, including the conditions of stationarity, ergodicity and invertibility, in a similar spirit of \cite{Comte2003} and \cite{Hafner2009} for the multivariate conditional variance models by \cite{baba1990} and \cite{Engle1995}. These results provide the basis for further generalisations, such as, for instance, the spatial model by \cite{Gasperoni2021}. As the conditional likelihood is available in close form, we estimate the static parameters with the method of maximum likelihood and prove strong consistency and asymptotic normality of the estimators. It is noteworthy to remark that when the degrees of freedom of the Student's $t$ distribution tend to infinity, we recover a linear Gaussian state-space model.    

The second contribution is the development of an estimation scheme grounded on Fisher's scoring method, based on closed-form analytic expressions, which can be directly implemented into any statistical or matrix-friendly software. 

The third contribution of the paper is an innovative application, dealing with estimation of regional consumer prices based on home scanner data. The use of scanner data to compute official consumer price indices (CPIs) is gaining popularity, because of their timeliness and a high level of product and geographical detail \cite{Feenstra2003}. However, they also suffer from a variety of shortcomings, which make time series of scanner data prices (SDPs) potentially very noisy, especially when they are estimated for population sub-groups, or at the regional level \cite{Silver1995}. There is extensive research and a lively debate on the issues related to the computation and use of scanner data based CPIs. In a dedicated session of the 2019 meeting of the the Ottawa Group on Price Indices, it has been suggested\footnote{See Jens Mehroff presentation at \url{https://eventos.fgv.br/sites/eventos.fgv.br/files/arquivos/u161/towards_a_new_paradigm_for_scanner_data_price_indices_0.pdf}} to adopt model-based filtering techniques to extract the signal from scanner-based time series of price data. These filtered estimates lose the classical price index formula interpretation, but are expected to deliver the same information content with a better signal-to-noise ratio. We show that our robust multivariate model, applied to SDPs, provides information on the dynamics of the time series and on their interrelations without being affected from outlying observations, which are naturally downweighted in the updating mechanism.

The paper is organised as follows. In Section \ref{The_Multivariate_DCS-t_Location Model} the  filter is specified. Section \ref{Stochastic_Properties} deals with the stochastic properties of the filter, while in Section \ref{Maximum_Likelihood_Estimation}, likelihood inference is discussed. The empirical analysis is reported in section \ref{Empirical_Applications}. Some concluding remarks are drawn in Section \ref{Concluding_Remarks}. The proofs of the results stated in the paper are collected in Appendix A. 
Online supplementary materials contain the details of a Monte Carlo study designed to assess the finite sample properties of the estimators, the relevant quantities for the implementation of the Fisher scoring algorithm as well as the proofs of some auxiliary Lemmata.

\section{The Multivariate Student's \emph{t} Location Filter}
\label{The_Multivariate_DCS-t_Location Model}

Let us consider a $\mathbb{R}^N$-vector of stochastic processes $\{ \boldsymbol{y}_t\}_{t\in\mathbb{Z}}$, $N \geq 1$, and let $\mathcal{F}_{t-1} = \sigma\{ \boldsymbol{y}_{t-1}, \boldsymbol{y}_{t-2}, $ $\boldsymbol{y}_{t-3}, \dots  \}$ be its filtration at time $t-1$. The following stochastic representation of $\boldsymbol{y}_t$ is considered,
\begin{equation}
\label{location_scale_model}
\boldsymbol{y}_t 
=
\boldsymbol{\mu}_t + \boldsymbol{\Omega}^{1/2} \boldsymbol{\epsilon}_t,
\end{equation}
where $\boldsymbol{\mu}_t$ is a time varying location vector of $\mathbb{R}^N$, $\boldsymbol{\Omega}$ is a $N\times N$ scale matrix that we assume to be static and $\boldsymbol{\epsilon}_t \sim \boldsymbol{t}_\nu (\boldsymbol{0}_N, \boldsymbol{I}_N)$ is an  independent identically distributed (\emph{IID}) multivariate standard \emph{t}-variate. With $\boldsymbol{0}_N$ we denote the null vector of $\mathbb{R}^N$ and with $\boldsymbol{I}_N$ the $N \times N$ identity matrix. 



Our interest is in recovering $\boldsymbol{\mu}_t$ based on a set of observed time series from $\boldsymbol{y}_t$, for $t=1, \dots, T$, where $T \in \mathbb{N}$. With no distributional assumptions on the dynamics of $\boldsymbol{\mu}_t$,  a filter can be specified, 
\begin{equation}
\label{phi_sre}
\boldsymbol{\mu}_{t+1|t} = \phi (\boldsymbol{\mu}_{t|t-1}, \boldsymbol{y}_t, \boldsymbol{\theta}),
\end{equation}
that is a stochastic recurrence equation (SRE), where $\boldsymbol{\theta}\in \boldsymbol{\Theta} \subset \mathbb{R}^p$ is a vector of unknown static parameters, $\boldsymbol{\mu}_{t|t-1}$ is a $\mathbb{R}^N$-random vector that takes values in $\boldsymbol{\mathcal{M}} \subset \mathbb{R}^N$ and $\phi: \boldsymbol{\mathcal{M}} \times \mathbb{R}^N \times \boldsymbol{\Theta} \mapsto\boldsymbol{\mathcal{M}}$ is a Lipschitz function. The subscript notation $t|t-1$ is used to emphasize the fact that $\boldsymbol{\mu}_{t|t-1}$ is an approximation of the dynamic location process at time $t$ given the past, that is equivalent to say that $\boldsymbol{\mu}_{t|t-1}$ is $\mathcal{F}_{t-1}$-measurable. Therefore, based on past observations and a starting value $\boldsymbol{\mu}_{1|0} \in \boldsymbol{\mathcal{M}}$, one can approximate the unobserved path of $\boldsymbol{\mu}_t$ in \eqref{location_scale_model} by mimicking the recursion in \eqref{phi_sre}. It is typically assumed that a parameter value $\boldsymbol{\theta}_0$ exists, at which the true location can be recovered, i.e. $\boldsymbol{\mu}_{t|t-1}(\boldsymbol{\theta}_0) = \boldsymbol{\mu}_t$ (assumption \ref{ass:cs} of correct specification).


In this paper, we approximate the temporal changes of the dynamic location by relying on the score-driven framework of \cite{CrealKoopLucas2013} and \cite{Harvey2013}. Specifically, we assume that, conditional on the past, the distribution of $\boldsymbol{y}_t$ is Student's \emph{t}, with $\nu>0$ degrees of freedom and conditional location  equal to $\boldsymbol{\mu}_{t|t-1}$, i.e. 
\begin{equation}
\label{multivariate_t}
f(\boldsymbol{y}_t | \mathcal{F}_{t-1})
=
\frac{\Gamma \big( \frac{\nu + N}{2} \big)}{ \Gamma \big( \frac{\nu}{2} \big) (\pi \nu)^{N/2}}
| \boldsymbol{\Omega} |^{-1/2}
\bigg[
1 +
\frac{(\boldsymbol{y}_t - \boldsymbol{\mu}_{t|t-1})^\top 
\boldsymbol{\Omega}^{-1} 
(\boldsymbol{y}_t - \boldsymbol{\mu}_{t|t-1})}{\nu}
\bigg]^{-(\nu + N)/2}
\end{equation}
and  specify the SRE in \eqref{phi_sre} as follows, 
\begin{align}
\label{score_driven_recursion}
\boldsymbol{\mu}_{t+1|t} - \boldsymbol{\omega} 
=&
\boldsymbol{\Phi} ( \boldsymbol{\mu}_{t|t-1} - \boldsymbol{\omega} ) 
+
\boldsymbol{K} \boldsymbol{u}_t,
\end{align}
where $\boldsymbol{\omega}$ is a $\mathbb{R}^N$ vector of unconditional means, $\boldsymbol{\Phi}$ and $\boldsymbol{K}$ are $\mathbb{R}^{N \times N}$ matrices of coefficients and the driving force $\boldsymbol{u}_t$ is 
proportional to the score of conditional density in \eqref{multivariate_t}. Indeed,  the conditional score with respect to the time varying location filter  is
\begin{equation*}
\frac{\partial \ln f(\boldsymbol{y}_t | \mathcal{F}_{t-1})}{ \partial \boldsymbol{\mu}_{t|t-1} }
=
 \boldsymbol{\Omega}^{-1}
\frac{\nu + N}{\nu }  \boldsymbol{u}_{t}.
\end{equation*}
where
\begin{equation}
\label{u_t}
\boldsymbol{u}_t =
(\boldsymbol{y}_t - \boldsymbol{\mu}_{t|t-1})/ w_t,
\end{equation}
with $w_t = 1 + (\boldsymbol{y}_t - \boldsymbol{\mu}_{t|t-1})^\top 
 \boldsymbol{\Omega}^{-1} 
(\boldsymbol{y}_t - \boldsymbol{\mu}_{t|t-1})/\nu$, is a martingale difference sequence, i.e. $\mathbb{E}_{t-1}[\boldsymbol{u}_t]=\boldsymbol{0}_N$, under correct specification, where the shorthand notation $\mathbb{E}_{t-1}[X]$ is used for the conditional expectation $\mathbb{E}[X | \mathcal{F}_{t-1}]$. The score as the driving force in an updating equation for a time varying parameter is the key feature of score-driven models. The rationale behind the recursion \eqref{score_driven_recursion} is very intuitive. Analogously to the Gauss-Newton algorithm, it improves the model fit by pointing in the direction of the greatest increase of the likelihood. Optimality of score driven updates in observation-driven models is discussed by {\cite{Blasques_Koopman_Lucas2015}. 


In the context of location estimation under the Student's \emph{t} assumption, a further relevant motivation for the score-driven methodology lies in the robustness of the implied filters. 
Indeed, 
the positive scaling factors $w_t$ in equation \eqref{u_t} are scalar weights that involve the Mahalanobis distance. They possess the role of re-weighting the large deviation from the mean incorporated in the innovation error 
\begin{equation}
\label{eq:v}
\boldsymbol{v}_t = \boldsymbol{y}_t - \boldsymbol{\mu}_{t|t-1}.
\end{equation} 
Robustness comes precisely from winsorizing the innovation error $\boldsymbol{v}_t$. 
Note that when $\nu\rightarrow \infty$, $\boldsymbol{u}_t$ converges to $\boldsymbol{v}_t$ and equations  (\ref{score_driven_recursion}) and (\ref{eq:v}) coincide with the steady state innovation form of a linear Gaussian state-space model. 

A formal proof of the robustness of the method is in the following Lemma, which provides  sufficient conditions for a filter to be robust, in line with \cite{Calvet_Czellar_Ronchetti2015}.  We first enounce the correct specification assumption. 

\begin{assumption}\label{ass:cs}
The filter in \eqref{phi_sre} is correctly specified, i.e. when $\boldsymbol{\theta} = \boldsymbol{\theta}_0$, where $\boldsymbol{\theta}_0$ is the true parameter vector, $\boldsymbol{\mu}_{t|t-1}(\boldsymbol{\theta}_0) = \boldsymbol{\mu}_t$. 
\end{assumption}


\begin{lemma}
\label{uniformly_boundedness_ut}
Under assumption \ref{ass:cs}, for $0 < \nu < \infty$, the vector sequence $\{ \boldsymbol{u}_t \}_{t\in\mathbb{Z}}$ is uniformly bounded, that is $\sup_{t} \mathbb{E}[\| \boldsymbol{u}_t \|] < \infty$ and  possesses all the even moments
\begin{equation*}
\mathbb{E}[\| \boldsymbol{u}_t \|^{2s}] = \| \boldsymbol{\Omega} \|^{s} 
\frac{B\big( \frac{N+2s}{2} , \frac{\nu+2s}{2} \big)}
{B\big( \frac{N}{2} , \frac{\nu}{2} \big)}\Big( \frac{\nu}{N} \Big)^{s},
\end{equation*}
for $s=1,2,\dots $ and where $B(\alpha,\beta)=\Gamma(\alpha)\Gamma(\beta)/\Gamma(\alpha+\beta)$ is the beta function and $\| \boldsymbol{\Omega} \| = \sqrt{\tr(\boldsymbol{\Omega}^\top\boldsymbol{\Omega})}$. The odd moments of $\boldsymbol{u}_t$ are all equal to zero. 
\end{lemma}
The moment structure reveals important features of the driving force $\boldsymbol{u}_t$, that turns out to be an an \emph{IID} sequence with zero mean vector and $(\vect)$-variance covariance matrix,
\begin{equation*}
\mathbb{E}[\boldsymbol{u}_t \otimes \boldsymbol{u}_t] 
= 
\vect \mathbb{E}[\boldsymbol{u}_t \boldsymbol{u}_t^\top ] 
= 
\frac{\nu^2}{(\nu+N)(\nu + N +2)} \, \vect \boldsymbol{\Omega}.
\end{equation*}

\section{Properties of the Filter}
\label{Stochastic_Properties}

Let us combine equations \eqref{score_driven_recursion} and \eqref{u_t} and write the filter explicitly, as follows, 
\begin{equation}
\label{score_driven_filter}
\boldsymbol{\mu}_{t+1|t} 
=
\boldsymbol{\omega} + \boldsymbol{\Phi} (\boldsymbol{\mu}_{t|t-1} - \boldsymbol{\omega})
+
\boldsymbol{K} 
\frac{\boldsymbol{y}_t - \boldsymbol{\mu}_{t|t-1}}
{1 + (\boldsymbol{y}_t - \boldsymbol{\mu}_{t|t-1})^\top 
\boldsymbol{\Omega}^{-1} 
(\boldsymbol{y}_t - \boldsymbol{\mu}_{t|t-1})/ \nu}.
\end{equation}
By starting at some initial value, $\boldsymbol{\mu}_{1|0}\in \boldsymbol{\mathcal{M}}$, and using equation \eqref{score_driven_filter} for $t=1, \dots, T$, with $T \in \mathbb{N}$, one can recover a unique filtered path $\{ \hat{\boldsymbol{\mu}}_{t|t-1} \}_{t \in \mathbb{N}}$ for every $\boldsymbol{\theta} \in \boldsymbol{\Theta}$. A desirable property is that the values used to initialise the process are asymptotically negligible, in the sense that as the time $t$ increases, the impact of the chosen $\boldsymbol{\mu}_{1|0}$ eventually vanishes and the process will converge to a unique stationary and ergodic sequence. 
This stability property of the filtered sequence $\{ \hat{\boldsymbol{\mu}}_{t|t-1}\}_{t \in \mathbb{N}}$ is known as invertibility, see \cite{Straumann_Mikosh2006} and \cite{Blasques_Gorgi_etal2018}. 
Existence of the unique stationary and ergodic solution to the SRE  \eqref{score_driven_filter} is established by Lemma \ref{SE_Dynamic_Location}. Invertibility of the filter is proved in Lemma \ref{INV_Dynamic_location}.

\begin{lemma}
\label{SE_Dynamic_Location} 
Let us consider equation  \eqref{score_driven_filter}, evaluated at the $\boldsymbol{\theta}=\boldsymbol{\theta}_0$. Assume that $0 < \nu < \infty$ and $\varrho(\boldsymbol{\Phi}) < 1$, where $\varrho(\boldsymbol{\Phi})$ denotes the spectral radius of $\boldsymbol{\Phi}$. Then, there exists a unique vector sequence $\{ \tilde{\boldsymbol{\mu}}_{t|t-1} \}_{t\in\mathbb{Z}}$ which is strictly stationary and ergodic with $\mathbb{E}[\|{\tilde{\boldsymbol{\mu}}}_{t|t-1} \|^m ] < \infty$ for every $m > 0$.

\end{lemma}
The stability condition $\varrho(\boldsymbol{\Phi}) < 1$ is a well-known condition in the theory of linear systems, see \cite{Hannan1970}, \cite{Hannan_Deistler1987} or \cite{Lutkepohl2007}, which extends to the case of the present nonlinear model.  

With the next Lemma, the relevant conditions under which the SRE in \eqref{score_driven_filter} is contractive on average are given so that the convergence of the filtered sequence $\{ \hat{\boldsymbol{\mu}}_{t|t-1} \}_{t \in \mathbb{N}}$ to a unique $\mathcal{F}_{t-1}$-measurable stationary and ergodic solution $\{ \tilde{\boldsymbol{\mu}}_{t|t-1} \}_{t \in \mathbb{Z}}$, irrespective of the initialization $\boldsymbol{\mu}_{1|0}$, is obtained as a corollary of Theorem 3.1 of \cite{Bougerol1993} or, equivalently, of Theorem 2.8 of \cite{Straumann_Mikosh2006}. Moreover, as a consequence of Lemma \ref{uniformly_boundedness_ut}, both $\{ \hat{\boldsymbol{\mu}}_{t|t-1} \}_{t \in \mathbb{N}}$ and $\{\tilde{\boldsymbol{\mu}}_{t|t-1} \}_{t \in \mathbb{Z}}$ have bounded moments.


\begin{lemma}
\label{INV_Dynamic_location}
Let the conditions of Lemma \ref{SE_Dynamic_Location} hold and assume that
\begin{align}
\label{inv_contraction}
\mathbb{E} \bigg[
\ln
\sup_{\boldsymbol{\theta} \in \boldsymbol{\Theta}} 
\sup_{\boldsymbol{\mu} \in \boldsymbol{\mathcal{M}}}
\bigg\| 
\prod_{j=1}^k
\boldsymbol{X}_{k-j+1}
\bigg\| \bigg] < 0,
\end{align}
for $k\geq 1$, where $\boldsymbol{\Theta}$ is a compact parameter space and $\boldsymbol{X}_{t} =
\boldsymbol{\Phi} + \boldsymbol{K}\partial \boldsymbol{u}_t/\partial \boldsymbol{\mu}_{t|t-1}^\top$. Then, the filtered location vector $\{ \hat{\boldsymbol{\mu}}_{t|t-1} \}_{t \in \mathbb{N}}$ is invertible and converges exponentially fast almost surely (e.a.s.) to the unique stationary ergodic sequence $\{ \tilde{\boldsymbol{\mu}}_{t|t-1} \}_{t \in \mathbb{Z}}$ for any initialization of the filtering recursion, $\boldsymbol{\mu}_{1|0} \in \boldsymbol{\mathcal{M}}$, that is,
\begin{align}
\label{eas_dynamic_location}
\sup_{\boldsymbol{\theta} \in \boldsymbol{\Theta}}  \|
\hat{\boldsymbol{\mu}}_{t|t-1} 
-
\tilde{\boldsymbol{\mu}}_{t|t-1}
\|\xrightarrow[]{\text{e.a.s.}} 0 
\tab \textit{as} \tab  t \rightarrow \infty,
\end{align}
Furthermore, 
$\sup_t \mathbb{E}[\sup_{\boldsymbol{\theta}\in \boldsymbol{\Theta}} \| \hat{\boldsymbol{\mu}}_{t|t-1} \|^m ] < \infty$ and $\mathbb{E}[\sup_{\boldsymbol{\theta}\in \boldsymbol{\Theta}} \|\tilde{\boldsymbol{\mu}}_{t|t-1} \|^m ] < \infty, \forall m \geq 1$.
\end{lemma}


The contraction condition in equation \eqref{inv_contraction} imposes restrictions on the parameter space $\boldsymbol{\Theta}$ that cannot be checked directly. Also, the expectation in the same equation cannot be verified in practice, since it depends on the unconditional, unknown,  distribution of $\boldsymbol{y}_t$, see also the discussion in \cite{Blasques_Gorgi_etal2018}. Thus, one can rely on sufficient conditions which are typically more restrictive than \eqref{inv_contraction} and that we discuss in the following, similarly to  \cite{Linton2020}.
Specifically, the contraction condition in \eqref{inv_contraction} is satisfied if
\begin{align}
\label{suff_inv_contraction}
\mathbb{E}
\bigg[\ln \sup_{\boldsymbol{\theta}\in \boldsymbol{\Theta}} 
\sup_{\boldsymbol{\mu}_{1|0} \in \boldsymbol{\mathcal{M}}}
\bigg\|\boldsymbol{X}_1 \bigg\|\bigg]<0.
\end{align}
Motivated by Example 3.8 of \cite{Straumann_Mikosh2006}, we rewrite $\boldsymbol{X}_1$ at $\boldsymbol{\theta}_0$, so that equation \eqref{suff_inv_contraction} becomes
\begin{align}
\label{X1_MC}
\mathbb{E}
\bigg[\ln
\bigg\|
\boldsymbol{\Phi}_0 + 
\frac{\boldsymbol{K}_0}{1 + \boldsymbol{\epsilon}_1^\top\boldsymbol{\epsilon}_1/\nu_0}
\bigg(
\frac{2\boldsymbol{\Omega}_0^{1/2}\boldsymbol{\epsilon}_1\boldsymbol{\epsilon}^\top_1\boldsymbol{\Omega}_0^{-1/2}/\nu_0}{1 + \boldsymbol{\epsilon}_1^\top\boldsymbol{\epsilon}_1/\nu_0 } -  
\boldsymbol{I}_N
\bigg)
\bigg\|\bigg]<0.
\end{align}
Since $ \boldsymbol{\epsilon}_1 \sim \boldsymbol{t}_{\nu_0} (\boldsymbol{0}_N, \boldsymbol{I}_N)$, based on Monte Carlo simulations, Figure \ref{MC_invertibility} displays a region for a bivariate model ($N=2$) that satisfies the condition \eqref{X1_MC} on a grid of values $(\|\boldsymbol{\Phi}_0\|, \|\boldsymbol{K}_0\|)\in (0,1)^2$, with $\nu_0 = 7$ and $\boldsymbol{\Omega}_0=\boldsymbol{I}_2$.

\begin{figure}[H]
\centering
\includegraphics[width=0.7\textwidth]{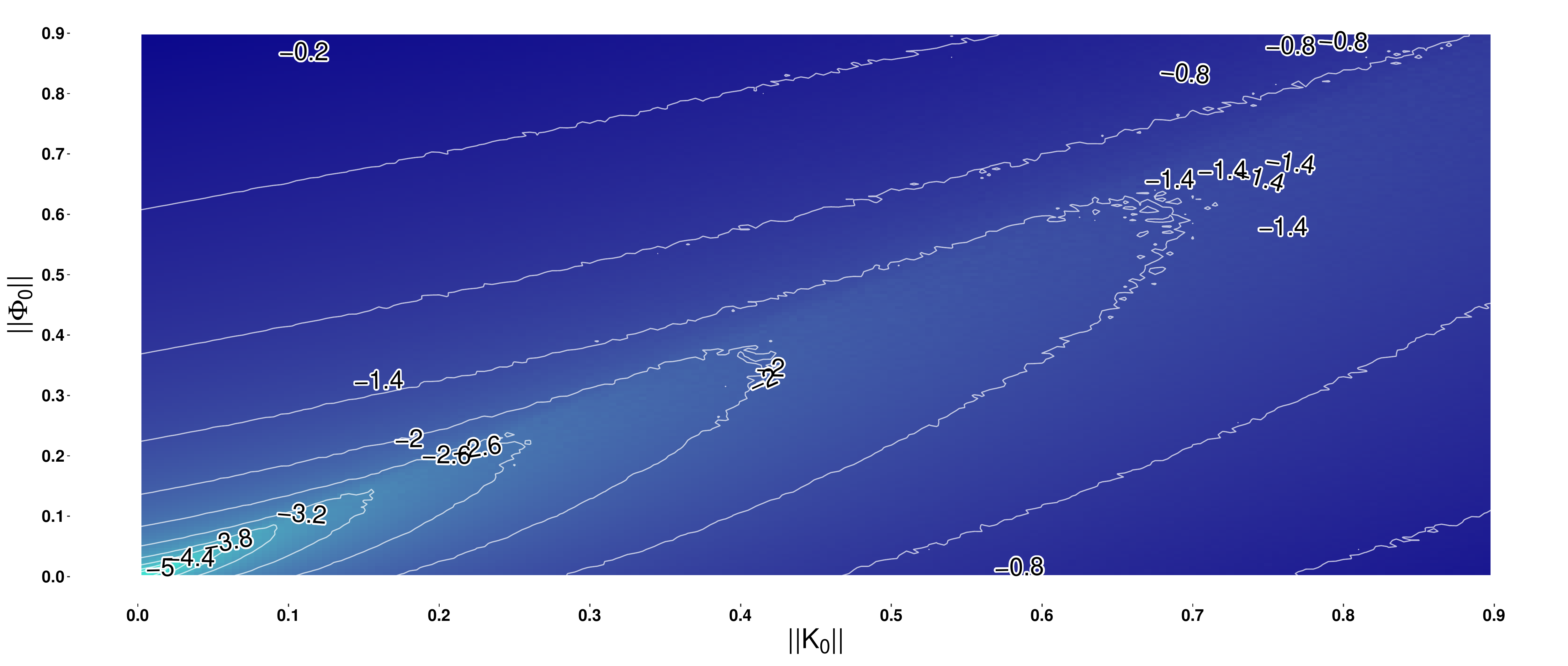} 
\caption{{Contour plot of the domain for invertibility.}}
\label{MC_invertibility}
\end{figure}

As expected, the restrictions that need to be imposed on $\| \boldsymbol{\Phi}_0 \|$ and $\| \boldsymbol{K}_0 \|$ are always stronger than those required for strict stationarity and ergodicity, see Lemma \ref{SE_Dynamic_Location}. Neverthless, the region depicted in Figure \ref{MC_invertibility} shows that a subset $\boldsymbol{\Theta^*}$ of the parameter space $\boldsymbol{\Theta}$ exists, with $\|\boldsymbol{\Phi}\| < 1 $ and $\|\boldsymbol{K}\|$ sufficiently small such that \eqref{suff_inv_contraction} is satisfied $\forall \boldsymbol{\theta} \in \boldsymbol \Theta^* \subset \boldsymbol{\Theta}$, producing a non degenerate invertibility region. 

In alternative to the simulation-based method, one can restrict the estimation procedure to the empirical version of the invertibility constraint in \eqref{inv_contraction} as in \cite{Wintenberger2013} and \cite{Blasques_Gorgi_etal2018}. The empirical counterpart of \eqref{inv_contraction} with $k=1$ is
\begin{align*}
\frac{1}{T}\sum_{t=1}^T
\ln
\bigg\|
\boldsymbol{\Phi} +&
\frac{\boldsymbol{K}}{1 + \boldsymbol{v}_t^\top \boldsymbol{\Omega}^{-1} \boldsymbol{v}_t/\nu}
\bigg(
\frac{2\boldsymbol{v}_t
\boldsymbol{v}_t^\top 
\boldsymbol{\Omega}^{-1}/\nu}{1 + \boldsymbol{v}_t^\top \boldsymbol{\Omega}^{-1} \boldsymbol{v}_t/\nu } -  
\boldsymbol{I}_N
\bigg)
\bigg\|
< -\delta,
\end{align*}
for some $\delta>0$ arbitrarily small.

%
%
%
%

To conclude, we note that the process $\{ \boldsymbol{y}_t \}_{t\in\mathbb{Z}}$  inherits some properties from those of the filter evaluated at the true parameter value. As a consequence of Lemma \ref{uniformly_boundedness_ut} and Lemma \ref{SE_Dynamic_Location}, we obtain the following result.

\begin{lemma}
\label{bounded_moments} 
Under the conditions of Lemma \ref{uniformly_boundedness_ut}  and Lemma \ref{SE_Dynamic_Location}, $\{ \boldsymbol{y}_t \}_{t\in\mathbb{Z}}$ is  stationary and ergodic. Moreover, $\forall m > \nu - \delta$, $\delta>0$, $\mathbb{E}[ \| \boldsymbol{y}_t \|^m ] < \infty$.
\end{lemma}

Finally,  the multi-step forecasts can be straightforwardly obtained as 
\begin{align*}
\mathbb{E}_T[\boldsymbol{y}_{T+l}] = 
\mathbb{E}_T[ \boldsymbol{\mu}_{T+l|T+l-1}] 
&= \boldsymbol{\omega} + 
\sum_{j=1}^{l-1} \boldsymbol{\Phi}^{\,j} (\boldsymbol{\mu}_{T+1|T} - \boldsymbol{\omega}).
\end{align*}

\section{Maximum Likelihood Estimation}
\label{Maximum_Likelihood_Estimation}
Let ${\ell}_t(\boldsymbol{\theta})$ denote the conditional $\log$-likelihood function for a single observation, obtained by taking the logarithm of \eqref{multivariate_t} considered as a function of the parameter
%
%
$\boldsymbol{\theta} = (\boldsymbol{\xi}^\top, \boldsymbol{\psi}^\top)^\top \in \boldsymbol{\Theta} \subset \mathbb{R}^p$, $\boldsymbol{\xi} = (\nu, (\vecth(\boldsymbol{\Omega}))^\top, \boldsymbol{\omega}^\top)^\top  \in \mathbb{R}^{s}$, with $s = 1 + \frac{1}{2}N(N+1) + N$ and $\boldsymbol{\psi} = ( (\vect\boldsymbol{\Phi})^\top, (\vect\boldsymbol{K})^\top)^\top\in \mathbb{R}^{d}$, with $d = (N \times N) + (N \times N)$ and hence, $p = s + d$. 


Lemma \ref{INV_Dynamic_location}, ensures that any choices of the initial condition  $\boldsymbol{\mu}_{1|0} \in \boldsymbol{\mathcal{M}}$ used for starting the filtering process are asymptotically equivalent, such that, once an initial value has been fixed, it is possible to obtain an approximated version of the conditional $\log$-likelihood, $\hat{\ell}_t(\boldsymbol{\theta})$, by replacing $\boldsymbol{\mu}_{t|t-1}$ in ${\ell}_t(\boldsymbol{\theta})$ by the filtered dynamic location $\hat{\boldsymbol{\mu}}_{t|t-1}$. Thus, for the whole sample, we obtain
$\hat{\ell}_T(\boldsymbol{\theta}) = \sum_{t=1}^{T} \hat{\ell}_t(\boldsymbol{\theta})$
and the MLE of $\boldsymbol{\theta}$ is  
\begin{equation*}
\widehat{\boldsymbol{\theta}}_T = \underset{\boldsymbol{\theta} \in \boldsymbol{\Theta}}{\arg\max} \, \hat{\ell}_T(\boldsymbol{\theta}).
\end{equation*}
We now discuss strong consistency and asymptotic normality of the MLE. 
%
The following assumptions are standard in the likelihood theory of non linear observation driven models.
\begin{assumption}\label{consistency_theorem_assumptions}
$\;$
\begin{enumerate}
\item \label{as1} The data generating process $\{ \boldsymbol{y}_t \}_{t \in \mathbb{Z}}$ is stationary and ergodic.
\item \label{as2} $\mathbb{E} [
\ln
\sup_{\boldsymbol{\theta} \in \boldsymbol{\Theta}} 
\sup_{\boldsymbol{\mu} \in \boldsymbol{\mathcal{M}}}
\| 
\prod_{j=1}^k
\boldsymbol{X}_{k-j+1}
\| ] < 0$ for $k \geq 1$.
\item \label{as3} The parameter space $\boldsymbol{\Theta}$ is compact with $0 < \nu < \infty$ and $\det \boldsymbol{K} \neq 0$.
\item \label{as4} The true parameter vector $\boldsymbol{\theta}_0$ belongs to the interior of $\boldsymbol{\Theta}$, i.e. $\boldsymbol{\theta}_0 \in \textit{int}(\boldsymbol{\Theta})$.
\item \label{as5} $\mathbb{E}[\| \boldsymbol{X}_t \otimes \boldsymbol{X}_t \|] < 1$.
\end{enumerate}
\end{assumption}
Assumption $4.1.1$ can be replaced by the conditions of Lemma \ref{bounded_moments}. Assumption $4.1.2$ ensures that the filtered sequence $\{\hat{\mu}_{t|t-1}\}_{t\in\mathbb{N}}$ converges to a stationary ergodic limit sequence, irrespective of the initial conditions. Assumptions $4.1.3$ and $4.1.4$ ensure the existence of the MLE and the validity of first order asymptotics. Assumption $4.1.5$ guarantees the existence of the information matrix.  

\begin{theorem}
\label{consistency_theorem}
Under conditions \ref{as1}--\ref{as4} in Assumption \ref{consistency_theorem_assumptions},
\begin{align*}
\hat{\boldsymbol{\theta}}_T 
\xrightarrow[]{\text{a.s.}} 
\boldsymbol{\theta}_0 
\tab
\text{as} \tab T \rightarrow \infty.
\end{align*}
\end{theorem}


\begin{theorem}
\label{asy_norm}
Under conditions \ref{as1}--\ref{as5} in Assumption \ref{consistency_theorem_assumptions},
\begin{align*}
\sqrt{T} ( \hat{\boldsymbol{\theta}}_T - \boldsymbol{\theta}_0 )
\xRightarrow[]{} 
\mathcal{N}(\boldsymbol{0}, \boldsymbol{\mathcal{I}}(\boldsymbol{\theta}_0)^{-1}),
\end{align*}
where,
\begin{align*}
\boldsymbol{\mathcal{I}}(\boldsymbol{\theta}_0) 
= 
- \mathbb{E} \bigg[ 
\frac{d^2 \ell_t(\boldsymbol{\theta})}{
d \boldsymbol{\theta} d \boldsymbol{\theta}^\top} 
\bigg|_{\boldsymbol{\theta} = \boldsymbol{\theta}_0}
\bigg]
\end{align*}
is the Fisher's Information matrix evaluated at the true parameter vector $\boldsymbol{\theta}_0$.
\end{theorem}


By Theorem \ref{consistency_theorem}, $\boldsymbol{\mathcal{I}}(\boldsymbol{\theta}_0)$ can be consistently estimated by
\begin{align}
\label{consistent_estimator_Fisher1}
\widehat{\boldsymbol{\mathcal{I}}}(\hat{\boldsymbol{\theta}}_T) 
=
- \frac{1}{T} \sum_{t=1}^{T}
\bigg[ 
\frac{d^2 \hat{\ell}_t(\boldsymbol{\theta})}{
d \boldsymbol{\theta} d \boldsymbol{\theta}^\top} 
\bigg|_{\boldsymbol{\theta} = \widehat{\boldsymbol{\theta}}_T}
\bigg].
\end{align}
As the dynamic location and its derivatives are nonlinear functions of the parameter $\boldsymbol{\theta}$, the general formula for the second  derivatives in \eqref{consistent_estimator_Fisher1} has the form below
\begin{align}
\label{chainrule_hessian}
\frac{d^2 \ell_t(\boldsymbol{\theta})}{
d \boldsymbol{\theta} d \boldsymbol{\theta}^\top} 
=&
\frac{\partial^2 \ell_t(\boldsymbol{\theta})}{\partial \boldsymbol{\theta} \partial \boldsymbol{\theta}^\top}
+ 
\bigg(
\frac{d (\boldsymbol{\mu}_{t|t-1} - \boldsymbol{\omega})}{d \boldsymbol{\theta}^\top}
\bigg)^\top
\frac{\partial^2 \ell_t(\boldsymbol{\theta})}{\partial \boldsymbol{\mu}_{t|t-1}
 \partial \boldsymbol{\mu}_{t|t-1}^\top}
\bigg(
\frac{d (\boldsymbol{\mu}_{t|t-1} - \boldsymbol{\omega})}{d \boldsymbol{\theta}^\top}
\bigg)\nonumber\\
&+
\frac{\partial \ell_t(\boldsymbol{\theta})}{\partial \boldsymbol{\mu}_{t|t-1}^\top}
\frac{d^2 (\boldsymbol{\mu}_{t|t-1} - \boldsymbol{\omega})}{d \boldsymbol{\theta}d \boldsymbol{\theta}^\top}.
\end{align}
To avoid the recursive evaluation of the second derivatives of the dynamic location vector, a simpler consistent estimator can be obtained based on the analytical form of the conditional information matrix $\boldsymbol{\mathcal{I}}_{t}(\boldsymbol{\theta})$, as in \cite{Fiorentini_Sentana_Calzolari2003}, defined as
\begin{align}
\label{cond_Fisher1}
\boldsymbol{\mathcal{I}}_{t}(\boldsymbol{\theta})
=
- \mathbb{E}_{t-1}
  \bigg[
\frac{d^2 \ell_t(\boldsymbol{\theta})}{
d \boldsymbol{\theta} d \boldsymbol{\theta}^\top}
\bigg].
\end{align}
Indeed, by the law of iterated expectations, one has 
\begin{align*}
\boldsymbol{\mathcal{I}}(\boldsymbol{\theta}) 
= 
\mathbb{E}
[\boldsymbol{\mathcal{I}}_{t}(\boldsymbol{\theta}) ]
=
-
 \mathbb{E}
  \bigg[
 \mathbb{E}_{t-1}
  \bigg[
\frac{d^2 \ell_t(\boldsymbol{\theta})}{
d \boldsymbol{\theta} d \boldsymbol{\theta}^\top} 
\bigg]
\bigg].
\end{align*}
Given the assumption of correct specification, the score vector evaluated at the true parameter vector $\boldsymbol{\theta}_0$ forms a martingale difference sequence, so that, under the assumptions of Theorem \ref{asy_norm}, asymptotic results for martingale difference sequences can be applied. In addition, the dynamic location (and its derivatives) are $\mathcal{F}_{t-1}$-measurable functions and therefore, after taking the conditional expectation, the last term in the right-hand-side of equation \eqref{chainrule_hessian} will cancel out.

It follows that, by Theorem \ref{consistency_theorem}, $\boldsymbol{\mathcal{I}}(\boldsymbol{\theta}_0)$ can be consistently estimated by
\begin{align*}
\widehat{\boldsymbol{\mathcal{I}}}( \widehat{\boldsymbol{\theta}}_T) 
= 
\frac{1}{T} \sum_{t=1}^{T}
 \widehat{\boldsymbol{\mathcal{I}}}_t( \widehat{\boldsymbol{\theta}}_T) ,
\end{align*}
where $\widehat{\boldsymbol{\mathcal{I}}}_t( \widehat{\boldsymbol{\theta}}_T) $ is the conditional information matrix in \eqref{cond_Fisher1} evaluated at the filtered dynamic location $\hat{\boldsymbol{\mu}}_{t|t-1}$ and at the MLE $\widehat{\boldsymbol{\theta}}_T$. The analytical form of $\boldsymbol{\mathcal{I}}_{t}(\boldsymbol{\theta})$ is derived in section S2.3.

\subsection{Computational Aspects}
\label{Fisher_Scoring_Section}
ML estimation and inference are carried out by means of Fisher's scoring method. A strongly reliable algorithm based on analytical formulae for the score vector and the Hessian matrix (reported in Appendix S2) is developed, which can be directly implemented in any statistical package through the following steps:
\begin{enumerate}
\item Choose a starting value $\widehat{\boldsymbol{\theta}}_T^{(0)} = ( \nu^{(0)}, (\vecth(\boldsymbol{\Omega}^{(0)}))^{\top}, (\boldsymbol{\omega}^{(0)})^\top, (\vect(\boldsymbol{\Phi}^{(0)}))^{\top}, (\vect(\boldsymbol{K}^{(0)}))^{\top} )^\top$
\item For $h>0$, update $\widehat{\boldsymbol{\theta}}_T^{(h)}$ using the scoring rule
$
\widehat{\boldsymbol{\theta}}_T^{(h+1)} = \widehat{\boldsymbol{\theta}}_T^{(h)} 
+ 
\big[
\widehat{\boldsymbol{\mathcal{I}}}_T(\widehat{\boldsymbol{\theta}}_T^{(h)})\big]
^{-1}
\widehat{\boldsymbol{s}}_T(\widehat{\boldsymbol{\theta}}_T^{(h)}),
$
where 
$
\boldsymbol{s}_T(\boldsymbol{\theta})
 = 
 \sum_{t=1}^{T}
\frac{d \ell_t (\boldsymbol{\theta})}{d \boldsymbol{\theta}}
\,\,\,\,\, \textit{and} \,\,\,\,\, 
\boldsymbol{\mathcal{I}}_{T}(\boldsymbol{\theta}) 
=
-\sum_{t=1}^{T}
 \mathbb{E}_{t-1} \left[ 
\frac{d^2 \ell_t(\boldsymbol{\theta})}{
d \boldsymbol{\theta} d \boldsymbol{\theta}^\top} \right].
$
\item Repeat until convergence, i.e.,
$
 \big\| \widehat{\boldsymbol{\theta}}_T^{(h+1)} - \widehat{\boldsymbol{\theta}}_T^{(h)}  \big\| /
\big\| \widehat{\boldsymbol{\theta}}_T^{(h)} \big\| < \delta
$
for some fixed $\delta > 0$.
\end{enumerate}
The analytical expressions for the score vector and the conditional information matrix used in step $2$ are in Section S2.

\subsection{Initial conditions}
To initialise the estimation procedure, we  follow the approach suggested in \cite{Fiorentini_Sentana_Calzolari2003}. First, a consistent estimator of the restricted version of the parameter vector $\tilde{\boldsymbol{\theta}}_T$ is obtained by the Gaussian quasi-ML procedure in \cite{Bollerslev_Wooldridge1992}. Second, a consistent method of moments is adopted for the degrees of freedom $\nu$, by making use of the empirical coefficient of excess kurtosis $\tilde{\kappa}$ on the standardized residuals and of the relation $\tilde{\nu} = (4\tilde{\kappa} + 6)/\tilde{\kappa}$. Convergence is fast in that usually few iterations of that procedure are needed, which makes scoring methods particularly appealing for estimation purposes.

\subsection{Monte Carlo analysis}
In section S1, we report the details of a Monte Carlo study aimed to assess the finite sample properties of the MLE based on the Fisher's scoring method detailed in the above section. In summary,  our approach performs well in terms of bias and root mean square errors for a wide range of time series, from the most severe heavy-tailed case (i.e., $\nu$ very small) to the Gaussian case (i.e. for $\nu\rightarrow \infty$), thus covering also the case of potential misspecification. In addition, we note that it delivers satisfactory results even when the number of iterations of the algorithm is limited to ten rounds.

\section{Empirical Analysis of Homescan Data Consumer Prices}
\label{Empirical_Applications}


In order to demonstrate a potential use of the robust score-driven filter, we show  an innovative application to the estimation of consumer prices from homescan data. This field of application is gaining interest, due to the growing availability of high frequency and high detail purchase data collected through scanner technologies at the retail point (retail scan) or household level (homescan). The latter of type of data allows one to obtain cost-of-living measures for vulnerable sub-groups of the population, and to explore the distributional effects of fiscal measures. While being a valuable source for detailed price information, post-purchase homescan price data are affected by a measurement noise that can be potentially large in small samples, and the application of filtering techniques may help to mitigate such noise and control for outliers. 

Scanner data are collected either at the retail level, e.g. supermarket data, or from households in consumer panels, i.e. homescan data. Retail scanner data are widely used to estimate prices, both for continuity with the traditional price survey methodology, and because they are expected to suffer less from the substitution (unit value) bias (\cite{Silver2001}). This bias is due to the fact that scanner data are based on actual transactions, i.e. prices are only observed after the consumer purchases the good. This implies that the observed price embodies a quality choice component, as consumers confronted with a price increase may opt for a cheaper option (or a cheaper retailer) and information on non-purchased items is missing. The bias can be particularly important for aggregated goods, such as those goods commonly represented by category-level prices like food and drinks. Thus, a wide body of research has been devoted to improve sampling strategies and the choice of weights in aggregation. A well-documented problem is the change in the composition of the consumption basket over time, an issue that can be exacerbated by high-frequency data \cite{Feenstra2003b}. For example, stockpiling of goods during promotion periods generate bias in price indices, as the purchased quantities are not independent over subsequent time periods \cite{Ivancic2011,Melser2018}. 

Although supermarket-level scanner data allow to mitigate the problem, as one expects a wide range of products to be purchased across the population of customers within a given time period, the use of homescan data to estimate prices and price indices has potentially major advantages. These advantages lie in the possibility to exploit household-level heterogeneity. Most importantly, it becomes feasible to estimate prices faced by particular population sub-groups whose consumption basket differs from the average one, as elderly households or low-income groups \cite{Kaplan2017,Broda2009}. However, the unit value issue is heavier with homescan data, as individual households buy a small range of products. Thus, variable shopping frequencies and zero purchases make it necessary to rely on very large samples of households to control the bias. The problem becomes even more conspicuous for prices at the regional level, for products that are not frequently purchased and for products whose demand is highly seasonal. 

Robust filtering techniques may constitute a powerful solution to the above mentioned problems, and may perform well even with relatively small samples of household as the one used in our application. 

To illustrate the potential contribution of the proposed method, we exploit a data-set that has been recently used to evaluate the effects of a tax on sugar-sweetened beveraged introduced in France in 2012 \cite{Capacci2019}. Our data consists of weekly scanner price data for food and non-alcoholic drinks. The data were collected in a single region, within the Italian GfK homescan consumer panel, based on purchase information on 318 households surveyed in the Piedmont region, over the period between January 2011 and December 2012. The regional scope and the relatively small sample provide an ideal setting to test the applicability and effectiveness of the multivariate filtering approach.

\begin{table}[h]
\caption{Average unit values, \euro \; per kilogram, Piedmont homescan data (standard deviations in brackets)}     			
\begin{center}													
\begin{tabular}{ l c c c c }  									
\toprule
\midrule
	&	\multicolumn{2}{c}{$2011$}		&	\multicolumn{2}{c}{$2012$}		\\		
\midrule									
Food	&	$4.343$	&	$(0.234)$	&	$4.226$	&	$(0.255)$	\\
Non-alcoholic drinks	&	$0.434$	&	$(0.047)$	&	$0.426$	&	$(0.052)$	\\
Coca-Cola	&	$1.000$	&	$(0.096)$	&	$1.100$	&	$(0.172)$	\\
\bottomrule									
\end{tabular}       
\end{center}							
\label{tab:descriptives}									
\end{table}	

\subsection{Data}

The data for our application consist of three time series of weekly unit values for food items, non-alcoholic drinks and Coca-Cola purchased by a sample of 318 households residing in the Piedmont region, Italy, over the period 2011-2012, and collected within the GfK Europanel homescan survey. The data-set provides information on weekly expenditures and purchased quantities for each of the three aggregated items, and unit values are obtained as expenditure-quantity ratios. 

Average unit values are shown in Table \ref{tab:descriptives}. Food and non-alcoholic drinks are composite aggregates, hence they are potentially subject to fluctuations in response to changes in the consumer basket even when prices are stable. Instead, Coca-Cola is a relatively homogeneous good, with little variability due to different packaging sizes.

\subsection{Results}

We fit the multivariate score-driven model developed in the paper to the considered vector of time series. ML estimation produces the following multivariate dynamic system of time varying locations for Drinks (D), Food (F) and Coca-Cola (C),

\begin{align*}
\hat{\boldsymbol{\omega}}=
\begin{bmatrix}
0.443   \\
\scriptstyle{(0.000)} \\
4.394   \\
\scriptstyle{(0.000)} \\
- 1.070   \\
\scriptstyle{(0.000)} 
\end{bmatrix}
\quad 
\hat{\boldsymbol{\Phi}}=
\begin{bmatrix}
     0.839 &    0.015    &  0.007  \\
        \scriptstyle{(0.011)}  &     \scriptstyle{(0.002)}  &     \scriptstyle{(0.005)}\\
     -0.528 &    0.912    &  0.342  \\
        \scriptstyle{(0.059)}  &     \scriptstyle{(0.009)}  &     \scriptstyle{(0.025)}\\
     0.222 &    0.023    &  0.847   \\
        \scriptstyle{(0.020)}  &     \scriptstyle{(0.003)}  &     \scriptstyle{(0.009)}\\
\end{bmatrix}
\quad 
\hat{\boldsymbol{K}}=
\begin{bmatrix}
     0.442 &    -0.023    &  0.007  \\
        \scriptstyle{(0.017)}  &     \scriptstyle{(0.003)}  &     \scriptstyle{(0.007)}\\
     0.334 &    0.216    &  -0.631  \\
        \scriptstyle{(0.079)}  &     \scriptstyle{(0.014)}  &     \scriptstyle{(0.038)}\\
     -0.290 &    -0.098    &  -0.014   \\
        \scriptstyle{(0.030)}  &     \scriptstyle{(0.005)}  &     \scriptstyle{(0.014)}\\
\end{bmatrix}
\end{align*}
where the values in parenthesis are the standard errors and with
\[
\hat{\nu}= 6.921 
\,\,\,{(0.229)} ,
\tab
\hat{\boldsymbol{\Omega}}=
\begin{bmatrix}
     0.162 &    \cdot    &  \cdot  \\
\scriptstyle{(0.138)}  & &  				\\
     0.348 &    53.258    &  \cdot  \\
\scriptstyle{(0.913)} & \scriptstyle{(0.327)} & \\
     -0.134 &    -0.579    &  9.086   \\
\scriptstyle{(0.057)} & \scriptstyle{(0.327)} & \scriptstyle{(0.155)} 
\end{bmatrix}
\times 10^{-3}.
\]

The estimated degrees of freedom are approximately $7$. We remark that the assumption of a (conditional) multivariate Student's \emph{t} distribution implies that all the univariate marginal distributions are tail equivalent, see \cite{Resnick2004}. This requires the implicit underlying assumption that the level of heavy-tailedness across the observed time series vector is fairly homogeneous. To investigate this issue, and for the sake of comparisons, we have carried out a univariate analysis, as in \cite{HarveyLuati2014}, from which it resulted that the estimated degrees of freedom were very low for Coca-Cola (about $4$) and medium size (smaller than $30$) for the other two series, as expected. Hence, the multivariate score-driven model developed in the paper reveals to be a good compromise between a multivariate non-robust filter, based on a linear Gaussian model, and a robust univariate estimator. Indeed, a multivariate Portmanteau test on the residuals obtained from the three univariate models is carried out to test the null hypothesis $H_0 : \boldsymbol{R}_1 = \dots = \boldsymbol{R}_m = \boldsymbol{0}$, where $\boldsymbol{R}_i$ is the sample cross-correlation matrix for some $i \in \{1, \dots, m\}$ against the alternative $H_1 : \boldsymbol{R}_i \neq \boldsymbol{0}$. The results of Table \ref{Portmanteau_Piedmont} indicate rejection of the null hypothesis of absence of of serial dependence in the trivariate series at the $5\%$ significance level. 

\begin{table}[h]
\caption{\emph{Multivariate Portmanteau test.}}
\begin{center}
\begin{tabularx}{0.5\linewidth}{ @{\extracolsep{\fill}} cc @{\extracolsep{\fill}} cc}
\toprule
\midrule
 \multicolumn{1}{c}{{$m$}} 
& \multicolumn{1}{c}{{$Q(m)$}}
& \multicolumn{1}{c}{{\emph{df}}}
&\multicolumn{1}{c}{{\emph{p-value}}}  \\
\midrule
$1$     &$13.7$ 	&$9$     &$0.000$       \\
$2$     &$40.8$ 	&$18$     &$0.000$      \\
$3$     &$58.6$ 	&$27$    &$0.000$       \\
$4$     &$89.6$ 	&$36$    &$0.000$      \\
$5$     &$105.9$ 	&$45$    &$0.000$       \\
\bottomrule
\end{tabularx}
\label{Portmanteau_Piedmont}
\end{center}
\end{table}

We also remark that the estimated degrees of freedom close to $7$ rule out the hypothesis that the data come from a linear Gaussian state-space model, in which case the estimated degrees of freedom would be definitely higher. Nevertheless, we have fitted a misspecified linear Gaussian state-space model estimated with the Kalman filter and, as expected, along with a higher sensitivity to extreme values, in particular in the last period of the Coca-Cola series, likelihood and information criteria are in favour of the  multivariate model based on the conditional Student's $t$ distribution. 

\begin{table}[!h]
\caption{\emph{{Likelihood, Akaike and Bayesian information criteria.}}}
\begin{center}
\label{MisTest}
\begin{tabularx}{\linewidth}{ @{\extracolsep{\fill}} lccc}
\toprule
\midrule
& \multicolumn{1}{c}{{$\log$-\emph{Lik}}} 
& \multicolumn{1}{c}{{\emph{AIC}}} 
& \multicolumn{1}{c}{{\emph{BIC}}}\\
\midrule
\emph{KF} &$241.16$    &-$434.32$ 	&-$370.85$  \\
\emph{DCS-t }   &$257.93$     &-$465.69$ 	&-$402.23$      \\
\bottomrule
\end{tabularx}
\end{center}
\end{table}

The matrix of the estimated autoregressive coefficients $\hat{\boldsymbol{\Phi}}$ measures the dependence across the filtered dynamic locations $\hat{\boldsymbol{\mu}}_{t|t-1}$, while the estimated scale matrix $\hat{\boldsymbol{\Omega}}$ measures the concurrent relationship between the three series under investigation, i.e. drink, food and Coca-Cola prices. For these matrices, we report the estimates of the coefficients and, in parenthesis, the relative standard errors. The diagonal elements of $\hat{\boldsymbol{\Phi}}$ show that each variable of interest is highly persistent. In order to explore the relation among the series, we implement an impulse response analysis. Figure  \ref{Piedmont_IRF_filters} shows the estimated impulse response functions.

\begin{figure}[h]
\centering
\includegraphics[width=0.96\textwidth]{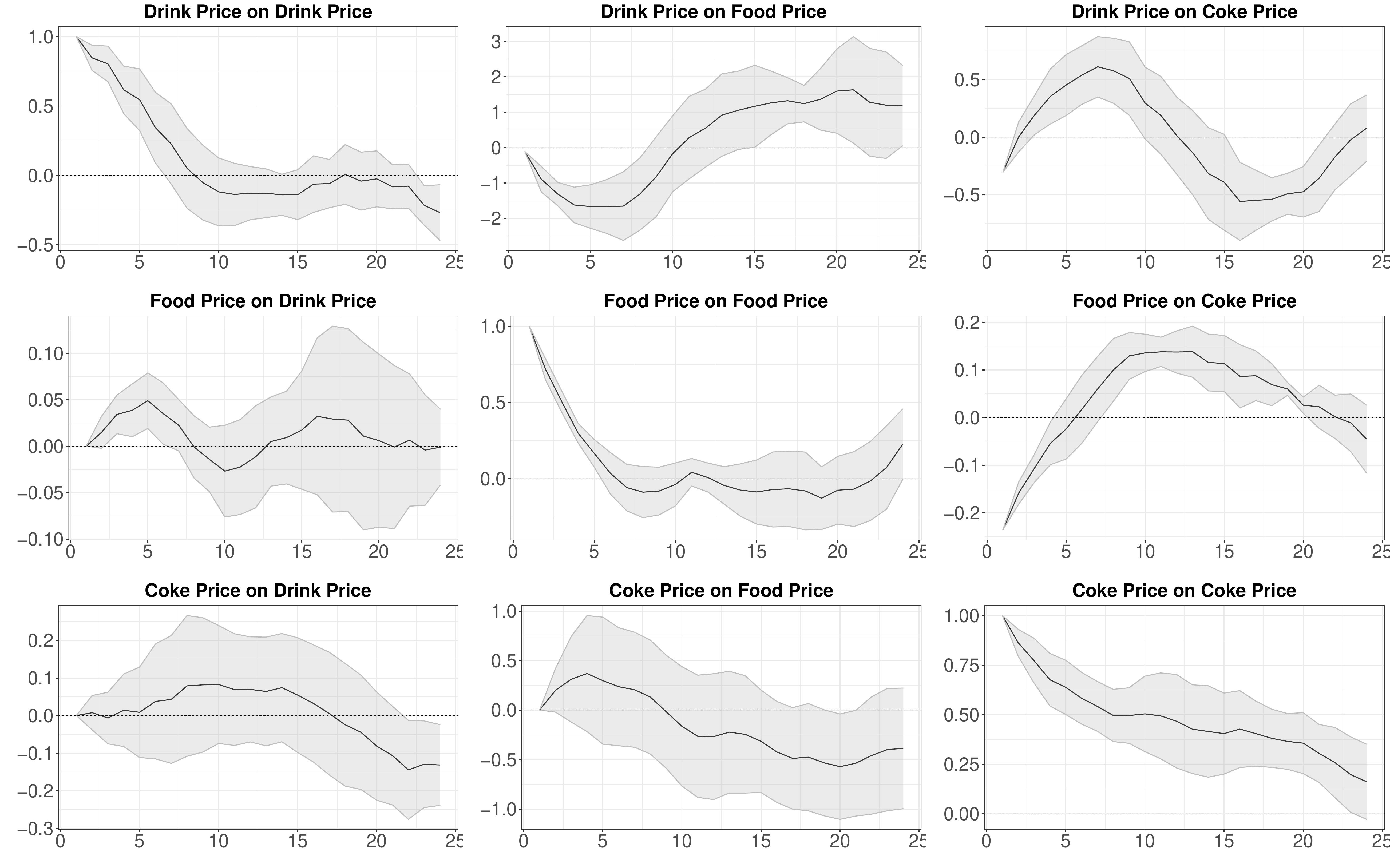} 
\caption{{Estimated impulse response functions of the filtered $\hat{\boldsymbol{\mu}}_{t|t-1}$ for a unit shock.}}
\label{Piedmont_IRF_filters}
\end{figure}
The nonlinear impulse are computed by using the local projections approach of \cite{Jorda2005}, and the confidence bands are obtained by using the Newey-West corrected standard-errors, see \cite{Newey1987}. What emerges is a negative relation between drink and food prices: a unit shock in drink prices will produce a negative shock in food prices. This may adjustments in purchasing decisions by the households aimed at mitigating the rising cost of their shopping basket. This would be evidence that univariate signals are likely to suffer from the unit value bias. Similarly, a non trivial negative relation exists between food and Coca-Cola prices. A unit shock on food prices yields a concurrent negative impact on Coca-Cola prices, which is also noted from the analysis of the cross-correlations. As one might expect, a positive correlation exists between Coca-Cola prices and drink prices, as the former product belongs to the latter category. Instead, unit shocks on food prices seem to have negligible correlation (if any) on drink prices.

\subsection{Interpretation}

Figure \ref{fig:DCSt} shows the original unit value time series and the corresponding signals extracted through the multivariate score-driven filter. Noise and outliers, as well as some irregular periodic pattern, are clearly visible in the drinks and food series. On the other hand, the Coca-Cola series is relatively regular, with the exception of few peaks, including a couple of large outliers in the second year. Given the homogeneous nature of the good, it is reasonable to believe that those extreme values are the results of measurement error. 

\begin{center}
\begin{figure}[h]
\centering
\includegraphics[width=0.8\textwidth]{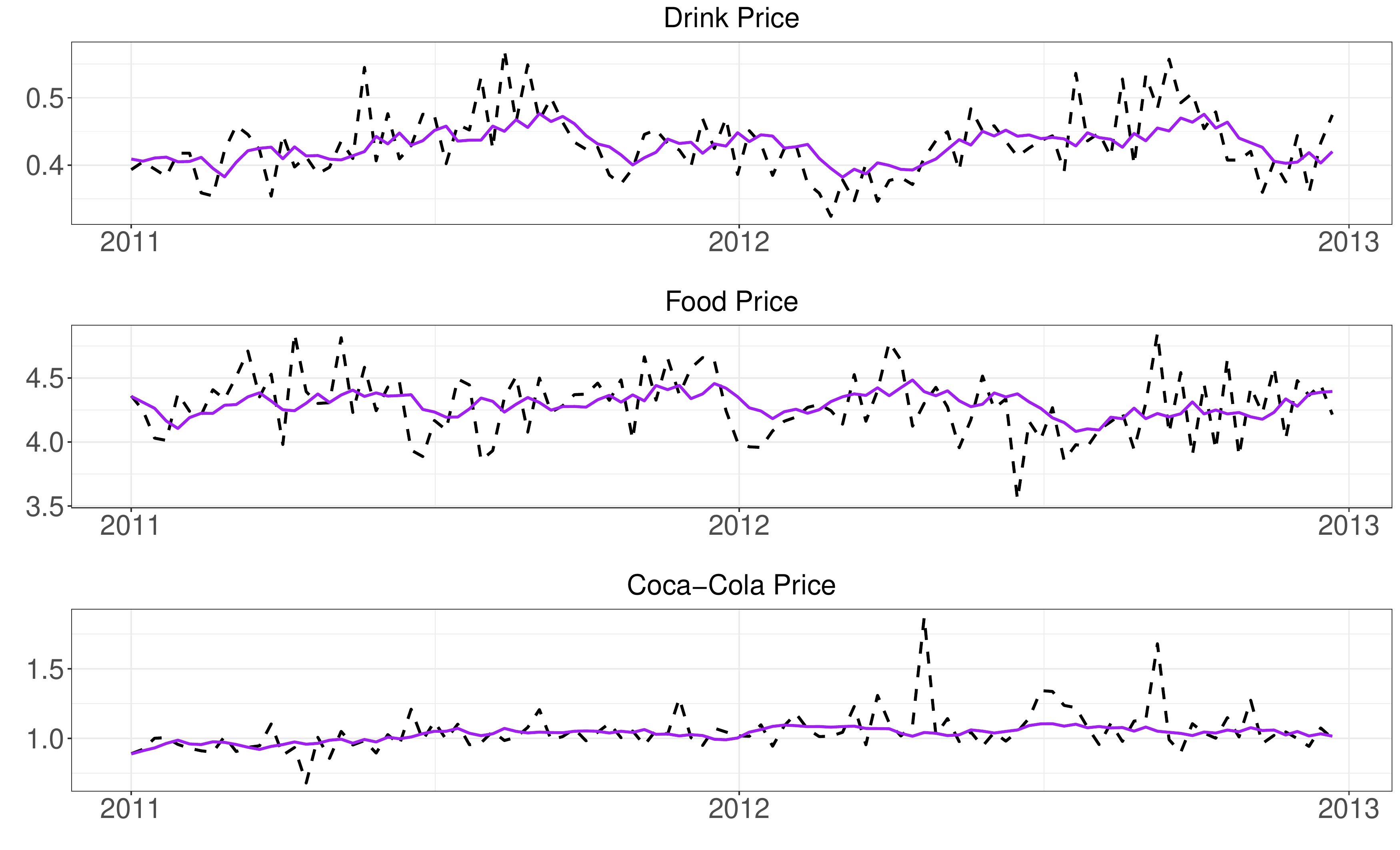}
\caption{{Original series (dotted line) and estimated signals}}
\label{fig:DCSt}
\end{figure}
\end{center}

\begin{center}
\begin{figure}[H]
\includegraphics[width=0.7\textwidth]{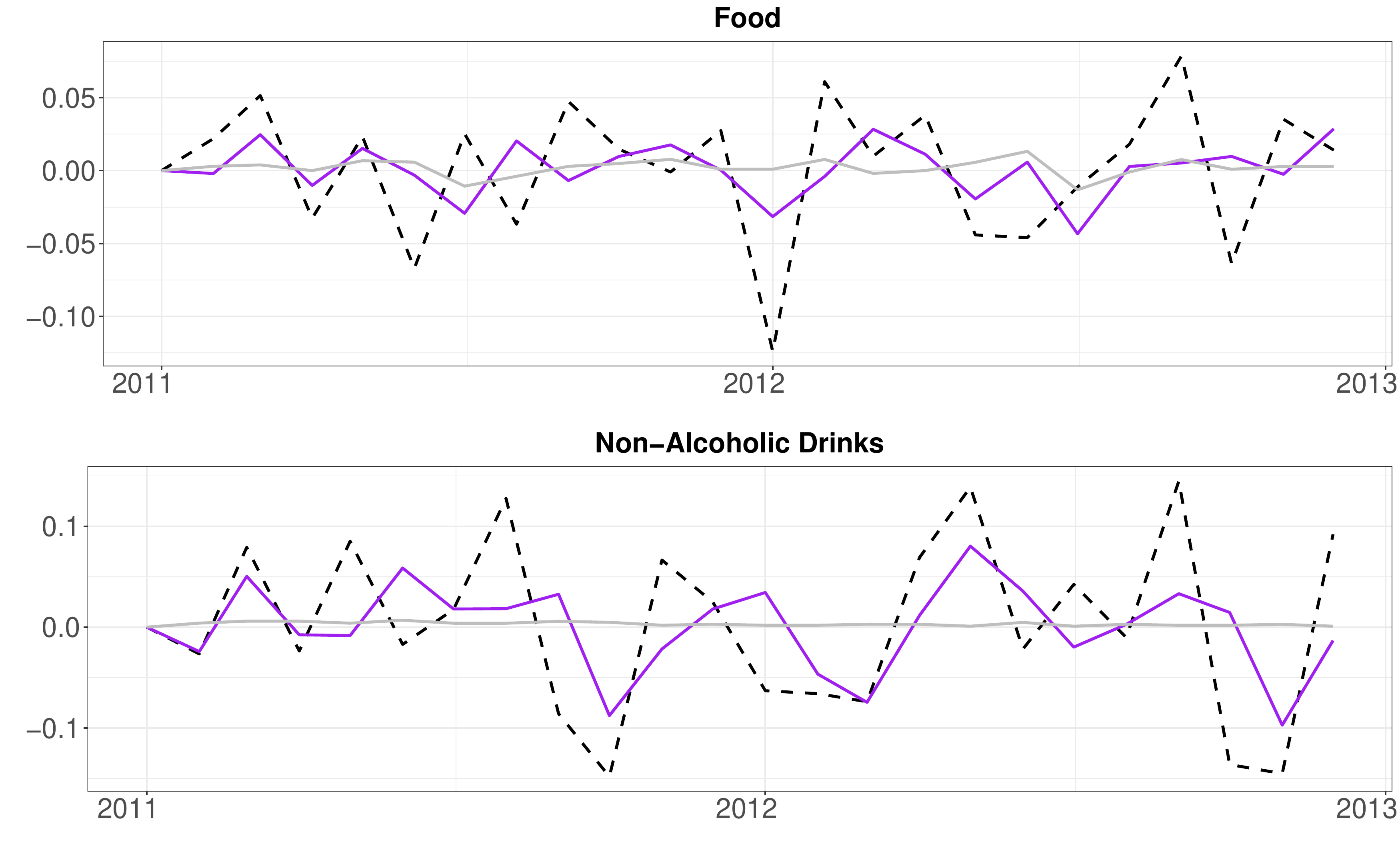}
\centering
\caption{{Raw unit value series (dotted line), estimated signal and Regional CPIs (log differences, grey line)}}
\label{fig:Figure2}
\end{figure}
\end{center}


The estimates illustrate an effective noise reduction and return patterns that are smoother and more consistent with a regular price time series. As one would expect, the Coca-Cola DCS-\emph{t} series is very flat, and suggests a relatively stable price over the two-years time window, with no outliers.

Figure \ref{fig:Figure2} shows the monthly natural logarithm differences of the raw homescan prices (HSP) and the estimated signals, together with changes in the official Regional CPIs (R-CPI) for food and non-alcoholic drinks, whereas no CPI to the brand detail is produced. The R-CPIs are provided by the National Statistical Institute (ISTAT). They have a monthly frequency and are built with a traditional survey-based approach on retailers. The comparison between the score-driven filtered values and the R-CPIs is purely indicative, as the unit values from the homescan data are weekly, whereas the official CPIs are monthly. This frequency difference may lead to biased comparisons \cite{Diewert2016}. Nevertheless, the graphs confirm that the score-driven signals are effective in reducing the noise in the data. This is especially true for the food series, whose CPIs are more volatile compared to drinks. The correlation between the raw homescan log-differenced unit value and the log-differenced food CPI is 0.05, against 0.44 when the filtered time series is considered. For the non-alcoholic drinks price series the gain is less conspicuous, as prices evolve very regularly over the time window. Still, an inexistent correlation between the HSP and the R-CPI (-0.02) turns into a positive one (+0.11) when considering the score-driven estimates and the R-CPI.

In essence, the empirical evidence suggests that a robust multivariate approach to model-based signal extraction produce meaningful price series from homescan data, especially when noise and outliers in the original data are relevant. We find the approach to perform reasonably well even with a low number of sampled households (318) and price time series (3), and with a relatively short time window (104 weeks). Future research might shed further light on the implications of dealing with a larger number of price series and longer time series.

\section{Concluding Remarks}
\label{Concluding_Remarks}

We developed a nonlinear and multivariate dynamic location filter which enables the extraction of reliable signals from vector processes affected by outliers and possibly non-Gaussian errors. Its peculiarity lies in the specification of a score-robust updating equation for the time-varying conditional location vector. Compared to the existing literature on observation driven models for time varying parameters, the model has two innovative features: (a) it extends the univariate first-order dynamic conditional location score by \cite{HarveyLuati2014} to the multivariate setting; and (b) it extends the dynamic model for time varying volatilities and correlations by \cite{CrealKoopLucas2011} to the location case. 

We derived the stochastic properties of the filter and, under correct specification, of the data generating process: bounded moments, stationarity, ergodicity, and filter invertibility. Parameters are estimated by ML and we provided closed formulae for the score vector and the Hessian matrix, which can be directly used for a scoring procedure. Consistency and asymptotic normality have been proved and a Monte-Carlo study showed good and reliable finite sample properties. In the case when the degrees of freedom tend to infinity, or, in practice, their estimate is of the order of hundreds, our specification converges to a linear and Gaussian model.  

The empirical application showed that robust filtering may lead to satisfactory estimates of price signals from homescan data, in the case when the multivariate dimension is low. We contribute to research in this area with two promising results. First, we show that robust  modeling allowing for heavy tails is more effective in dealing with noisy series affected by outliers or extreme observations. Second, the multivariate extension of the DCS-\emph{t} model has shown more appropriate than the robust univariate filtering approach in the case of scanner price data, as price time series are expected to have a good degree of correlation. This proves to be valuable information to reduce the noise across the modelled price time series.

\bibliography{winnower_template}
\bibliographystyle{chicago}


\bigskip
\begin{center}
{\large\bf SUPPLEMENTARY MATERIAL}
\end{center}

Additional supporting information may be found in the online appendix for this article at the publisher's website.

    \section*{Appendix A: Main Proofs}
    \label{appendix_main_proofs}


\subsubsection*{Proof of Lemma \ref{uniformly_boundedness_ut}} 
%
The score $\boldsymbol{u}_t$ in equation (\ref{u_t}) can be written as
\begin{equation}
\label{u_t_pred_err}
\boldsymbol{u_t} = \boldsymbol{v}_t(1-b_t)
\end{equation}
with $b_t = 1 - 1/w_t$ and where, conditional to $\mathcal F_{t-1}$, 
\begin{equation}
\label{b_t}
b_t = \frac{ 
\boldsymbol{v}_t^\top 
\boldsymbol{\Omega}^{-1}
\boldsymbol{v}_t / \nu}
{1 + \boldsymbol{v}_t^\top 
\boldsymbol{\Omega}^{-1} 
\boldsymbol{v}_t/ \nu}, 
\,\,\,\,\,  0 \leq b_t \leq 1,
\,\,\,\,\, \text{with} \,\,\,\,\,
 b_t \sim \mathcal{B}\textit{eta}
\Big(\frac{N}{2}, \frac{\nu}{2} \Big),
\end{equation}
i.e. the driving force $\boldsymbol{u}_t$ is a continuous function of a beta distributed random variable, see Pag. 19 of \cite{Kotz_Nadarajah2004} or Proposition 39 of \cite{Harvey2013}.
For $0 < \nu < \infty$, $\| \boldsymbol{u}_t \| = 0$ if $ \| \boldsymbol{v}_t \| = 0$, while $\| \boldsymbol{u}_t \| \rightarrow 0$ if $\| \boldsymbol{v}_t \| \rightarrow \infty$ because $b_t \rightarrow 1$. Therefore, we achieve that $\sup_t \mathbb{E}[\| \boldsymbol{u}_t \| ] < \infty$.

Second, we retrieve the moment structure of $\boldsymbol{u}_t$. Under assumption \ref{ass:cs}, the following stochastic representation is valid for the driving force 
\begin{align}
\label{stochastic_rep2}
\boldsymbol{u}_t =\sqrt{\nu} \sqrt{b_t (1-b_t) } 
 \boldsymbol{\Omega}^{1/2}  \mathbf{z}_t,
\end{align}
where $\mathbf{z}_t$ is uniformly distributed on the unit sphere in $\mathbb{R}^N$ independently of $b_t$, see \cite{Fang1990}. It follows that for even integers $m =2s, s=1,2, \dots$, the moments of $\boldsymbol{u}_t$ can be expressed as 
\begin{align*}
\mathbb{E}
\Big[ \| \boldsymbol{u}_t \|^m \Big] 
=&
\nu^{m/2}
\| \boldsymbol{\Omega} \|^{m/2} 
\mathbb{E} 
\Big[  b_t^{m/2} (1-b_t)^{m/2} \Big]
\mathbb{E}
\Big[ \| \mathbf{z}_t \|^{m} \Big] \nonumber\\
=&
\frac{\| \boldsymbol{\Omega} \|^{m/2} }
{B\big( \frac{N}{2} , \frac{\nu}{2} \big)}
\Big( \frac{\nu}{N} \Big)^{m/2}  \int b_t^{\frac{N+m}{2}-1} (1-b_t)^{\frac{\nu+m}{2}-1} \mathrm{d} b_t \nonumber\\
=&
\| \boldsymbol{\Omega} \|^{m/2} 
\Big( \frac{\nu}{N} \Big)^{m/2}
\frac{B\big( \frac{N+m}{2} , \frac{\nu+m}{2} \big)}
{B\big( \frac{N}{2} , \frac{\nu}{2} \big)}. \, \square
\end{align*}


\subsubsection*{Proof of Lemma \ref{SE_Dynamic_Location}}
It follows from Lemma \ref{uniformly_boundedness_ut}, that, at $\boldsymbol{\theta} = \boldsymbol{\theta}_0$,  the score $\boldsymbol{u}_t$ forms a martingale difference sequence with zero mean and time-invariant covariance matrix. This implies that the process $\{ \boldsymbol{u}_t \}_{t\in\mathbb{Z}}$ is \emph{IID} and hence, independently distributed of $\boldsymbol{\mu}_{t|t-1}$. Therefore, by using recursive arguments, for each starting value $\boldsymbol{\mu}_{s|s-1}$, where $s$ is a fixed time point, one has that 
$
\boldsymbol{\mu}_{t+1|t} - \boldsymbol{\omega}
=
\boldsymbol{\Phi}^{t-s}(\boldsymbol{\mu}_{s|s-1} - \boldsymbol{\omega})
+
\sum_{j=0}^{t-1} \boldsymbol{\Phi}^{j}\boldsymbol{K} \boldsymbol{u}_{t-j}.
$
Consequently, according to the theory of linear systems, see \cite{Hannan_Deistler1987}, the condition $\varrho(\boldsymbol{\Phi}) < 1$ is sufficient for the existence and uniqueness of a strictly stationary and ergodic solution $\{ \tilde{\boldsymbol{\mu}}_{t|t-1} \}_{t \in \mathbb{Z}}$.

Then, when the process starts from the infinite past, we can write
$
\tilde{\boldsymbol{\mu}}_{t+1|t} - \boldsymbol{\omega}
=
\sum_{j=0}^\infty  
\boldsymbol{\Phi}^j \boldsymbol{K} \boldsymbol{u}_{t-j},
$
so that, from Lemma \ref{uniformly_boundedness_ut}, by taking the unconditional expectation and applying the triangle, H{\"o}lder and Minkowsky inequalities, we get 
\begin{align*}
\mathbb{E} \bigg[ 
\| \tilde{\boldsymbol{\mu}}_{t+1|t} - \boldsymbol{\omega} \|^m  \bigg] 
&=
\mathbb{E} \bigg[ 
\bigg\| 
\sum_{j=0}^\infty  
\boldsymbol{\Phi}^j \boldsymbol{K} \boldsymbol{u}_{t-j} 
\bigg\|^m  \bigg]  
\leq
\bigg\{
\bar{c}
\sum_{j=0}^\infty
\bar{\rho}^j 
\bigg(
\mathbb{E} \Big[ 
\| 
\boldsymbol{u}_{t-j} 
\|^{m}
 \Big]
\bigg)^{1/m} \bigg\}^{m} 
< \infty,
\end{align*}
where $\bar{c}= {N} \| \boldsymbol{K} \|$ and $\bar{\rho} < 1$. The first inequality follows from a standard result in linear algebra, as $\|\boldsymbol{\Phi}\| = \|\boldsymbol{P} \boldsymbol{\Lambda} \boldsymbol{P}^{-1} \| = \tr( \boldsymbol{\Lambda}) = \sum_{i=1}^N \rho_i $ where $\rho_i$ are the eigenvalues of $\boldsymbol{\Phi}$. $\square$

\subsubsection*{Proof of Lemma \ref{INV_Dynamic_location}}
The stationary and ergodic solution of equation \eqref{score_driven_filter} can be embedded in a first order nonlinear dynamic system.
$
\tilde{\boldsymbol{\mu}}_{t+1|t} = \phi (\tilde{\boldsymbol{\mu}}_{t|t-1}, \boldsymbol{y}_t, \boldsymbol{\theta}),
 t \in \mathbb{Z}.
$
Let us define inductively, for $k \geq 1$ and any initialization $\hat{\boldsymbol{\mu}}_{1|0} \in \boldsymbol{\mathcal{M}}$, a sequence of Lipschitz maps  $\phi^{(k+1)}: \boldsymbol{\mathcal{M}} \times \mathbb{R}^N \times \boldsymbol{\Theta} \mapsto \boldsymbol{\mathcal{M}}$ for $k \geq 1$ such that
$
\phi^{(k+1)}(\hat{\boldsymbol{\mu}}_{1|0}, \boldsymbol{y}_{1}, \dots, \boldsymbol{y}_{k+1}, \boldsymbol{\theta})
=
\phi \, (\phi^{(k)}(\hat{\boldsymbol{\mu}}_{1|0}, \boldsymbol{y}_{1}, \dots, \boldsymbol{y}_{k}, \boldsymbol{\theta}), 
\boldsymbol{y}_{k+1}, \boldsymbol{\theta} ).
$
By applying the mean value theorem to $\phi (\hat{\boldsymbol{\mu}}_{t|t-1} , \boldsymbol{y}_{t}, \boldsymbol{\theta})$, that is, the nonstationary Lipschitz map,  we obtain
\begin{equation}
\label{zeroMSREs}
\hat{\boldsymbol{\mu}}_{t+1|t} 
=
\widehat{\boldsymbol{X}}^\star_{t} \hat{\boldsymbol{\mu}}_{t|t-1}
+
\varphi(\hat{\boldsymbol{\mu}}^\star_{t|t-1}, \boldsymbol{y}_{t}, \boldsymbol{\theta}),
\end{equation}
where  $\hat{\boldsymbol{\mu}}_{t|t-1}^\star$ denotes a set of points between $\hat{\boldsymbol{\mu}}_{t|t-1}$ and $\tilde{\boldsymbol{\mu}}_{t|t-1}$. Moreover, we have that $\widehat{\boldsymbol{X}}^\star_{t} = \phi^\prime(\hat{\boldsymbol{\mu}}_{t|t-1}^\star , \boldsymbol{y}_t, \boldsymbol{\theta}) $, where $\phi^\prime$ denotes the first partial derivatives of $\phi$ with respect to the transpose of the vector $\hat{\boldsymbol{\mu}}_{t|t-1}^\star$, and $\varphi(\hat{\boldsymbol{\mu}}_{t|t-1}^\star, \boldsymbol{y}_{t}, \boldsymbol{\theta}) = \phi (\tilde{\boldsymbol{\mu}}_{t|t-1} , \boldsymbol{y}_t, \boldsymbol{\theta}) 
- \widehat{\boldsymbol{X}}^\star_{t} {\boldsymbol{\mu}}_{t|t-1} $. Equation \eqref{zeroMSREs} is a multivariate SRE, that can be viewed as vector autoregressive process with random coefficients. 
The sufficient conditions for invertibility given by \cite{Bougerol1993} and \cite{Straumann_Mikosh2006} then become 
\begin{align}
\label{inv1}
\mathbb{E}
\bigg[ \ln^{+} \sup_{\boldsymbol{\theta} \in \boldsymbol{\Theta}}
&\big\|  \phi (\tilde{\boldsymbol{\mu}}_{1|0} , \boldsymbol{y}_1, \boldsymbol{\theta}) 
- \tilde{\boldsymbol{\mu}}_{1|0}
\big\| \bigg] 
< \infty,
\tab
\mathbb{E}
\bigg[ \ln^{+} \sup_{\boldsymbol{\theta} \in \boldsymbol{\Theta}}
\bigg\| {\boldsymbol{X}}_{1}
\bigg\| \bigg] 
< \infty,
\end{align}
for any $\tilde{\boldsymbol{\mu}}_{1|0} \in \boldsymbol{\mathcal{M}}$ and
\begin{align}
\label{inv2}
\mathbb{E} \bigg[
\ln
\sup_{\boldsymbol{\theta} \in \boldsymbol{\Theta}}&
\sup_{\boldsymbol{\mu}\in \boldsymbol{\mathcal{M}}}  
 \bigg\| 
 \prod_{j=1}^k
 \boldsymbol{X}_{k-j+1} \bigg\| \bigg]
< 0,
\end{align}
for $k \geq 1$ and where $\ln^{+} x = \max\{ 0, \ln x \}$.

Let us consider condition \eqref{inv1}. One has  
\begin{align*}
\mathbb{E}
\bigg[& \ln^{+} \sup_{\boldsymbol{\theta} \in \boldsymbol{\Theta}}
 \big\|\phi (\tilde{\boldsymbol{\mu}}_{1|0} , \boldsymbol{y}_1, \boldsymbol{\theta}) 
- \tilde{\boldsymbol{\mu}}_{1|0}
 \big\| \bigg] 
 \leq
 2\ln 2  + 
\ln^{+} \sup_{\boldsymbol{\theta} \in \boldsymbol{\Theta}}
 \big\| \boldsymbol{\Phi} \big\|
 +
  2\ln^{+} \sup_{\boldsymbol{\theta} \in \boldsymbol{\Theta}}
 \big\| \tilde{\boldsymbol{\mu}}_{1|0} - \boldsymbol{\omega} \big\| 
 \nonumber\\
 &\tab \tab \tab \tab \tab +
\ln^{+} \sup_{\boldsymbol{\theta} \in \boldsymbol{\Theta}}
 \big\| \boldsymbol{K} \big\|
 +
 \mathbb{E}
\bigg[ \ln^{+} \sup_{\boldsymbol{\theta} \in \boldsymbol{\Theta}}
 \big\| \boldsymbol{u}_1 \big\| \bigg] < \infty
\end{align*}
by compactness of $\boldsymbol{\Theta}$ and since $\boldsymbol{u}_t$ is uniformly bounded $\forall t$ in both $\boldsymbol{\mu}_{t|t-1} \in \boldsymbol{\mathcal{M}}$ and $\boldsymbol{y}_t \in \mathbb{R}^{N}$. In particular, for any  $\boldsymbol{\mu}_{t|t-1} \in \boldsymbol{\mathcal{M}}$, as $\| \boldsymbol{y}_t \| \rightarrow \infty$ we obtain that $\| \boldsymbol{u}_t \|\rightarrow 0$. Thus, $\sup_t \mathbb{E}[\sup_{\boldsymbol{\theta} \in \boldsymbol{\Theta}}\| \boldsymbol{u}_t  \|]<\infty$ which clearly implies $\mathbb{E}
[ \ln^{+} \sup_{\boldsymbol{\theta} \in \boldsymbol{\Theta}}
\| \boldsymbol{u}_1 \| ] < \infty$.

Moreover, note that $\mathbb{E}
\big[ \ln^{+} \sup_{\boldsymbol{\theta} \in \boldsymbol{\Theta}}
\big\| {\boldsymbol{X}}_{1}
\big\| \big] 
< \infty$ directly follows from the contraction condition $\mathbb{E} \left[
\ln
\sup_{\boldsymbol{\theta} \in \boldsymbol{\Theta}}
\sup_{\tilde{\boldsymbol{\mu}}_{1|0} \in \boldsymbol{\mathcal{M}}}  
 \left\| 
 \boldsymbol{X}_{1} \right\| \right]
< 0$. Therefore, condition \eqref{inv1} is fulfilled.

As far as condition \eqref{inv2} is concerned, the exponentially fast almost sure convergence of the filtered $\{ \hat{\boldsymbol{\mu}}_{t|t-1} \}_{t \in \mathbb{N}}$ is obtained as an application of Theorem 3.1 in \cite{Bougerol1993} or Theorem 2.8 in \cite{Straumann_Mikosh2006}, since the contraction condition \eqref{inv2} implies that 
$
\sup_{\boldsymbol{\theta} \in \boldsymbol{\Theta}} 
\|  \hat{\boldsymbol{\mu}}_{t+1|t} - \tilde{\boldsymbol{\mu}}_{t+1|t} \|
=
\sup_{\boldsymbol{\theta} \in \boldsymbol{\Theta}}
\left\| 
\left(\prod_{i=0}^{t-1} \widehat{\boldsymbol{X}}_{t-i}^\star \right)  \,
\left( \hat{\boldsymbol{\mu}}_{1|0} - \tilde{\boldsymbol{\mu}}_{1|0} \right) \right\| \nonumber
\leq
\, \varrho^t \, c,
$
where $c > 0$ and $0 < \varrho < 1$ are constants.

Therefore, all the requirements of \cite{Bougerol1993}'s Theorem are satisfied. Additionally, the claim that the moments are bounded follow from the fact that, as noted above, $\boldsymbol{u}_t$ is uniformly bounded. $\square$

\subsubsection*{Proof of Lemma \ref{bounded_moments}}
Under the correct specification assumption \ref{ass:cs}, for $\boldsymbol{\theta}=\boldsymbol{\theta}_0$ the stationary and ergodic solution $\{\tilde{\boldsymbol{\mu}}_{t|t-1} \}_{t \in \mathbb{Z}}$ coincide with $\{{\boldsymbol{\mu}}_{t|t-1} \}_{t \in \mathbb{Z}}$ in \eqref{score_driven_recursion}, and, consequently, with $\boldsymbol{\mu}_t$, since Lemma \ref{SE_Dynamic_Location} ensures that the SE solution is unique. As a consequence of Lemma \ref{uniformly_boundedness_ut} and Lemma \ref{SE_Dynamic_Location},  the process $\{\boldsymbol{y}_t\}_{t\in\mathbb Z}$ is stationary by continuity and its moments are bounded. Ergodicity of $\{\boldsymbol{y}_t\}_{t\in\mathbb Z}$ under the same assumptions follows by Proposition 4.3 of \cite{Krengel1985} $\square$ \\


To prove consistency and asymptotic normality of the MLE, additional quantities are introduced. Let us define the empirical average $\log$-likelihood function based on the chosen initial value $\boldsymbol{\mu}_{1|0}$ and on the filtered sequence $\{ \hat{\boldsymbol{\mu}}_{t|t-1} \}_{t \in \mathbb{N}}$
\begin{equation}
\label{started_likelihood} 
\widehat{{\mathcal{L}}}_T (\boldsymbol{\theta}) = \frac{1}{T} \sum_{t=1}^{T} \widehat{\ell}_t(\boldsymbol{\theta}),
\end{equation}
and the likelihood based on the stationary sequence $\{ {\tilde{\boldsymbol{\mu}}}_{t|t-1} \}_{t \in \mathbb{Z}}$
\begin{equation}
\label{Stationary_Eergodic_likelihood_equation}
{\mathcal{L}}_T (\boldsymbol{\theta}) = \frac{1}{T} \sum_{t=1}^{T} \ell_t(\boldsymbol{\theta}),
\end{equation}
with the following limit
\begin{equation}
\label{Limit_likelihood}
{\mathcal{L}} (\boldsymbol{\theta}) = \mathbb{E}[ \ell_t(\boldsymbol{\theta})].
\end{equation}
The first and second derivatives of the above quantities with respect of the parameter will be denoted as $\widehat{\mathcal{L}}^\prime_T(\boldsymbol{\theta}),{\mathcal{L}}^\prime_T(\boldsymbol{\theta}), {\mathcal{L}}^\prime(\boldsymbol{\theta})$ and as $\widehat{\mathcal{L}}^{\prime\prime}_T(\boldsymbol{\theta}), {\mathcal{L}}^{\prime\prime}_T(\boldsymbol{\theta}), {\mathcal{L}}^{\prime\prime}(\boldsymbol{\theta})$, respectively.


The proof of consistency is based on some Lemmata that we report here for sake of clarity. The proofs of the Lemmata are in the online appendix.

\begin{lemma}
\label{Identifiability}
Assume that conditions \ref{as1}, \ref{as2} and \ref{as3} in Assumption \ref{consistency_theorem_assumptions} are satisfied.
Then $\mathbb{E}
\left[ 
\sup_{\boldsymbol{\theta} \in \boldsymbol{\Theta}}   
| \ell_t(\boldsymbol{\theta} )|
\right] < \infty
$
and 
$
\mathbb{E}
\big[ 
| \ell_t(\boldsymbol{\theta}_0 )|
\big] < \infty.
$
Furthermore, under condition \ref{as4}, for every $\boldsymbol{\theta} \neq \boldsymbol{\theta}_0 \in \boldsymbol{\Theta}$,
$
\mathbb{E}
\big[ 
| \ell_t(\boldsymbol{\theta} )|
\big] 
<
\mathbb{E}
\big[ 
| \ell_t(\boldsymbol{\theta}_0 )|
\big].
$
\end{lemma}

\begin{lemma}
\label{ULLN_likelihood}
Assume that conditions \ref{as1}, \ref{as2} and \ref{as3} in Assumption \ref{consistency_theorem_assumptions} are satisfied. Then,
$
\sup_{\boldsymbol{\theta} \in \boldsymbol{\Theta}}
| 
\widehat{\mathcal{L}}_T(\boldsymbol{\theta} )
-
\mathcal{L}_T(\boldsymbol{\theta} )
|
\xrightarrow[]{\text{a.s.}} 0 
 \text{ as }   t \rightarrow \infty,
$
and 
$
\sup_{\boldsymbol{\theta} \in \boldsymbol{\Theta}}
| 
\mathcal{L}_T(\boldsymbol{\theta} )
-
\mathcal{L}(\boldsymbol{\theta} )
|
\xrightarrow[]{\text{a.s.}} 0 
 \text{ as }   t \rightarrow \infty,
$
where $\widehat{\mathcal{L}}_T(\boldsymbol{\theta}) $, $\mathcal{L}_T(\boldsymbol{\theta} )$ and  $\mathcal{L}(\boldsymbol{\theta} )$ and are defined in \eqref{started_likelihood}, \eqref{Stationary_Eergodic_likelihood_equation} and \eqref{Limit_likelihood}, respectively.
\end{lemma}

\subsubsection*{Proof of Theorem \ref{consistency_theorem}}
One has, 
\begin{align*}
\sup_{\boldsymbol{\theta} \in \boldsymbol{\Theta}}
| 
\widehat{\mathcal{L}}_T(\boldsymbol{\theta} )
-
\mathcal{L}(\boldsymbol{\theta} )
|
\leq
\sup_{\boldsymbol{\theta} \in \boldsymbol{\Theta}}
| 
\widehat{\mathcal{L}}_T(\boldsymbol{\theta} )
-
{\mathcal{L}}_T(\boldsymbol{\theta} )
|
+
\sup_{\boldsymbol{\theta} \in \boldsymbol{\Theta}}
|
{\mathcal{L}}_T(\boldsymbol{\theta} )
-
{\mathcal{L}}(\boldsymbol{\theta} )
|.
\end{align*}
By Lemma  \ref{ULLN_likelihood} 
and the Ergodic Theorem, 
$
\lim_{T \to \infty} \widehat{\mathcal{L}}_T(\boldsymbol{\theta}_0 )
=
\lim_{T \to \infty} \mathcal{L}_T(\boldsymbol{\theta}_0 )
=
\mathcal{L}(\boldsymbol{\theta}_0 ),
$ 
 and, by Lemma \ref{Identifiability},
$ 
\mathcal{L}(\boldsymbol{\theta} ) < \mathcal{L}(\boldsymbol{\theta}_0), \forall  \boldsymbol{\theta} \neq \boldsymbol{\theta}_0. $ 
Following similar arguments of Theorem 3.4 in \cite{White1994}, one can show that strong consistency holds if $\forall$ $\boldsymbol{\theta} \neq \boldsymbol{\theta}_0$, $\exists$ $\mathcal{B}_\eta(\boldsymbol{\theta})$, where 
$\mathcal{B}_\eta(\boldsymbol{\theta}) 
=
\{ \boldsymbol{\theta} : \| \boldsymbol{\theta} - \boldsymbol{\theta}_0 \| > \eta, \eta>0 \}$ s.t. for any $\boldsymbol{\theta}^\star \in \mathcal{B}_\eta(\boldsymbol{\theta})$,  
\begin{align*}
\limsup_{T \to \infty}
\sup_{\boldsymbol{\theta}^\star \in \mathcal{B}_\eta(\boldsymbol{\theta})}
\widehat{\mathcal{L}}_T(\boldsymbol{\theta})
<
\lim_{T \to \infty}
\widehat{\mathcal{L}}_T(\boldsymbol{\theta}_0)
\tab \textit{a.s.}
\end{align*}
With a similar reasoning, by the reverse Fatou's Lemma and the Ergodic Theorem 
\begin{align*}
\limsup_{T \to \infty} 
\sup_{\boldsymbol{\theta}^\star \in \mathcal{B}_\eta(\boldsymbol{\theta})}&
\widehat{\mathcal{L}}_T(\boldsymbol{\theta})
=
\limsup_{T \to \infty} 
\sup_{\boldsymbol{\theta}^\star \in \mathcal{B}_\eta(\boldsymbol{\theta})}
\mathcal{L}_T(\boldsymbol{\theta})
=
\limsup_{T \to \infty} 
\sup_{\boldsymbol{\theta}^\star \in \mathcal{B}_\eta(\boldsymbol{\theta})}
\frac{1}{T} \sum_{t=1}^T 
\ell_t (\boldsymbol{\theta}) \\
\leq&
\limsup_{T \to \infty} 
\frac{1}{T} \sum_{t=1}^T 
\sup_{\boldsymbol{\theta}^\star \in \mathcal{B}_\eta(\boldsymbol{\theta})}
\ell_t (\boldsymbol{\theta}) 
=
\mathbb{E} 
\left[
\sup_{\boldsymbol{\theta}^\star \in \mathcal{B}_\eta(\boldsymbol{\theta})}
\ell_t (\boldsymbol{\theta})
\right],
\end{align*}
and therefore, $\forall$ $\varepsilon > 0$ $\exists$ $\eta > 0$ s.t.
$
\mathbb{E} 
\left[
\sup_{\boldsymbol{\theta}^\star \in \mathcal{B}_\eta(\boldsymbol{\theta})}
\ell_t (\boldsymbol{\theta})
\right]
<
\mathbb{E} 
\left[
\ell_t (\boldsymbol{\theta})
\right] + \varepsilon
=
\mathcal{L}(\boldsymbol{\theta} ) + \varepsilon.
$
Note that $\varepsilon$ can be made arbitrarily small. Therefore, the  uniqueness and identifiability of the maximizer $\boldsymbol{\theta}_0 \in \boldsymbol{\Theta}$, is ensured by the uniqueness of $\boldsymbol{\theta}_0$ as the maximizer of the likelihood, see Lemma \ref{Identifiability}, the compactness of the parameter space $\boldsymbol{\Theta}$ and finally, the continuity of the limit $\mathcal{L}(\boldsymbol{\theta} )$ in $\boldsymbol{\theta} \in \boldsymbol{\Theta}$ which is ensured from the continuity of $\mathcal{L}_T(\boldsymbol{\theta} )$ in $\boldsymbol{\theta} \in \boldsymbol{\Theta}$, $\forall T \in \mathbb{N}$ and the uniform convergence in Lemma \ref{ULLN_likelihood}. Then, strong consistency follows by Theorem 3.4 in \cite{White1994}. $\square$ \\

The proof of asymptotic normality requires the following Lemmata, proved in the online appendix.

\begin{lemma}
\label{CLT_first_differential_Likelihood}
Assume that conditions \ref{as1}, \ref{as2} and \ref{as3} in Assumption \ref{consistency_theorem_assumptions} are satisfied. Then, the first derivatives of the $\log$-likelihood $ \mathcal{L}^{\prime}_T(\boldsymbol{\theta}_0)$ obeys the CLT for martingale difference sequences, that is
$
\sqrt{T} \mathcal{L}^{\prime}_T(\boldsymbol{\theta}_0)
\xRightarrow[]{} \mathcal{N}(\boldsymbol{0}, \boldsymbol{V})
 \text{ as }   t \rightarrow \infty,
$
where 
$
\boldsymbol{V}
=
\mathbb{E}
\left[
(\mathcal{L}_T^{\prime}(\boldsymbol{\theta}_0))
(\mathcal{L}_T^{\prime}(\boldsymbol{\theta}_0))^\top
\right].
$
\end{lemma}

\begin{lemma}
\label{STARTED_diff_likelihood_effect}
Assume that conditions \ref{as1}, \ref{as2}, \ref{as3} and \ref{as4} in Assumption \ref{consistency_theorem_assumptions} are satisfied. Then, 
$
{\sqrt{T}}
\| 
\widehat{\mathcal{L}}^\prime_T(\boldsymbol{\theta}_0) 
-
{\mathcal{L}}^\prime_T(\boldsymbol{\theta}_0) 
\|
\xrightarrow[]{\text{P}} 0 
 \text{ as }   T \rightarrow \infty.
$
\end{lemma}

\begin{lemma}
\label{STARTED_likelihood_second_diff_effect}
Assume that conditions \ref{as1}, \ref{as2} and \ref{as3}, in Assumption \ref{consistency_theorem_assumptions} are satisfied. Then,
$
\sup_{\boldsymbol{\theta} \in \boldsymbol{\Theta}}
| 
\widehat{\mathcal{L}}^{\prime\prime}_T(\boldsymbol{\theta} )
-
\mathcal{L}^{\prime\prime}_T(\boldsymbol{\theta} )
|
\xrightarrow[]{\text{a.s.}} 0 
 \text{ as }   t \rightarrow \infty.
$
\end{lemma}

\begin{lemma}
\label{ULLN_likelihood_second_diff}
Assume that conditions \ref{as1}, \ref{as2}, \ref{as3} and \ref{as4} in Assumption \ref{consistency_theorem_assumptions} are satisfied. Then, 
$
\sup_{\boldsymbol{\theta} \in \boldsymbol{\Theta}}
| 
\mathcal{L}^{\prime\prime}_T(\boldsymbol{\theta} )
-
\mathcal{L}^{\prime\prime}(\boldsymbol{\theta} )
|
\xrightarrow[]{\text{a.s.}} 0 
 \text{ as }   t \rightarrow \infty,
$
\end{lemma}

\begin{lemma}
\label{Properties_second_diff_likelihood}
Assume that conditions \ref{as1}, \ref{as2}, \ref{as3}, \ref{as4} and \ref{as5} in Assumption \ref{consistency_theorem_assumptions} are satisfied. Then, the second derivative processes of the likelihood $\big\{ \frac{d^2 \ell_t(\boldsymbol{\theta})}{d \boldsymbol{\theta} d\boldsymbol{\theta}^\top} \big\}_{t \in \mathbb{Z}}$ are stationary ergodic with bounded moments. In particular, 
$
\mathbb{E}\big[ \frac{d^2 \ell_t(\boldsymbol{\theta})}{d \boldsymbol{\theta} d\boldsymbol{\theta}^\top}  \big] < \infty,
$ and is nonsingular.

\end{lemma}

\subsubsection*{Proof of Theorem \ref{asy_norm} (Asymptotic Normality)}
Standard arguments for the proof of asymptotic normality and the Taylor's theorem lead to the expansion of the conditional likelihood's score function around a neighborhood of $\boldsymbol{\theta}_0$, which yields
\begin{align}
\label{taylorNORMALITY}
\boldsymbol{0} =&
\sqrt{T} \widehat{\mathcal{L}}^\prime_T(\hat{\boldsymbol{\theta}}_T)
=
\sqrt{T}
 \Big[ \widehat{\mathcal{L}}^\prime_T(\boldsymbol{\theta}_0) - \mathcal{L}^\prime_T(\boldsymbol{\theta}_0) \Big] 
 + \sqrt{T} \mathcal{L}^\prime_T(\boldsymbol{\theta}_0) \nonumber\\
&+ 
\Big[ \big( 
\mathcal{L}^{\prime\prime}_T(\boldsymbol{\theta}_0) - \mathcal{L}^{\prime\prime}(\boldsymbol{\theta}_0) \big)
+ \big(
\widehat{\mathcal{L}}^{\prime \prime}_T(\boldsymbol{\theta}^\star) - \mathcal{L}^{\prime\prime}_T(\boldsymbol{\theta}_0) \big)
+ \mathcal{L}^{\prime\prime}(\boldsymbol{\theta}_0) \Big] 
\sqrt{T} (\hat{\boldsymbol{\theta}}_T - \boldsymbol{\theta}_0) ,
\end{align}
where $\boldsymbol{\theta}^\star$ lies on the chord between $\hat{\boldsymbol{\theta}}_T$ and $ \boldsymbol{\theta}_0$, componentwise. 

First, the fact that $\sqrt{T} \mathcal{L}^\prime_T(\boldsymbol{\theta}_0)$ obeys the CLT for martingales is entailed in Lemma \ref{CLT_first_differential_Likelihood}. Convergence of the first difference in square brackets of equation \eqref{taylorNORMALITY} is ensured by Lemma \ref{STARTED_diff_likelihood_effect}. Thus, by the asymptotic equivalence (see Lemma 4.7 in \cite{White2001}) $\widehat{\mathcal{L}}^\prime_T(\boldsymbol{\theta}_0)$ has the same asymptotic distribution of $\sqrt{T} \mathcal{L}^\prime_T(\boldsymbol{\theta}_0)$. As regards the second line, we have that the middle term vanishes almost surely and exponentially fast, since Lemma \ref{STARTED_likelihood_second_diff_effect} demonstrates that the initial conditions for the likelihood's second derivatives are asymptotically irrelevant and the consistency theorem further ensures the convergence in the same point by continuity arguments of the likelihood's second derivatives. In addition, the first term in the brackets of the second line vanishes as well by the Uniform Law of Large Numbers discussed in Lemma \ref{ULLN_likelihood_second_diff}. Finally, with Lemma \ref{Properties_second_diff_likelihood} at hand, we can easily solve equation \eqref{taylorNORMALITY}, since $\mathcal{L}^{\prime\prime}(\boldsymbol{\theta}_0)$ is nonsingular.
Slusky's Lemma (see Lemma 2.8 \textit{(iii)} of \cite{VanDerVaart1998}) completes the proof. $\square$

\end{document}


\def\spacingset#1{\renewcommand{\baselinestretch}%
{#1}\small\normalsize} \spacingset{1}


\if0\blind
{\title{\bf Online supplement for the paper:\\ A Robust Score-Driven Filter for Multivariate Time Series}
  \author{
  Enzo D'Innocenzo\thanks{Corresponding author.}\\
  Department of Econometrics and Data Science, Vrije Universiteit Amsterdam\\
   De Boelelaan 1105, 1081 HV Amsterdam, Netherlands\\
   e-mail: e.dinnocenzo@vu.nl\\
   \\
   Alessandra Luati\hspace{.2cm}\\
Department of Statistical Sciences, University of Bologna,\\
 Via delle Belle Arti 41, 40126 Bologna, Italy,\\
  e-mail: alessandra.luati@unibo.it\\
  and\\
   \\
   Mario Mazzocchi\hspace{.2cm}\\
Department of Statistical Sciences, University of Bologna,\\
 Via delle Belle Arti 41, 40126 Bologna, Italy,\\
  e-mail: mario.mazzocchi@unibo.it}
  \maketitle
} \fi

\bigskip

  \begin{abstract}
In this online supplementary materials, we provide details of a Monte Carlo study aimed to assess the finite sample properties of the MLE derived in the paper (Section \ref{Simulation_Study}), the relevant quantities for the implementation of the Fisher scoring algorithm (Section \ref{Computational_Aspects}), the proofs of Lemma \ref{Identifiability}-\ref{Properties_second_diff_likelihood} in the main paper, as well as some additional auxiliary Lemmata (Section \ref{appendix_lemmata}).  


    \end{abstract}

    \section{Monte Carlo}
\label{Simulation_Study}
The finite-sample properties of the MLE are investigated via Monte-Carlo simulations. 
%
We assume that the distribution of the heavy-tailed \emph{IID} errors will be $\boldsymbol{\epsilon}_t \sim \boldsymbol{t}_{\nu_{0}}(\boldsymbol{0}_2, \boldsymbol{I}_2)$, where $\nu_{0} \in \{ 3, 5, 10, 100 \}$, that is, a standard bivariate Student's \emph{t} with three, five and ten degrees of freedom, while the case when $\nu_0=100$ we cover the case when $\boldsymbol{\epsilon}_t$ tends to behave like a standard Gaussian noise. A property of the multivariate model introduced so far is that it estimates a linear Gaussian model when the errors are actually Gaussian. In this sense, the filter is robust to misspecification if normality holds. On the other hand, it is important to remark that we are assuming that all the time series share the same degrees of freedom $\nu_0$. It is well-known that estimating the degrees of freedom in Student's \emph{t} distributions can be quite challenging, since the implied profile likelihood is remarkably flat, see the discussion in \cite{Breusch_Robertson_Welsh1997}. 

In practice, we simulate data from the different specifications of the standard bivariate Student's \emph{t} and for each of the realized paths of the time series we consider the recursion in \eqref{score_driven_filter}, which satisfies the conditions of Lemma \eqref{SE_Dynamic_Location}. During the process which generates the data, we use a burn-in period of $1,000$ replications and we store $T=250$, $500$ and $1,000$ observations. This ensures that the collected $\{ \boldsymbol{y}_t \}_{t \in \mathbb{Z}}$ are stationary ergodic.

With this simulated data at hand, we start the Fisher's scoring algorithm based on the analytical formulae described in Section \ref{Fisher_Scoring_Section}. We repeat this simulation scheme $M=1,000$ times for each case and we use the empirical measures of bias and root mean square error to quantify the accuracy of our proposed estimators. Formally, the empirical bias measure and the empirical root mean square error of $\hat{\nu}$ over the $M = 1,000$ replications are computed as 
\begin{align*}
\textit{Bias}(\hat{\nu}) = \frac{1}{M} \sum_{m=1}^{M} (\hat{\nu}_{m} - \nu_{0}), \quad \textit{RMSE}(\hat{\nu}) = 
\sqrt{\frac{1}{M} \sum_{m=1}^{M} (\hat{\nu}_{m} - \nu_{0})^2}.
\end{align*}

In the bivariate case, the vector of parameters assumes the following form
\begin{align*}
\boldsymbol{\theta} = (\nu, \Omega_{11}, \Omega_{21}, \Omega_{22}, \omega_{1}, \omega_{2},
\Phi_{11}, \Phi_{21}, \Phi_{12}, \Phi_{22}, \kappa_{11}, \kappa_{21}, \kappa_{12}, \kappa_{22} )^{\top},
\end{align*}
thus $\boldsymbol{\theta} \in \mathbb{R}^{14}$, which means that  a complete bivariate system is characterized by $14$ parameters. The true parameters of the considered DGP are
\[
\nu_{0} \in \{ 3, 5, 10, 100 \},
\,\,
\boldsymbol{\Omega}_0
= \boldsymbol{I}_2,
\,\,  
\boldsymbol{\omega}_0
=
\begin{bmatrix}
    -3             &        5
\end{bmatrix},
\,\,
\boldsymbol{\Phi}_0
=
\begin{bmatrix}
    0.85  &  0.00     \\
    0.00  &  0.80
\end{bmatrix},
\,
\boldsymbol{K}_0
=
\begin{bmatrix}
    0.95  &  0.05     \\
    0.05  &  0.90
\end{bmatrix}.
\]
The Monte-Carlo results are reported in Tables \ref{MCsimulation_Bivariate_Table_nu3}
to \ref{MCsimulation_Bivariate_Table_nu100} according to the values of the degrees of freedom $\nu_0 \in \{ 3, 5, 10, 100 \}$, respectively. Each table contains three columns: \emph{Estimate} reports the Monte-Carlo average of the point estimates obtained from the simulations, while \emph{Bias} and \emph{RMSE} report the Monte-Carlo deviations from the true values as described above, which are associated with the time series dimensions, that is $T = 250$, $500$ and $1,000$. 

The first evident result, common to all the tables, is that as the time series dimension increases, the values of the empirical \emph{Bias} and \emph{RMSE} tend to reduce sharply, which is line with the consistency Theorem \ref{consistency_theorem}. In particular, we note that even if the value of $\nu_0$ is very low, namely $\nu_0 = 3$, the results are still satisfactory. 

In general, estimation of the number of degrees of freedom is rather accurate. In the approximately Gaussian case when $\nu_0=100$ the filter collapses to the steady state form of the Kalman filter and the degrees of freedom parameter is recovered already in the case of the smallest sample size. 
Moreover, the decreasing bias and \emph{RMSE} patterns may be due to the fixed initial value of the dynamic location vector $\boldsymbol{\mu}_{1|0}$ which was used to start the filter recursions. However, the invertibility conditions introduced in Lemma \ref{INV_Dynamic_location} ensure that for $T\rightarrow \infty$, this initial estimation bias will eventually tapers off. 
In conclusion, the \emph{ML} estimations deliver satisfactory results in terms of bias and root mean square error, hence the reliability of the Fisher-scoring method.

\newpage

\vfill

\begin{table}
\caption{Monte-Carlo Simulation results for $\nu_0 = 3$.}
\centering
\resizebox{\linewidth}{!}{%
\label{MCsimulation_Bivariate_Table_nu3}
\begin{tabular}{ @{\extracolsep{\fill}} l ccc ccc ccc}
\toprule
\midrule
& 
&\multicolumn{1}{c}{$T=250$} 
& 
&
&\multicolumn{1}{c}{$T=500$}
&
&
&\multicolumn{1}{c}{$T=1000$}
&
\\
\cmidrule(r){2-4}  \cmidrule(l){5-7} \cmidrule(l){8-10} 
&  \multicolumn{1}{c}{\emph{Estimate}}
&  \multicolumn{1}{c}{\emph{Bias}}
&  \multicolumn{1}{c}{\emph{RMSE}}
&  \multicolumn{1}{c}{\emph{Estimate}}
&  \multicolumn{1}{c}{\emph{Bias}}
&  \multicolumn{1}{c}{\emph{RMSE}}
&  \multicolumn{1}{c}{\emph{Estimate}}
&  \multicolumn{1}{c}{\emph{Bias}}
&  \multicolumn{1}{c}{\emph{RMSE}}
\\
\midrule
$\nu$ &$2.987$ &$0.013$ &$0.457$ &$2.979$ &$0.021$ &$0.473$ &$3.016$ &-$0.016$ &$0.321$ \\
$\Omega_{11}$ &$0.972$ &$0.028$ &$0.132$  &$0.972$ &$0.028$ &$0.130$  &$0.988$ &$0.011$ &$0.093$\\
$\Omega_{12}$ &-$0.000$ &$0.000$ &$0.073$ &-$0.004$ &$0.004$ &$0.074$ &$0.000$ &$0.000$ &$0.053$ \\
$\Omega_{22}$ &$0.971$ &$0.029$ &$0.135$ &$0.972$ &$0.028$ &$0.136$ &$0.991$ &$0.008$ &$0.090$ \\
$\omega_1$ &-$2.996$ &-$0.004$ &$0.288$  &-$2.990$ &-$0.009$ &$0.257$ &-$3.008$ &$0.009$ &$0.183$ \\
$\omega_2$ &$5.002$ &-$0.003$ &$0.215$   &$5.004$ &-$0.004$ &$0.207$ &$5.002$ &-$0.002$ &$0.145$ \\
$\Phi_{11}$ &$0.831$ &$0.019$ &$0.063$ &$0.836$ &$0.014$ &$0.062$ &$0.840$ &$0.010$ &$0.040$ \\
$\Phi_{12}$ &$0.001$ &-$0.001$ &$0.082$ &$0.000$ &$0.000$ &$0.083$ &-$0.000$ &$0.001$ &$0.047$ \\
$\Phi_{21}$ &$0.000$ &$0.000$ &$0.069$ &-$0.001$ &$0.001$ &$0.066$ &-$0.000$ &$0.000$ &$0.041$\\
$\Phi_{22}$ &$0.768$ &$0.032$ &$0.091$ &$0.771$ &$0.029$ &$0.084$ &$0.789$ &$0.011$ &$0.048$\\
$\kappa_{11}$ &$0.955$ &-$0.005$ &$0.184$  &$0.941$ &$0.009$ &$0.117$ &$0.954$ &-$0.004$ &$0.086$ \\
$\kappa_{12}$ &$0.052$ &-$0.002$ &$0.142$ &$0.054$ &-$0.004$ &$0.145$ &$0.049$ &$0.000$ &$0.097$\\
$\kappa_{21}$ &$0.054$ &-$0.004$ &$0.149$ &$0.057$ &-$0.007$ &$0.152$ &$0.051$ &-$0.001$ &$0.098$\\
$\kappa_{22}$ &$0.898$ &$0.002$ &$0.182$ &$0.894$ &$0.006$ &$0.186$ &$0.899$ &$0.001$ &$0.121$\\
\midrule 
\bottomrule
\end{tabular}%
}
\end{table}

\vfill

\begin{table}
\caption{Monte-Carlo Simulation results for $\nu_0 = 5$.}
\centering
\resizebox{\linewidth}{!}{%
\label{MCsimulation_Bivariate_Table_nu5}
\begin{tabular}{ @{\extracolsep{\fill}} l ccc ccc ccc}
\toprule
\midrule
& 
&\multicolumn{1}{c}{$T=250$} 
& 
&
&\multicolumn{1}{c}{$T=500$}
&
&
&\multicolumn{1}{c}{$T=1000$}
&
\\
\cmidrule(r){2-4}  \cmidrule(l){5-7} \cmidrule(l){8-10} 
&  \multicolumn{1}{c}{\emph{Estimate}}
&  \multicolumn{1}{c}{\emph{Bias}}
&  \multicolumn{1}{c}{\emph{RMSE}}
&  \multicolumn{1}{c}{\emph{Estimate}}
&  \multicolumn{1}{c}{\emph{Bias}}
&  \multicolumn{1}{c}{\emph{RMSE}}
&  \multicolumn{1}{c}{\emph{Estimate}}
&  \multicolumn{1}{c}{\emph{Bias}}
&  \multicolumn{1}{c}{\emph{RMSE}}
\\
\midrule
$\nu$ &$5.089$ &-$0.090$ &$1.084$ &$5.075$ &-$0.075$ &$0.693$ &$5.012$ &-$0.012$ &$0.573$ \\
$\Omega_{11}$ &$0.978$ &$0.220$ &$0.121$ &$0.993$ &$0.007$ &$0.086$ &$0.997$ &$0.003$ &$0.068$\\
$\Omega_{12}$ &$0.000$ &-$0.001$ &$0.075$ &-$0.002$ &$0.002$ &$0.050$ &-$0.001$ &$0.001$ &$0.046$\\
$\Omega_{22}$ &$0.974$ &$0.025$ &$0.123$ &$0.988$ &$0.012$ &$0.084$ &$0.992$ &$0.008$ &$0.038$\\
$\omega_1$ &-$2.973$ &-$0.027$ &$0.326$ &-$2.994$ &-$0.006$ &$0.219$ &-$3.002$ &$0.002$ &$0.127$\\
$\omega_2$ &$5.011$ &-$0.011$ &$0.268$ &$4.995$ &$0.005$ &$0.156$ &$4.997$ &$0.003$ &$0.133$\\
$\Phi_{11}$ &$0.831$ &$0.019$ &$0.055$ &$0.831$ &$0.019$ &$0.055$ &$0.844$ &$0.006$ &$0.055$ \\
$\Phi_{12}$ &-$0.000$ &$0.001$ &$0.068$ &-$0.000$ &$0.001$ &$0.068$ &$0.000$ &$0.000$ &$0.044$ \\
$\Phi_{21}$ &-$0.000$ &$0.001$ &$0.056$ &-$0.001$ &$0.001$ &$0.056$ &-$0.001$ &$0.001$ &$0.024$ \\
$\Phi_{22}$ &$0.776$ &$0.023$ &$0.069$ &$0.777$ &$0.023$ &$0.069$ &$0.788$ &$0.012$ &$0.039$ \\
$\kappa_{11}$ &$0.974$ &$0.002$ &$0.154$ &$0.949$ &$0.001$ &$0.103$ &$0.950$ &-$0.001$ &$0.083$\\
$\kappa_{12}$ &$0.047$ &$0.003$ &$0.115$ &$0.050$ &$0.000$ &$0.075$ &$0.050$ &$0.000$ &$0.027$\\
$\kappa_{21}$ &$0.055$ &-$0.005$ &$0.112$ &$0.055$ &-$0.005$ &$0.073$ &$0.052$ &-$0.002$ &$0.055$\\
$\kappa_{22}$ &$0.896$ &$0.004$ &$0.138$ &$0.900$ &-$0.001$ &$0.099$ &$0.900$ &-$0.000$ &$0.049$\\
\midrule 
\bottomrule
\end{tabular}%
}
\end{table}

\vfill

\clearpage

\vfill

\begin{table}[H]
\caption{Monte-Carlo Simulation results for $\nu_0 = 10$.}
\centering
\resizebox{\linewidth}{!}{%
\label{MCsimulation_Bivariate_Table_nu10}
\begin{tabular}{ @{\extracolsep{\fill}} l ccc ccc ccc}
\toprule
\midrule
& 
&\multicolumn{1}{c}{$T=250$} 
& 
&
&\multicolumn{1}{c}{$T=500$}
&
&
&\multicolumn{1}{c}{$T=1000$}
&
\\
\cmidrule(r){2-4}  \cmidrule(l){5-7} \cmidrule(l){8-10} 
&  \multicolumn{1}{c}{\emph{Estimate}}
&  \multicolumn{1}{c}{\emph{Bias}}
&  \multicolumn{1}{c}{\emph{RMSE}}
&  \multicolumn{1}{c}{\emph{Estimate}}
&  \multicolumn{1}{c}{\emph{Bias}}
&  \multicolumn{1}{c}{\emph{RMSE}}
&  \multicolumn{1}{c}{\emph{Estimate}}
&  \multicolumn{1}{c}{\emph{Bias}}
&  \multicolumn{1}{c}{\emph{RMSE}}
\\
\midrule
$\nu$ &$10.529$ &-$0.529$ &$4.727$ &$10.383$ &-$0.384$ &$2.232$ &$10.226$ &-$0.226$ &$1.631$ \\
$\Omega_{11}$ &$0.989$ &$0.011$ &$0.097$ &$0.995$ &$0.005$ &$0.075$ &$0.995$ &$0.004$ &$0.057$\\
$\Omega_{12}$ &-$0.001$ &$0.002$ &$0.067$ &$0.000$ &$0.000$ &$0.045$ &-$0.000$ &$0.002$ &$0.035$\\
$\Omega_{22}$ &$0.991$ &$0.008$ &$0.108$ &$0.991$ &$0.009$ &$0.074$ &$0.993$ &$0.006$ &$0.057$\\
$\omega_1$ &-$3.014$ &-$0.015$ &$0.365$ &-$2.994$ &-$0.006$ &$0.234$ &-$2.996$ &-$0.004$ &$0.189$\\
$\omega_2$ &$5.013$ &-$0.013$ &$0.287$ &$4.994$ &$0.006$ &$0.204$ &$4.997$ &$0.002$ &$0.129$\\
$\Phi_{11}$ &$0.834$ &$0.016$ &$0.052$ &$0.838$ &$0.011$ &$0.032$ &$0.845$ &$0.005$ &$0.023$\\
$\Phi_{12}$ &-$0.005$ &$0.006$ &$0.064$ &$0.002$ &-$0.003$ &$0.040$ &-$0.002$ &$0.002$ &$0.027$\\
$\Phi_{21}$ &$0.002$ &-$0.002$ &$0.047$ &$0.001$ &-$0.001$ &$0.031$ &-$0.000$ &$0.000$ &$0.024$\\
$\Phi_{22}$ &$0.781$ &$0.019$ &$0.063$ &$0.789$ &$0.011$ &$0.040$ &$0.794$ &$0.006$ &$0.028$\\
$\kappa_{11}$ &$0.926$ &$0.024$ &$0.113$ &$0.946$ &$0.003$ &$0.089$ &$0.949$ &$0.001$ &$0.065$\\
$\kappa_{12}$ &$0.059$ &-$0.009$ &$0.083$ &$0.051$ &-$0.001$ &$0.065$ &$0.049$ &$0.001$ &$0.050$\\
$\kappa_{21}$ &$0.042$ &$0.007$ &$0.091$ &$0.048$ &$0.002$ &$0.064$ &$0.049$ &$0.000$ &$0.050$\\
$\kappa_{22}$ &$0.877$ &$0.023$ &$0.123$ &$0.896$ &$0.004$ &$0.083$ &$0.893$ &$0.007$ &$0.061$\\
\midrule 
\bottomrule
\end{tabular}%
}
\end{table}

\vfill

\vfill

\begin{table}[H]
\caption{Monte-Carlo Simulation results for $\nu_0 = 100$.}
\centering
\resizebox{\linewidth}{!}{%
\label{MCsimulation_Bivariate_Table_nu100}
\begin{tabular}{ @{\extracolsep{\fill}} l ccc ccc ccc}
\toprule
\midrule
& 
&\multicolumn{1}{c}{$T=250$} 
& 
&
&\multicolumn{1}{c}{$T=500$}
&
&
&\multicolumn{1}{c}{$T=1000$}
&
\\
\cmidrule(r){2-4}  \cmidrule(l){5-7} \cmidrule(l){8-10} 
&  \multicolumn{1}{c}{\emph{Estimate}}
&  \multicolumn{1}{c}{\emph{Bias}}
&  \multicolumn{1}{c}{\emph{RMSE}}
&  \multicolumn{1}{c}{\emph{Estimate}}
&  \multicolumn{1}{c}{\emph{Bias}}
&  \multicolumn{1}{c}{\emph{RMSE}}
&  \multicolumn{1}{c}{\emph{Estimate}}
&  \multicolumn{1}{c}{\emph{Bias}}
&  \multicolumn{1}{c}{\emph{RMSE}}
\\
\midrule
$\nu$ &$ 96.708$ &$3.290 $ &$25.729$ &$98.782 $ &$1.218 $ &$13.782 $ &$100.855 $ &-$0.855$ &$12.866 $ \\
$\Omega_{11}$ &$0.999 $ &$0.001 $ &$0.119 $ &$1.016 $ &-$0.017 $ &$0.085 $ &$1.002 $ &-$0.002 $ &$0.064 $\\
$\Omega_{12}$ &-$0.005 $ &$0.005 $ &$0.068 $ &$0.006 $ &-$0.006 $ &$0.056 $ &$0.000 $ &-$0.001 $ &$0.030 $\\
$\Omega_{22}$ &$0.991 $ &$0.009 $ &$0.106 $ &$1.008 $ &-$0.008 $ &$0.084 $ &$1.003 $ &-$0.003 $ &$0.061 $\\
$\omega_1$ &-$2.974 $ &-$0.026 $ &$0.380 $ &-$2.964 $ &-$0.035 $ &$0.302 $ &-$3.032 $ &$0.032 $ &$0.268 $\\
$\omega_2$ &$4.944$ &$0.056 $ &$0.304 $ &$5.042 $ &-$0.042 $ &$0.230 $ &$5.009 $ &-$0.009 $ &$0.118 $\\
$\Phi_{11}$ &$0.826 $ &$0.023 $ &$0.054 $ &$0.834 $ &$0.016 $ &$0.039 $ &$0.841 $ &$0.009 $ &$0.039 $\\
$\Phi_{12}$ &$0.004 $ &-$0.004 $ &$0.060 $ &$0.002 $ &-$0.002 $ &$0.040 $ &$0.001 $ &-$0.001 $ &$0.022 $\\
$\Phi_{21}$ &-$0.001 $ &$0.001 $ &$0.045 $ &-$0.004 $ &$0.004 $ &$0.033 $ &$0.001 $ &-$0.001 $ &$0.022 $\\
$\Phi_{22}$ &$0.780 $ &$0.019 $ &$0.061 $ &$0.785 $ &$0.015 $ &$0.043 $ &$0.785 $ &$0.015 $ &$0.034 $\\
$\kappa_{11}$ &$0.945 $ &$0.004 $ &$0.121 $ &$0.946 $ &$0.004 $ &$0.093 $ &$0.947 $ &$0.003 $ &$0.049 $\\
$\kappa_{12}$ &$0.048 $ &$0.002 $ &$0.094 $ &$0.052 $ &-$0.002 $ &$0.062 $ &$0.055$ &-$0.005 $ &$0.039 $\\
$\kappa_{21}$ &$0.064 $ &-$0.014 $ &$0.098 $ &$0.049 $ &$0.001 $ &$0.069 $ &$0.049 $ &$0.001 $ &$0.037 $\\
$\kappa_{22}$ &$0.910 $ &-$0.010 $ &$0.108 $ &$0.903 $ &-$0.003 $ &$0.090 $ &$0.907 $ &-$0.008 $ &$0.038 $\\
\midrule 
\bottomrule
\end{tabular}%
}
\end{table}

\vfill

	\newpage
\section{Computational Aspects}
\label{Computational_Aspects}
This Appendix is devoted to the construction of score vector and the Hessian matrix, essential for estimation and inference. Our approach to tackle this problem is based on the matrix differential calculus by \cite{Magnus_Neudecker2019}. As argued by the authors, one of the advantages to represent the conditional log-density in its differential form is that we can straightforwardly retrieve all the partial derivatives, thus avoiding the problem of dealing with the dimensions of the matrices and vectors involved. 

\subsection{The Score Vector}
\label{score_vector}
The expressions for the score might be collected in a single vector, 
\[
\boldsymbol{s}_t(\boldsymbol{\theta}) 
=
\left[\begin{array}{ccccc}
\boldsymbol{s}_t^{(\nu)}(\boldsymbol{\theta}) &
\boldsymbol{s}_t^{(\mathrm{v}(\boldsymbol{\Omega}))}(\boldsymbol{\theta}) &
\boldsymbol{s}_t^{(\boldsymbol{\omega})}(\boldsymbol{\theta}) &
\boldsymbol{s}_t^{(\mathrm{v}(\boldsymbol{\Phi}))}(\boldsymbol{\theta}) &
\boldsymbol{s}_t^{(\mathrm{v}(\boldsymbol{K}))}(\boldsymbol{\theta})
\end{array}\right]^\top,
\]
yielding the recursions for the static parameters
\begin{align*}
\boldsymbol{s}_t^{(\nu)}(\boldsymbol{\theta})
=&
\frac{1}{2} \bigg[ \psi \bigg( \frac{\nu + N}{2} \bigg) - \psi \bigg( \frac{\nu}{2}\bigg) - \frac{N}{\nu} + \frac{\nu + N }{\nu} \, b_t - \ln w_t \bigg]\nonumber\\
&   +
\frac{\nu + N}{\nu } \frac{1}{w_t} 
\bigg(
\frac{d(\boldsymbol{\mu}_{t|t-1} - \boldsymbol{\omega})}{d \nu}
\bigg)^\top
\boldsymbol{\Omega}^{-1} (\boldsymbol{y}_t - \boldsymbol{\mu}_{t|t-1})
,\\
\boldsymbol{s}_t^{(\mathrm{v}(\boldsymbol{\Omega}))}(\boldsymbol{\theta})
=&
\frac{1}{2} \boldsymbol{\mathcal{D}}_{N}^\top (\boldsymbol{\Omega}^{-1/2} \otimes \boldsymbol{\Omega}^{-1/2}) \bigg[
\frac{\nu + N}{ \nu } \frac{1}{w_t} 
(\boldsymbol{\epsilon}_t \otimes  \boldsymbol{\epsilon}_t) - \vect \boldsymbol{I}_N \bigg]\nonumber\\
&  +
\frac{\nu + N}{\nu } \frac{1}{w_t} 
\bigg(
\frac{d(\boldsymbol{\mu}_{t|t-1} - \boldsymbol{\omega})}{d (\vecth(\boldsymbol{\Omega}))^\top}
\bigg)^\top
\boldsymbol{\Omega}^{-1} (\boldsymbol{y}_t - \boldsymbol{\mu}_{t|t-1})
,
\end{align*}
for the unconditional mean
\begin{align*}
\boldsymbol{s}_t^{(\boldsymbol{\omega})}(\boldsymbol{\theta})
&=
\frac{\nu + N}{\nu } \frac{1}{w_t} 
\bigg(
\frac{d(\boldsymbol{\mu}_{t|t-1} - \boldsymbol{\omega})}{d \boldsymbol{\omega}^\top}
\bigg)^\top
\boldsymbol{\Omega}^{-1} (\boldsymbol{y}_t - \boldsymbol{\mu}_{t|t-1})
,
\end{align*}
and for the remaining parameters determining the dynamics of the location vector
\begin{align*}
\boldsymbol{s}_t^{(\mathrm{v}(\boldsymbol{\Phi}))}(\boldsymbol{\theta})
&=
\frac{\nu + N}{\nu } \frac{1}{w_t} 
\bigg(
\frac{d(\boldsymbol{\mu}_{t|t-1} - \boldsymbol{\omega})}{d (\vect \boldsymbol{\Phi})^\top}
\bigg)^\top
\boldsymbol{\Omega}^{-1} (\boldsymbol{y}_t - \boldsymbol{\mu}_{t|t-1})
,\\
\boldsymbol{s}_t^{(\mathrm{v}(\boldsymbol{K}))}(\boldsymbol{\theta})
&=
\frac{\nu + N}{\nu } \frac{1}{w_t} 
\bigg(
\frac{d(\boldsymbol{\mu}_{t|t-1} - \boldsymbol{\omega})}{d (\vect \boldsymbol{K})^\top}
\bigg)^\top
\boldsymbol{\Omega}^{-1} (\boldsymbol{y}_t - \boldsymbol{\mu}_{t|t-1})
.
\end{align*}
Similarly, the conditional information matrix may be represented as follows, 
\[
\scalemath{0.85}{
\boldsymbol{\mathcal{I}}_t(\boldsymbol{\theta}) 
=
\left[\begin{array}{ccccc}
\boldsymbol{\mathcal{I}}^{(\nu)}_t(\boldsymbol{\theta}) & 
\boldsymbol{\mathcal{I}}^{(\nu, \mathrm{v}(\boldsymbol{\Omega}))}_t(\boldsymbol{\theta}) & 
\underset{1 \times N}{\boldsymbol{0}} & 
\boldsymbol{\mathcal{I}}^{(\nu, \mathrm{v}(\boldsymbol{\Phi}))}_t(\boldsymbol{\theta}) & 
\boldsymbol{\mathcal{I}}^{(\nu, \mathrm{v}(\boldsymbol{K}))}_t(\boldsymbol{\theta}) \\
\boldsymbol{\mathcal{I}}^{(\mathrm{v}(\boldsymbol{\Omega}), \nu)}_t(\boldsymbol{\theta}) & 
\boldsymbol{\mathcal{I}}^{(\mathrm{v}(\boldsymbol{\Omega}))}_t(\boldsymbol{\theta}) & 
\underset{N^2 \times N}{\boldsymbol{0}} & 
\boldsymbol{\mathcal{I}}^{(\mathrm{v}(\boldsymbol{\Omega}), \mathrm{v}(\boldsymbol{\Phi}))}_t(\boldsymbol{\theta}) & 
\boldsymbol{\mathcal{I}}^{(\mathrm{v}(\boldsymbol{\Omega}), \mathrm{v}(\boldsymbol{K}))}_t(\boldsymbol{\theta}) \\
\underset{N \times 1}{\boldsymbol{0}} & 
\underset{N \times N^2}{\boldsymbol{0}} & 
\boldsymbol{\mathcal{I}}^{(\boldsymbol{\omega})}_t(\boldsymbol{\theta}) & 
\underset{N \times N^2}{\boldsymbol{0}} & 
\underset{N \times N^2}{\boldsymbol{0}} \\
\boldsymbol{\mathcal{I}}^{(\mathrm{v}(\boldsymbol{\Phi}), \nu)}_t(\boldsymbol{\theta}) & 
\boldsymbol{\mathcal{I}}^{(\mathrm{v}(\boldsymbol{\Phi}), \mathrm{v}(\boldsymbol{\Omega}))}_t(\boldsymbol{\theta}) & 
\underset{N^2 \times N}{\boldsymbol{0}} & 
\boldsymbol{\mathcal{I}}^{(\mathrm{v}(\boldsymbol{\Phi}))}_t(\boldsymbol{\theta}) & 
\boldsymbol{\mathcal{I}}^{(\mathrm{v}(\boldsymbol{\Phi}), \mathrm{v}(\boldsymbol{K}))}_t(\boldsymbol{\theta}) \\
\boldsymbol{\mathcal{I}}^{(\mathrm{v}(\boldsymbol{K}), \nu)}_t(\boldsymbol{\theta}) & 
\boldsymbol{\mathcal{I}}^{(\mathrm{v}(\boldsymbol{K}), \mathrm{v}(\boldsymbol{\Omega}))}_t(\boldsymbol{\theta}) & 
\underset{N^2 \times N}{\boldsymbol{0}} & 
\boldsymbol{\mathcal{I}}^{(\mathrm{v}(\boldsymbol{K}), \mathrm{v}(\boldsymbol{\Phi}))}_t(\boldsymbol{\theta}) & 
\boldsymbol{\mathcal{I}}^{(\mathrm{v}(\boldsymbol{K}))}_t(\boldsymbol{\theta}) \\
\end{array}\right]}.
\]
The four blocks of the matrix have the following expansions: the first block is composed by
\begin{align*}
\boldsymbol{\mathcal{I}}^{(\nu)}_t(\boldsymbol{\theta})
&=
\frac{1}{4} \bigg[ \psi^{\prime} \bigg( \frac{\nu}{2} \bigg) 
-  \psi^{\prime} \bigg( \frac{\nu + N}{2} \bigg) 
- \frac{2 N(\nu + N + 4)}{\nu(\nu +N)(\nu+N+2)}\bigg] \nonumber\\
&\tab \tab \tab   +
\frac{\nu + N}{\nu + N + 2}
\bigg( \frac{d (\boldsymbol{\mu}_{t|t-1} - \boldsymbol{\omega})}{d \nu} \bigg)^\top
\boldsymbol{\Omega}^{-1}
\bigg( \frac{d (\boldsymbol{\mu}_{t|t-1} - \boldsymbol{\omega})}{d \nu} \bigg), \\
\boldsymbol{\mathcal{I}}^{(\mathrm{v}(\boldsymbol{\Omega}), \nu)}_t(\boldsymbol{\theta})
&=
- \frac{1}{(\nu + N) (\nu + N + 2)} 
\boldsymbol{\mathcal{D}}_{N}^\top (\vecth (\boldsymbol{\Omega}^{-1})) \nonumber\\
&\tab \tab \tab   +
\frac{\nu + N}{\nu + N + 2}
\bigg( \frac{d (\boldsymbol{\mu}_{t|t-1} - \boldsymbol{\omega})}{d (\vecth (\boldsymbol{\Omega}))^\top} \bigg)^\top
\boldsymbol{\Omega}^{-1}
\bigg( \frac{d (\boldsymbol{\mu}_{t|t-1} - \boldsymbol{\omega})}{d \nu} \bigg), \\
\boldsymbol{\mathcal{I}}^{(\mathrm{v}(\boldsymbol{\Omega}))}_t(\boldsymbol{\theta})
&=
\frac{\nu + N}{2(\nu + N + 2)}
\boldsymbol{\mathcal{D}}_N^\top 
(\boldsymbol{\Omega}^{-1} \otimes \boldsymbol{\Omega}^{-1})
\boldsymbol{\mathcal{D}}_N  
\nonumber\\
 &\tab \tab \tab   -
 \frac{1}{2(\nu + N + 2)}
\boldsymbol{\mathcal{D}}_N^\top 
 (\vecth (\boldsymbol{\Omega}^{-1})) (\vecth (\boldsymbol{\Omega}^{-1}))^\top
 \boldsymbol{\mathcal{D}}_N \nonumber\\
 &\tab \tab \tab   +
\frac{\nu + N}{\nu + N + 2}
\bigg( \frac{d (\boldsymbol{\mu}_{t|t-1} - \boldsymbol{\omega})}{d (\vecth (\boldsymbol{\Omega}))^\top} \bigg)^\top
\boldsymbol{\Omega}^{-1}
\bigg( \frac{d (\boldsymbol{\mu}_{t|t-1} - \boldsymbol{\omega})}{d (\vecth (\boldsymbol{\Omega}))^\top} \bigg).
\end{align*}
The second,
\begin{align*}
\boldsymbol{\mathcal{I}}^{(\mathrm{v}(\boldsymbol{\Phi}), \nu)}_t(\boldsymbol{\theta})
&=
\frac{\nu + N}{\nu + N + 2}
\bigg( \frac{d (\boldsymbol{\mu}_{t|t-1} - \boldsymbol{\omega})}{d (\vect \boldsymbol{\Phi})^\top} \bigg)^\top
\boldsymbol{\Omega}^{-1}
\bigg( \frac{d (\boldsymbol{\mu}_{t|t-1} - \boldsymbol{\omega})}{d \nu} \bigg),
\\
\boldsymbol{\mathcal{I}}^{(\mathrm{v}(\boldsymbol{\Phi}), \mathrm{v}(\boldsymbol{\Omega}))}_t(\boldsymbol{\theta})
&=
\frac{\nu + N}{\nu + N + 2}
\bigg( \frac{d (\boldsymbol{\mu}_{t|t-1} - \boldsymbol{\omega})}{d (\vect \boldsymbol{\Phi})^\top} \bigg)^\top
\boldsymbol{\Omega}^{-1}
\bigg( \frac{d (\boldsymbol{\mu}_{t|t-1} - \boldsymbol{\omega})}{d (\vecth(\boldsymbol{\Omega}))^\top} \bigg),
\\
\boldsymbol{\mathcal{I}}^{(\mathrm{v}(\boldsymbol{K}), \mathrm{v}(\boldsymbol{\Omega}))}_t(\boldsymbol{\theta})
&=
\frac{\nu + N}{\nu + N + 2}
\bigg( \frac{d (\boldsymbol{\mu}_{t|t-1} - \boldsymbol{\omega})}{d (\vect \boldsymbol{K})^\top} \bigg)^\top
\boldsymbol{\Omega}^{-1}
\bigg( \frac{d (\boldsymbol{\mu}_{t|t-1} - \boldsymbol{\omega})}{d (\vecth(\boldsymbol{\Omega}))^\top} \bigg),
\\
\boldsymbol{\mathcal{I}}^{(\mathrm{v}(\boldsymbol{K}), \nu)}_t(\boldsymbol{\theta})
&=
\frac{\nu + N}{\nu + N + 2}
\bigg( \frac{d (\boldsymbol{\mu}_{t|t-1} - \boldsymbol{\omega})}{d (\vect \boldsymbol{K})^\top} \bigg)^\top
\boldsymbol{\Omega}^{-1}
\bigg( \frac{d (\boldsymbol{\mu}_{t|t-1} - \boldsymbol{\omega})}{d \nu} \bigg).
\end{align*}
Third, the unconditional mean
\begin{align*}
\boldsymbol{\mathcal{I}}^{(\boldsymbol{\omega})}_t(\boldsymbol{\theta})
=
\frac{\nu + N}{\nu + N + 2}
\bigg( \frac{d (\boldsymbol{\mu}_{t|t-1} - \boldsymbol{\omega})}{d \boldsymbol{\omega}^\top} \bigg)^\top
\boldsymbol{\Omega}^{-1}
\bigg( \frac{d (\boldsymbol{\mu}_{t|t-1} - \boldsymbol{\omega})}{d \boldsymbol{\omega}^\top} \bigg).
\end{align*}
By symmetry, the fourth and last block are composed by
\begin{align*}
\boldsymbol{\mathcal{I}}^{(\mathrm{v}(\boldsymbol{\Phi}))}_t(\boldsymbol{\theta})
&=
\frac{\nu + N}{\nu + N + 2}
\bigg( \frac{d (\boldsymbol{\mu}_{t|t-1} - \boldsymbol{\omega})}{d (\vect \boldsymbol{\Phi})^\top} \bigg)^\top
\boldsymbol{\Omega}^{-1}
\bigg( \frac{d (\boldsymbol{\mu}_{t|t-1} - \boldsymbol{\omega})}{d (\vect \boldsymbol{\Phi})^\top} \bigg),
\\
\boldsymbol{\mathcal{I}}^{(\mathrm{v}(\boldsymbol{\Phi}), \mathrm{v}(\boldsymbol{K}))}_t(\boldsymbol{\theta})
&=
\frac{\nu + N}{\nu + N + 2}
\bigg( \frac{d (\boldsymbol{\mu}_{t|t-1} - \boldsymbol{\omega})}{d (\vect \boldsymbol{\Phi})^\top} \bigg)^\top
\boldsymbol{\Omega}^{-1}
\bigg( \frac{d (\boldsymbol{\mu}_{t|t-1} - \boldsymbol{\omega})}{d (\vect \boldsymbol{K})^\top} \bigg),
\\
\boldsymbol{\mathcal{I}}^{(\mathrm{v}(\boldsymbol{K}))}_t(\boldsymbol{\theta})
&=
\frac{\nu + N}{\nu + N + 2}
\bigg( \frac{d (\boldsymbol{\mu}_{t|t-1} - \boldsymbol{\omega})}{d (\vect \boldsymbol{K})^\top} \bigg)^\top
\boldsymbol{\Omega}^{-1}
\bigg( \frac{d (\boldsymbol{\mu}_{t|t-1} - \boldsymbol{\omega})}{d (\vect \boldsymbol{K})^\top} \bigg).
\end{align*}
Notably, $\boldsymbol{\mathcal{I}}^{(\boldsymbol{\omega}, \boldsymbol{\xi})}_t(\boldsymbol{\theta}) = \boldsymbol{0}$ and $\boldsymbol{\mathcal{I}}^{(\boldsymbol{\omega}, \boldsymbol{\psi})}_t(\boldsymbol{\theta})= \boldsymbol{0}$, i.e. $\boldsymbol{\omega}$ is asymptotically independent of the other parameters. Moreover, none of the terms of the conditional information matrix involves the second derivatives of the dynamic location. This result is a direct consequence of the asymptotic properties of the proposed MLE under the assumption of correct specification of the model and some regularity conditions.

To construct the score vector, we take the first differential of the likelihood function 
\begin{align}
\label{diffloglik}
\mathrm{d}\ell_t(\boldsymbol{\theta}) =& \, \frac{1}{2} \bigg[ \psi \bigg( \frac{\nu + N}{2} \bigg) - \psi \bigg( \frac{\nu}{2}\bigg) - \frac{N}{\nu} + \frac{\nu + N }{\nu} \, b_t - \ln w_t \bigg] (\mathrm{d} \nu) \nonumber\\
&+ \frac{1}{2} 
(\mathrm{d} \vecth (\boldsymbol{\Omega}))^\top \boldsymbol{\mathcal{D}}_{N}^\top
(\boldsymbol{\Omega}^{-1/2} \otimes \boldsymbol{\Omega}^{-1/2})
\bigg[ \frac{\nu + N}{ \nu } \frac{1}{w_t} 
(\boldsymbol{\epsilon}_t \otimes  \boldsymbol{\epsilon}_t) - \vect \boldsymbol{I}_N \bigg]\nonumber\\
&+ \frac{\nu + N}{\nu } \frac{1}{w_t} (\mathrm{d} \boldsymbol{\mu}_{t|t-1})^\top
\boldsymbol{\Omega}^{-1} (\boldsymbol{y}_t - \boldsymbol{\mu}_{t|t-1}),
\end{align}
where $\psi(x) = d \ln \Gamma (x) / d (x)$ is the digamma function and $\boldsymbol{\mathcal{D}}_{N}$ the duplication matrix, which allow us to write $\mathrm{d} \vect \boldsymbol{\Omega} = \boldsymbol{\mathcal{D}}_{N} (\mathrm{d} \vecth (\boldsymbol{\Omega}))$, since the scale matrix is symmetric. Secondly, we define
$
\boldsymbol{s}_t(\boldsymbol{\theta}) = d \ell_t (\boldsymbol{\theta})/d \boldsymbol{\theta}
$
and partition the parameter as $\boldsymbol{\theta} = (\boldsymbol{\xi}^\top, \boldsymbol{\psi}^\top)^\top$, so that the score vector can be partitioned into two blocks and two distinct applications of the chain rule are required. 
Specifically, for $\boldsymbol{\xi} = (\boldsymbol{\omega}^\top, (\vecth(\boldsymbol{\Omega}))^\top, \nu)^\top$, we have 
\begin{align*}
\boldsymbol{s}_t^{(\boldsymbol{\xi})}(\boldsymbol{\theta}) = 
\frac{d \ell_t(\boldsymbol{\theta})}{d \boldsymbol{\xi}} =
\frac{\partial \ell_t (\boldsymbol{\theta})}{\partial \boldsymbol{\xi}} +
\bigg(
 \frac{d(\boldsymbol{\mu}_{t|t-1} - \boldsymbol{\omega})}{d \boldsymbol{\xi}^\top}
\bigg)^\top
\frac{\partial \ell_t(\boldsymbol{\theta})}{\partial \boldsymbol{\mu}_{t|t-1}},
\end{align*}
while for $\boldsymbol{\psi} = ((\vect \boldsymbol{\Phi})^\top, (\vect \boldsymbol{K})^\top)^\top$, we have
\begin{align*}
\boldsymbol{s}_t^{(\boldsymbol{\psi})}(\boldsymbol{\theta}) =
\frac{d \ell_t(\boldsymbol{\theta})}{d \boldsymbol{\psi}} =
\bigg(
 \frac{d(\boldsymbol{\mu}_{t|t-1} - \boldsymbol{\omega})}{d \boldsymbol{\psi}^\top}
\bigg)^\top
\frac{\partial \ell_t(\boldsymbol{\theta})}{\partial \boldsymbol{\mu}_{t|t-1} }.
\end{align*}
Let us start by considering the first differential of the dynamic location 
\begin{align*}
\mathrm{d} (\boldsymbol{\mu}_{t+1|t} - \boldsymbol{\omega})  =&
 \boldsymbol{\Phi} \mathrm{d} (\boldsymbol{\mu}_{t|t-1} - \boldsymbol{\omega}) + 
 \big[(\boldsymbol{\mu}_{t|t-1} - \boldsymbol{\omega})^\top \otimes \boldsymbol{I}_N \big] 
 \mathrm{d} \vect \boldsymbol{\Phi} \nonumber\\
&+  \big[(\boldsymbol{u}_{t})^\top \otimes \boldsymbol{I}_N \big] \mathrm{d} \vect \boldsymbol{K} 
+\boldsymbol{K} (\mathrm{d} \boldsymbol{u}_t),
\end{align*}
where 
\begin{align*}
\mathrm{d} \boldsymbol{u}_t =&  (\boldsymbol{y}_t - \boldsymbol{\mu}_{t|t-1}) b_t (1-b_t) /\nu (\mathrm{d} \nu) \nonumber\\
&+ (\boldsymbol{y}_t - \boldsymbol{\mu}_{t|t-1})(1-b_t)^2 / \nu
(\boldsymbol{\epsilon}_t \otimes \boldsymbol{\epsilon}_t )^\top 
(\boldsymbol{\Omega}^{-1/2} \otimes \boldsymbol{\Omega}^{-1/2}) 
\boldsymbol{\mathcal{D}}_{N}  (\mathrm{d} \vecth (\boldsymbol{\Omega})) \nonumber\\
&+ 2 (\boldsymbol{y}_t - \boldsymbol{\mu}_{t|t-1})(1-b_t)^2 / \nu 
(\boldsymbol{y}_t - \boldsymbol{\mu}_{t|t-1})^\top 
\boldsymbol{\Omega}^{-1} (\mathrm{d} \boldsymbol{\mu}_{t|t-1})
- (1-b_t) (\mathrm{d} \boldsymbol{\mu}_{t|t-1}).
\end{align*}
Let us embed the dynamic differential as an SRE
\begin{equation*}
\mathrm{d}(\boldsymbol{\mu}_{t+1|t} - \boldsymbol{\omega} )
=
\boldsymbol{X}_t \mathrm{d} ( \boldsymbol{\mu}_{t|t-1} - \boldsymbol{\omega} ) 
+
\boldsymbol{R}_t, 
\end{equation*}
where
\begin{align}
\label{Xt}
\boldsymbol{X}_t = \boldsymbol{\Phi} + \boldsymbol{K} \boldsymbol{\mathcal{C}}_t,
\end{align}
and
\begin{equation}
\label{Rt}
\boldsymbol{R}_t 
=
\boldsymbol{K} \boldsymbol{a}_t \mathrm{d} \nu 
+ \boldsymbol{K} \boldsymbol{B}_t \mathrm{d} \vect \boldsymbol{\Omega}
+  \boldsymbol{D}_{t} \mathrm{d} \vect \boldsymbol{\Phi}
+  \boldsymbol{E}_t \mathrm{d} \vect \boldsymbol{K}.
\end{equation}
The terms of the latter equations are
\begin{align}
\boldsymbol{a}_t = 
\frac{\partial \boldsymbol{u}_t}{\partial \nu} 
&=
(\boldsymbol{y}_t - \boldsymbol{\mu}_{t|t-1}) b_t (1-b_t) /\nu, \nonumber
\\
\boldsymbol{B}_t = 
\frac{\partial \boldsymbol{u}_t}{\partial (\vecth(\boldsymbol{\Omega}))^\top}  
&=
(1-b_t)^2 / \nu \, (\boldsymbol{y}_t - \boldsymbol{\mu}_{t|t-1})
(\boldsymbol{\Omega}^{-1/2} \boldsymbol{\epsilon_t} \otimes 
\boldsymbol{\Omega}^{-1/2}\boldsymbol{\epsilon_t} )^\top \boldsymbol{\mathcal{D}}_{N}, \nonumber
\\
\boldsymbol{\mathcal{C}}_t =
\frac{\partial \boldsymbol{u}_t}{\partial \boldsymbol{\mu}_{t|t-1}^\top}  
&=
2 (1-b_t)^2 / \nu \, (\boldsymbol{y}_t - \boldsymbol{\mu}_{t|t-1})
(\boldsymbol{y}_t - \boldsymbol{\mu}_{t|t-1})^\top 
\boldsymbol{\Omega}^{-1} - (1-b_t) \boldsymbol{I}_N,\nonumber
\end{align}
which we write, for convenience, also in their vectorised form
\begin{align}
\boldsymbol{a}_t &= 
b_t^{3/2} (1- b_t)^{1/2} / \nu
\boldsymbol{\Omega}^{1/2} \mathbf{z}_t, \nonumber \\
\vect \boldsymbol{B}_t &=
 \nu b_t^{3/2} (1- b_t)^{1/2}
(\boldsymbol{\Omega}^{-1/2} \otimes \boldsymbol{\Omega}^{-1/2} \otimes \boldsymbol{\Omega}^{1/2})
(\mathbf{z}_t \otimes \mathbf{z}_t \otimes \mathbf{z}_t),\nonumber \\
\label{alternvecCt}
\vect \boldsymbol{\mathcal{C}}_t &=
2 b_t (1- b_t) \, (\boldsymbol{\Omega}^{-1/2} \otimes \boldsymbol{\Omega}^{1/2})
(\mathbf{z}_t \otimes \mathbf{z}_t) -
 (1-b_t) \vect \boldsymbol{I}_N.
\end{align}
The partial derivatives
\begin{align*}
\boldsymbol{C} =
\frac{\partial (\boldsymbol{\mu}_{t|t-1} - \boldsymbol{\omega}) }{\partial \boldsymbol{\omega}^\top}  
&=
(\boldsymbol{I}_N - \boldsymbol{\Phi}), \nonumber
\\
\boldsymbol{D}_t =
\frac{\partial (\boldsymbol{\mu}_{t|t-1} - \boldsymbol{\omega}) }{\partial (\vect \boldsymbol{\Phi})^\top} 
&=
 \big[(\boldsymbol{\mu}_{t|t-1} - \boldsymbol{\omega})^\top \otimes \boldsymbol{I}_N \big],
\\
\boldsymbol{E}_t =
\frac{\partial (\boldsymbol{\mu}_{t|t-1} - \boldsymbol{\omega}) }{\partial (\vect \boldsymbol{K})^\top}  
&=
 \big[(\boldsymbol{u}_{t})^\top \otimes \boldsymbol{I}_N \big],
\end{align*}
are required to obtain the final recursions, necessary for the iterative procedure
\begingroup
\allowdisplaybreaks
\begin{align}
\label{sysMSREs}
\frac{d(\boldsymbol{\mu}_{t+1|t} - \boldsymbol{\omega})}{d \nu} =& \boldsymbol{X}_t \frac{d(\boldsymbol{\mu}_{t|t-1} - \boldsymbol{\omega})}{d \nu} 
+ \boldsymbol{K} \boldsymbol{a}_t, \nonumber\\
\frac{d(\boldsymbol{\mu}_{t+1|t} - \boldsymbol{\omega})}{d (\vecth (\boldsymbol{\Omega}))^\top} =&
\boldsymbol{X}_t
\frac{d(\boldsymbol{\mu}_{t|t-1} - \boldsymbol{\omega})}{d (\vecth (\boldsymbol{\Omega}))^\top}
+ \boldsymbol{K} \boldsymbol{B}_t, \nonumber\\
\frac{d(\boldsymbol{\mu}_{t+1|t} - \boldsymbol{\omega})}{d \boldsymbol{\omega}^\top} =&
\boldsymbol{X}_t
\frac{d(\boldsymbol{\mu}_{t|t-1} - \boldsymbol{\omega})}{d \boldsymbol{\omega}^\top}
+ \boldsymbol{C}, \\
\frac{d (\boldsymbol{\mu}_{t+1|t} - \boldsymbol{\omega})}{d (\vect \boldsymbol{\Phi})^\top} =& \boldsymbol{X}_t \frac{d (\boldsymbol{\mu}_{t|t-1} - \boldsymbol{\omega})}{d (\vect \boldsymbol{\Phi})^\top} + \boldsymbol{D}_t, \nonumber\\
\frac{d (\boldsymbol{\mu}_{t+1|t} - \boldsymbol{\omega})}{d (\vect \boldsymbol{K})^\top} =& \boldsymbol{X}_t \frac{d (\boldsymbol{\mu}_{t|t-1} - \boldsymbol{\omega})}{d (\vect \boldsymbol{K})^\top} + \boldsymbol{E}_t. \nonumber
\end{align}
\endgroup
The discussion on the required partial derivatives of the $\log$-likelihood function is similarly tackled. 
From \eqref{diffloglik} the calculation are straightforward, we define
\begin{align*}
\alpha_t &= 
\frac{\partial \ell_t(\boldsymbol{\theta})}{\partial \nu } 
= 
\frac{1}{2} \bigg[ \psi \bigg( \frac{\nu + N}{2} \bigg) - \psi \bigg( \frac{\nu}{2}\bigg) - \frac{N}{\nu} + \frac{\nu + N }{\nu} \, b_t - \ln w_t \bigg], \\
\boldsymbol{\beta}_t &= 
\frac{\partial \ell_t(\boldsymbol{\theta})}{\partial (\vecth(\boldsymbol{\Omega})) } 
=
\frac{1}{2} \boldsymbol{\mathcal{D}}_{N}^\top (\boldsymbol{\Omega}^{-1/2} \otimes \boldsymbol{\Omega}^{-1/2}) \bigg[
\frac{\nu + N}{ \nu } \frac{1}{w_t} 
(\boldsymbol{\epsilon}_t \otimes  \boldsymbol{\epsilon}_t) - \vect \boldsymbol{I}_N \bigg],\\
\boldsymbol{\varsigma}_t &= 
\frac{\partial \ell_t(\boldsymbol{\theta})}{\partial \boldsymbol{\mu}_{t|t-1} } 
=
\frac{\nu + N}{\nu } \frac{1}{w_t} 
\boldsymbol{\Omega}^{-1} (\boldsymbol{y}_t - \boldsymbol{\mu}_{t|t-1}),
\end{align*}
which completes the construction of the score vector.

\subsection{The Hessian Matrix}
\label{hessian_Matrix}
Like in the previous section, we obtain the second differential of the conditional $\log$-likelihood by differentiating \eqref{diffloglik}, which yields
\begin{align}
\label{secdiffloglik}
\mathrm{d}^2 \ell_t ( \boldsymbol{\theta}) 
=&
\frac{1}{2} \bigg[ \frac{1}{2} \psi^{\prime} \bigg( \frac{\nu + N}{2} \bigg) 
- \frac{1}{2} \psi^{\prime} \bigg( \frac{\nu}{2} \bigg) + \frac{N}{\nu^2} 
- \frac{N}{\nu^2} b_t 
- \frac{\nu + N}{\nu^2} b_t (1-b_t) + \frac{1}{\nu} b_t \bigg] (\mathrm{d}^2 \nu) \nonumber\\
&+
\bigg[ \frac{\nu + N}{2\nu^2} (1- b_t)^2
(\mathrm{d} \vect \boldsymbol{\Omega})^\top 
(\boldsymbol{\Omega}^{-1/2} \otimes \boldsymbol{\Omega}^{-1/2})
\nonumber\\
&\tab \times (\boldsymbol{\epsilon}_t  \boldsymbol{\epsilon}_t^\top \otimes
 \boldsymbol{\epsilon}_t  \boldsymbol{\epsilon}_t^\top)
(\boldsymbol{\Omega}^{-1/2} \otimes \boldsymbol{\Omega}^{-1/2})
(\mathrm{d} \vect \boldsymbol{\Omega}) \bigg] \nonumber\\
&- 
\bigg[ \frac{\nu + N}{\nu}  (1- b_t) (\mathrm{d} \boldsymbol{\mu}_{t|t-1})^\top
\boldsymbol{\Omega}^{-1} (\mathrm{d} \boldsymbol{\mu}_{t|t-1}) \bigg] 
\nonumber\\
&-  
\bigg[ \frac{\nu + N}{\nu}  (1- b_t) (\mathrm{d}^2 \boldsymbol{\mu}_{t|t-1})^\top
\boldsymbol{\Omega}^{-1/2} \boldsymbol{\epsilon}_t \bigg] \nonumber\\
 &+
\bigg[ \frac{\nu + N}{\nu^2} (1- b_t)^2 (\mathrm{d}\boldsymbol{\mu}_{t|t-1})^\top
(\boldsymbol{\Omega}^{-1/2} \boldsymbol{\epsilon}_t 
 \boldsymbol{\epsilon}_t^\top  \boldsymbol{\Omega}^{-1/2}
\otimes \boldsymbol{\epsilon}_t^\top \boldsymbol{\Omega}^{-1/2} )
(\mathrm{d} \vect \boldsymbol{\Omega}) \bigg] \nonumber\\
&-
\bigg[ \frac{\nu + N}{\nu} (1- b_t) (\mathrm{d} \vect \boldsymbol{\Omega})^\top 
(\boldsymbol{\Omega}^{-1} \otimes \boldsymbol{\Omega}^{-1/2}
\boldsymbol{\epsilon}_t \boldsymbol{\epsilon}_t^\top \boldsymbol{\Omega}^{-1/2})  
(\mathrm{d} \vect \boldsymbol{\Omega}) \bigg] \nonumber\\
&+ 
\bigg[ 2 \frac{\nu + N}{\nu} (1- b_t) (\mathrm{d} \boldsymbol{\mu}_{t|t-1})^\top
(\boldsymbol{\epsilon}_t^\top \boldsymbol{\Omega}^{-1/2} \otimes 
\boldsymbol{\Omega}^{-1}) (\mathrm{d} \vect \boldsymbol{\Omega}) \bigg] \nonumber\\
&+
\bigg[ 2 \frac{\nu + N}{\nu^2} (1- b_t)^2 (\mathrm{d} \boldsymbol{\mu}_{t|t-1})^\top
\boldsymbol{\Omega}^{-1/2} \boldsymbol{\epsilon}_t
\boldsymbol{\epsilon}_t^\top \boldsymbol{\Omega}^{-1/2}
 (\mathrm{d} \boldsymbol{\mu}_{t|t-1}) \bigg] \nonumber\\
&+
\bigg[ \frac{1}{2} (\mathrm{d} \vect \boldsymbol{\Omega})^\top
(\boldsymbol{\Omega}^{-1} \otimes \boldsymbol{\Omega}^{-1})
 (\mathrm{d} \vect \boldsymbol{\Omega}) \bigg] \nonumber\\
&+
\bigg[(\mathrm{d}\boldsymbol{\mu}_{t|t-1})^\top \boldsymbol{\Omega}^{1/2} \boldsymbol{\epsilon}_t
+ \frac{1}{2}(\mathrm{d} \vect \boldsymbol{\Omega})^\top
(\boldsymbol{\Omega}^{-1/2} \otimes \boldsymbol{\Omega}^{-1/2}) 
(\boldsymbol{\epsilon}_t \otimes \boldsymbol{\epsilon}_t )\bigg] \nonumber\\
&\times 
\bigg[ 
  \frac{\nu + N}{\nu^2} b_t (1 - b_t) 
  - \frac{N}{\nu^2} (1 - b_t)
  \bigg](\mathrm{d}\nu),
 \end{align}
where $\psi^{\prime}(x) = d^2 \ln \Gamma (x) / d (x)^2$ is the trigamma function.

We thus define the Hessian matrix 
\begin{equation*}
\boldsymbol{\mathcal{H}}_t(\boldsymbol{\theta}) = 
\frac{d^2 \ell_t (\boldsymbol{\theta})}
{d \boldsymbol{\theta} d \boldsymbol{\theta}^\top},
\end{equation*}
Similar arguments as those used in the computation of the score vector lead us to decompose the Hessian into four blocks and then apply the chain rule separately to each block.
The first set is $\boldsymbol{\xi} = (\boldsymbol{\omega}^\top, (\vecth(\boldsymbol{\Omega}))^\top, \nu)^\top$,
\begin{align*}
\boldsymbol{\mathcal{H}}_t^{(\boldsymbol{\xi})}&(\boldsymbol{\theta})=
\frac{d^2 \ell_t(\boldsymbol{\theta})}{d \boldsymbol{\xi}d \boldsymbol{\xi}^\top} \nonumber\\
&=
\frac{\partial^2 \ell_t(\boldsymbol{\theta})}{\partial \boldsymbol{\xi} \partial \boldsymbol{\xi}^\top}
+ 
\bigg(
\frac{d (\boldsymbol{\mu}_{t|t-1} - \boldsymbol{\omega})}{d \boldsymbol{\xi}^\top}
\bigg)^\top
\frac{\partial^2 \ell_t(\boldsymbol{\theta})}{\partial \boldsymbol{\mu}_{t|t-1}
 \partial \boldsymbol{\mu}_{t|t-1}^\top}
\bigg(
\frac{d (\boldsymbol{\mu}_{t|t-1} - \boldsymbol{\omega})}{d \boldsymbol{\xi}^\top}
\bigg)\nonumber\\
&\tab +
\frac{\partial \ell_t(\boldsymbol{\theta})}{\partial \boldsymbol{\mu}_{t|t-1}^\top}
\frac{d^2 (\boldsymbol{\mu}_{t|t-1} - \boldsymbol{\omega})}{d \boldsymbol{\xi}d \boldsymbol{\xi}^\top}.
\end{align*}
As regards the second vector of parameters $\boldsymbol{\psi} = ((\vect \boldsymbol{\Phi})^\top, (\vect \boldsymbol{K})^\top)^\top$, we have
\begin{align*}
\boldsymbol{\mathcal{H}}_t^{(\boldsymbol{\psi})}&(\boldsymbol{\theta})=
\frac{d^2 \ell_t(\boldsymbol{\theta})}{d \boldsymbol{\psi}d \boldsymbol{\psi}^\top} \nonumber\\
&=
\bigg(
\frac{d (\boldsymbol{\mu}_{t|t-1} - \boldsymbol{\omega})}{d \boldsymbol{\psi}^\top}
\bigg)^\top
\frac{\partial^2 \ell_t(\boldsymbol{\theta})}{\partial \boldsymbol{\mu}_{t|t-1}
 \partial \boldsymbol{\mu}_{t|t-1}^\top}
 \bigg(
\frac{d (\boldsymbol{\mu}_{t|t-1} - \boldsymbol{\omega})}{d \boldsymbol{\psi}^\top}
\bigg)\nonumber\\
&\tab +
\frac{\partial \ell_t(\boldsymbol{\theta})}{\partial \boldsymbol{\mu}_{t|t-1}^\top}
\frac{d^2 (\boldsymbol{\mu}_{t|t-1} - \boldsymbol{\omega})}{d \boldsymbol{\psi}d \boldsymbol{\psi}^\top},
\end{align*}
and finally, by  symmetry, we get the remaining blocks 
\begin{align*}
\boldsymbol{\mathcal{H}}_t^{(\boldsymbol{\xi}, \boldsymbol{\psi})}&(\boldsymbol{\theta}) =
\frac{d^2 \ell_t(\boldsymbol{\theta})}{d \boldsymbol{\xi}d \boldsymbol{\psi}^\top} \nonumber\\
&= 
\bigg( 
\frac{d (\boldsymbol{\mu}_{t|t-1} - \boldsymbol{\omega})}{d \boldsymbol{\xi}^\top}
\bigg)^\top
\frac{\partial^2 \ell_t(\boldsymbol{\theta})}{\partial \boldsymbol{\mu}_{t|t-1}
 \partial \boldsymbol{\mu}_{t|t-1}^\top}
 \bigg(
\frac{d (\boldsymbol{\mu}_{t|t-1} - \boldsymbol{\omega})}{d \boldsymbol{\psi}^\top}
\bigg)\nonumber\\
&\tab +
\frac{\partial \ell_t(\boldsymbol{\theta})}{\partial \boldsymbol{\mu}_{t|t-1}^\top}
\frac{d^2 (\boldsymbol{\mu}_{t|t-1} - \boldsymbol{\omega})}{d \boldsymbol{\xi}d \boldsymbol{\psi}^\top}.
\end{align*}
As far as the second differentials of the dynamic equation are concerned, we have
\begin{align*}
\mathrm{d}^2 \boldsymbol{\mu}_{t+1|t} =&
 \boldsymbol{\Phi} \mathrm{d}^2 \boldsymbol{\mu}_{t|t-1} + 2 [\mathrm{d}(\boldsymbol{\mu}_{t|t-1} - \boldsymbol{\omega})^\top \otimes \boldsymbol{I}_N] 
\mathrm{d} \vect \boldsymbol{\Phi} \nonumber\\
&+ 2[\mathrm{d}(\boldsymbol{u}_{t})^\top \otimes \boldsymbol{I}_N]  \vect \boldsymbol{K}
+ \boldsymbol{K} (\mathrm{d}^2 \boldsymbol{u}_t),
\end{align*}
that, in turn, implies expanding $\mathrm{d}^2 \boldsymbol{u}_t$ with respect to the parameters of the  Student's \emph{t}.

After some algebra we get the second differential of the driving-force
\begingroup
\allowdisplaybreaks
\begin{align*}
\mathrm{d}^2 \boldsymbol{u}_t =& 
2(\boldsymbol{y}_t - \boldsymbol{\mu}_{t|t-1})/\nu
 \big[ b_t^2 (1-b_t)/\nu - b_t (1-b_t) \big] (\mathrm{d}^2\nu)\nonumber\\
&+ 
2(1-b_t)^3/\nu^2 \Big\{ \big[
(\mathrm{d} \vect \boldsymbol{\Omega})^\top \otimes (\boldsymbol{y}_t - \boldsymbol{\mu}_{t|t-1})
(\boldsymbol{\epsilon}_t \otimes \boldsymbol{\epsilon}_t)^\top \big] 
\vect(\boldsymbol{\Omega}^{-1/2} \otimes \boldsymbol{\Omega}^{-1/2}) \Big\} \nonumber\\
&\tab \tab \tab \tab \tab    \times 
\big[ (\boldsymbol{\epsilon}_t \otimes \boldsymbol{\epsilon}_t)^\top
(\boldsymbol{\Omega}^{-1/2} \otimes \boldsymbol{\Omega}^{-1/2})
(\mathrm{d} \vect \boldsymbol{\Omega}) \big] \nonumber\\
&-
 2(1-b_t)^2/\nu \Big\{ \big[
(\mathrm{d} \vect \boldsymbol{\Omega})^\top 
(\boldsymbol{\Omega}^{-1/2} \boldsymbol{\epsilon}_t \otimes \boldsymbol{\Omega}^{-1}) \nonumber\\
&\tab \tab \tab \tab \tab \otimes
(\boldsymbol{y}_t - \boldsymbol{\mu}_{t|t-1})(\boldsymbol{y}_t - \boldsymbol{\mu}_{t|t-1})^\top
\boldsymbol{\Omega}^{-1} \big] (\mathrm{d} \vect \boldsymbol{\Omega}) \Big\} \nonumber\\
&+
8(1-b_t)^3/\nu^2 \Big\{ \big[
(\mathrm{d} \boldsymbol{\mu}_{t|t-1})^\top  \otimes 
(\boldsymbol{y}_t - \boldsymbol{\mu}_{t|t-1}) (\boldsymbol{y}_t - \boldsymbol{\mu}_{t|t-1})^\top
  \big] \vect \boldsymbol{\Omega}^{-1} \Big\}\nonumber\\
&\tab \tab \tab \tab \tab \tab \tab   \times  
 \big[ (\boldsymbol{y}_t - \boldsymbol{\mu}_{t|t-1})^\top \boldsymbol{\Omega}^{-1}
  (\mathrm{d} \boldsymbol{\mu}_{t|t-1}) \big] \nonumber\\
&-
2(1-b_t)^2/\nu \Big\{ \big[
(\mathrm{d} \boldsymbol{\mu}_{t|t-1})^\top \boldsymbol{\Omega}^{-1}  \otimes \boldsymbol{I}_N \big]
\nonumber\\
&\tab \tab \tab \tab \times
\big[ (\boldsymbol{y}_t - \boldsymbol{\mu}_{t|t-1}) \otimes \boldsymbol{I}_N 
+ \boldsymbol{I}_N  \otimes (\boldsymbol{y}_t - \boldsymbol{\mu}_{t|t-1})\big]
(\mathrm{d} \boldsymbol{\mu}_{t|t-1}) \Big\} \nonumber\\
&-
2(1-b_t)^2/\nu  \Big\{  \big[ 
(\mathrm{d} \boldsymbol{\mu}_{t|t-1})^\top \boldsymbol{\Omega}^{-1}  \otimes \boldsymbol{I}_N \big]
 \big[  (\boldsymbol{y}_t - \boldsymbol{\mu}_{t|t-1}) \otimes \boldsymbol{I}_N \big]
 (\mathrm{d} \boldsymbol{\mu}_{t|t-1})  \Big\} \nonumber\\
&+
2(1-b_t)^2/\nu  \Big\{  \big[
(\boldsymbol{y}_t - \boldsymbol{\mu}_{t|t-1}) (\boldsymbol{y}_t - \boldsymbol{\mu}_{t|t-1})^\top
\boldsymbol{\Omega}^{-1} (\mathrm{d}^2 \boldsymbol{\mu}_{t|t-1}) \big] \Big\}\nonumber\\
&\tab \tab \tab \tab \tab -  (1-b_t) \Big\{(\mathrm{d}^2 \boldsymbol{\mu}_{t|t-1}) \Big\} \nonumber\\
&+ 
4 (1-b_t)^3 / \nu^2 
\Big\{ \big[ (\mathrm{d} \boldsymbol{\mu}_{t|t-1})^\top \otimes 
\boldsymbol{\Omega}^{1/2}\boldsymbol{\epsilon}_t
\boldsymbol{\epsilon}_t^\top \boldsymbol{\Omega}^{1/2} \big]
\nonumber\\
&\tab \tab \tab \tab \tab \times(\vect \boldsymbol{\Omega}^{-1})
(\boldsymbol{\epsilon}_t \otimes \boldsymbol{\epsilon}_t)^\top
(\boldsymbol{\Omega}^{-1} \otimes \boldsymbol{\Omega}^{-1}) 
(\mathrm{d} \vect \boldsymbol{\Omega}) \Big\} \nonumber \\
&- (1-b_t)^2 / \nu 
\Big\{ \big[ (\mathrm{d} \boldsymbol{\mu}_{t|t-1})^\top \otimes \boldsymbol{I}_N \big]( \vect  \boldsymbol{I}_N)
\big[(\boldsymbol{\epsilon}_t \otimes \boldsymbol{\epsilon}_t)^\top
(\boldsymbol{\Omega}^{-1} \otimes \boldsymbol{\Omega}^{-1}) 
(\mathrm{d} \vect \boldsymbol{\Omega})\big]\Big\}  \nonumber \\
&- 2 (1-b_t)^2 / \nu 
\Big\{ \big[ (\mathrm{d} \boldsymbol{\mu}_{t|t-1})^\top \otimes 
\boldsymbol{\Omega}^{1/2}\boldsymbol{\epsilon}_t
\boldsymbol{\epsilon}_t^\top \boldsymbol{\Omega}^{1/2} \big]
(\boldsymbol{\Omega}^{-1} \otimes \boldsymbol{\Omega}^{-1}) 
(\mathrm{d} \vect \boldsymbol{\Omega})\Big\} \nonumber \\
&+
\Big\{ [(\mathrm{d} \vect \boldsymbol{\Omega})^\top \otimes 
(\boldsymbol{y}_t - \boldsymbol{\mu}_{t|t-1})
(\boldsymbol{\epsilon}_t \otimes \boldsymbol{\epsilon}_t)^\top]
\vect (\boldsymbol{\Omega}^{-1/2} \otimes \boldsymbol{\Omega}^{-1/2}) \Big\} \nonumber\\
&\tab \tab \tab \tab \tab \times
 \big[ 2b_t(1-b_t)^2/\nu^2 - (1-b_t)^2/\nu^2) \big](\mathrm{d} \nu) \nonumber\\
&+ 
2\Big\{ \big[
(\mathrm{d} \boldsymbol{\mu}_{t|t-1})^\top  \otimes 
(\boldsymbol{y}_t - \boldsymbol{\mu}_{t|t-1}) (\boldsymbol{y}_t - \boldsymbol{\mu}_{t|t-1})^\top
  \big]( \vect \boldsymbol{\Omega}^{-1} )\Big\}
  \nonumber\\
&\tab \tab \tab \tab \tab \times
\big[2 b_t (1- b_t) / \nu^2 - (1-b_t)^2 / \nu^2 \big](\mathrm{d} \nu) \nonumber\\
&-
\Big\{ \big[ (\mathrm{d} \boldsymbol{\mu}_{t|t-1})^\top \otimes \boldsymbol{I}_N \big]
( \vect  \boldsymbol{I}_N)\Big\}
 \big[  b_t (1- b_t) / \nu \big](\mathrm{d} \nu).
\end{align*}
\endgroup
Let us write
\begin{equation*}
\mathrm{d}^2(\boldsymbol{\mu}_{t+1|t} - \boldsymbol{\omega} )
=
\boldsymbol{X}_t \mathrm{d}^2 ( \boldsymbol{\mu}_{t|t-1} - \boldsymbol{\omega} )
+
\boldsymbol{K} 
\mathrm{d}( \boldsymbol{\mu}_{t|t-1} - \boldsymbol{\omega} )^\top
\boldsymbol{\mathcal{C}}^\prime_t 
\mathrm{d}( \boldsymbol{\mu}_{t|t-1} - \boldsymbol{\omega} ) 
+
\boldsymbol{Q}_t, 
\end{equation*}
where $\boldsymbol{X}_t$ is as in \eqref{Xt} and
\begin{align}
\label{Qt}
\boldsymbol{Q}_t 
=&
\boldsymbol{K} \boldsymbol{a}^\prime_t \mathrm{d}^2 \nu 
+ \boldsymbol{K} \boldsymbol{B}^\prime_t \mathrm{d}^2 \vect \boldsymbol{\Omega}
+ \boldsymbol{K} (\mathrm{d} \vect \boldsymbol{\Omega})^\top 
\widehat{\boldsymbol{aB}}^\prime_t   \mathrm{d} \nu \nonumber\\
&+  \boldsymbol{D}^\prime_{t} \mathrm{d}^2 \vect \boldsymbol{\Phi}
+  \boldsymbol{E}^\prime_t \mathrm{d}^2 \vect \boldsymbol{K}
+ (\mathrm{d} \vect \boldsymbol{\Phi})^\top 
\widehat{\boldsymbol{DE}}_t^\prime
(\mathrm{d} \vect \boldsymbol{K}).
\end{align}
We now derive the terms of recursion \eqref{Qt}.
We first need a set of partial derivative
\begingroup
\allowdisplaybreaks
\begin{align*}
\boldsymbol{a}^\prime_t =
\frac{\partial^2 \boldsymbol{u}_t}{\partial \nu^2}
=
2(\boldsymbol{y}_t - \boldsymbol{\mu}_{t|t-1})/\nu
 \big[ b_t^2 (1-b_t)/\nu - b_t (1-b_t) \big],
\end{align*}
\endgroup
\begingroup
\allowdisplaybreaks
\begin{align*}
\boldsymbol{B}^\prime_t =&
\frac{\partial^2 \boldsymbol{u}_t}{\partial (\vecth(\boldsymbol{\Omega})) \partial (\vecth(\boldsymbol{\Omega}))^\top} 
= 2(1-b_t)^3/\nu^2 \nonumber\\
&\times \Big\{ \big[
\boldsymbol{\mathcal{D}}_{N}^\top \otimes (\boldsymbol{y}_t - \boldsymbol{\mu}_{t|t-1})
(\boldsymbol{\epsilon}_t \otimes \boldsymbol{\epsilon}_t)^\top \big] 
\vect(\boldsymbol{\Omega}^{-1/2} \otimes \boldsymbol{\Omega}^{-1/2}) \Big\} 
\nonumber\\ 
&   \times 
\big[ (\boldsymbol{\epsilon}_t \otimes \boldsymbol{\epsilon}_t)^\top
(\boldsymbol{\Omega}^{-1/2} \otimes \boldsymbol{\Omega}^{-1/2})  \boldsymbol{\mathcal{D}}_{N} \big] \nonumber\\
&-
 2(1-b_t)^2/\nu  \Big\{ \big[
\boldsymbol{\mathcal{D}}_{N}^\top 
(\boldsymbol{\Omega}^{-1/2} \boldsymbol{\epsilon}_t  \otimes \boldsymbol{\Omega}^{-1})\nonumber\\
& \otimes
(\boldsymbol{y}_t - \boldsymbol{\mu}_{t|t-1})(\boldsymbol{y}_t - \boldsymbol{\mu}_{t|t-1})^\top
\boldsymbol{\Omega}^{-1} \big] \boldsymbol{\mathcal{D}}_{N} \Big\},
\end{align*}
\endgroup
\begingroup
\allowdisplaybreaks
\begin{align}
\label{partialscores_C_prime_t}
\boldsymbol{\mathcal{C}}^\prime_t =
\frac{\partial^2 \boldsymbol{u}_t}{\partial \boldsymbol{\mu}_{t|t-1} \partial \boldsymbol{\mu}_{t|t-1}^\top} 
=& 
8(1-b_t)^3/\nu^2 \Big\{ \big[
\boldsymbol{I}_N   \otimes 
(\boldsymbol{y}_t - \boldsymbol{\mu}_{t|t-1}) (\boldsymbol{y}_t - \boldsymbol{\mu}_{t|t-1})^\top
  \big] (\vect \boldsymbol{\Omega}^{-1}) \Big\}  \nonumber\\
& \times 
 \big[ (\boldsymbol{y}_t - \boldsymbol{\mu}_{t|t-1})^\top \boldsymbol{\Omega}^{-1}\big] 2(1-b_t)^2/\nu \Big\{ \big[
\boldsymbol{\Omega}^{-1}  \otimes \boldsymbol{I}_N \big]\nonumber\\
& 
\times\big[ (\boldsymbol{y}_t - \boldsymbol{\mu}_{t|t-1}) \otimes \boldsymbol{I}_N 
+ \boldsymbol{I}_N  \otimes (\boldsymbol{y}_t - \boldsymbol{\mu}_{t|t-1})\big]\Big\} \nonumber\\
&-
2(1-b_t)^2/\nu  \Big\{  \big[ 
\boldsymbol{\Omega}^{-1}  \otimes \boldsymbol{I}_N \big]
 \big[  (\boldsymbol{y}_t - \boldsymbol{\mu}_{t|t-1}) \otimes \boldsymbol{I}_N \big] \Big\}.
\end{align}
\endgroup
Secondly, a set of partial cross-derivatives
\begingroup
\allowdisplaybreaks
\begin{align}
\widehat{\boldsymbol{aB}}^\prime_t =
\frac{\partial^2 \boldsymbol{u}_t}{\partial (\vecth(\boldsymbol{\Omega})) \partial \nu } 
=&
[\boldsymbol{I}_N \otimes 
(\boldsymbol{y}_t - \boldsymbol{\mu}_{t|t-1})
(\boldsymbol{\epsilon}_t \otimes \boldsymbol{\epsilon}_t)^\top]
\vect (\boldsymbol{\Omega}^{-1/2} \otimes \boldsymbol{\Omega}^{-1/2}) \nonumber\\
&\times 
\big[ 2b_t(1-b_t)^2/\nu^2 - (1-b_t)^2/\nu^2) \big],\nonumber
\\
\label{partialscores_aC_prime_t}
\widehat{\boldsymbol{a\mathcal{C}}}^\prime_t =
\frac{\partial^2 \boldsymbol{u}_t}{\partial \boldsymbol{\mu}_{t|t-1} \partial \nu }
=&
2\Big\{ \big[
\boldsymbol{I}_N  \otimes 
(\boldsymbol{y}_t - \boldsymbol{\mu}_{t|t-1}) (\boldsymbol{y}_t - \boldsymbol{\mu}_{t|t-1})^\top
  \big](\vect \boldsymbol{\Omega}^{-1}) \Big\} \nonumber\\
&\times
\big[2 b_t (1- b_t) / \nu^2 - (1-b_t)^2 / \nu^2 \big] \nonumber\\
&-
\Big\{ \big[ (\mathrm{d} \boldsymbol{\mu}_{t|t-1})^\top \otimes \boldsymbol{I}_N \big]
( \vect  \boldsymbol{I}_N)\Big\}\nonumber\\
&\times
 \big[  b_t (1- b_t) / \nu \big],
\\
\label{partialscores_BC_prime_t}
\widehat{\boldsymbol{B\mathcal{C}}}^\prime_t =
\frac{\partial^2 \boldsymbol{u}_t}{\partial \boldsymbol{\mu}_{t|t-1} \partial (\vecth(\boldsymbol{\Omega}))^\top} 
=&
4 (1-b_t)^3 / \nu^2 
\Big\{ \big[ \boldsymbol{I}_N \otimes 
\boldsymbol{\Omega}^{1/2}\boldsymbol{\epsilon}_t
\boldsymbol{\epsilon}_t^\top \boldsymbol{\Omega}^{1/2} \big]\nonumber\\& \times
(\vect \boldsymbol{\Omega}^{-1})
(\boldsymbol{\epsilon}_t \otimes \boldsymbol{\epsilon}_t)^\top
(\boldsymbol{\Omega}^{-1} \otimes \boldsymbol{\Omega}^{-1}) 
\Big\} \nonumber\\
&- (1-b_t)^2 / \nu 
\Big\{ \big[ \boldsymbol{I}_N \otimes \boldsymbol{I}_N \big]( \vect  \boldsymbol{I}_N)\nonumber\\
& \times
\big[(\boldsymbol{\epsilon}_t \otimes \boldsymbol{\epsilon}_t)^\top
(\boldsymbol{\Omega}^{-1} \otimes \boldsymbol{\Omega}^{-1}) 
 \big]\Big\}  \nonumber \\
&- 2 (1-b_t)^2 / \nu 
\Big\{ \big[ \boldsymbol{I}_N \otimes 
\boldsymbol{\Omega}^{1/2}\boldsymbol{\epsilon}_t
\boldsymbol{\epsilon}_t^\top \boldsymbol{\Omega}^{1/2} \big]
\nonumber\\
&\times
(\boldsymbol{\Omega}^{-1} \otimes \boldsymbol{\Omega}^{-1}) 
 \Big\}.
\end{align}
\endgroup
In addition, we a new set of partial derivatives defined by
\begingroup
\allowdisplaybreaks
\begin{align*}
\boldsymbol{D}_t^\prime 
=
\frac{\partial [d (\boldsymbol{\mu}_{t|t-1} - \boldsymbol{\omega})]}{\partial (\vect \boldsymbol{\Phi})
d (\vect \boldsymbol{\Phi})^\top} 
=&
2 \bigg[ \bigg(
\frac{d(\boldsymbol{\mu}_{t|t-1} - \boldsymbol{\omega})}{ d (\vect \boldsymbol{\Phi})^\top} 
\bigg)^\top \otimes \boldsymbol{I}_N \bigg], \\
\boldsymbol{E}_t^\prime 
=
\frac{\partial [d(\boldsymbol{\mu}_{t|t-1} - \boldsymbol{\omega})]}{\partial (\vect \boldsymbol{K})
d (\vect \boldsymbol{K})^\top} 
=&
2 \bigg[ \bigg(
\boldsymbol{\mathcal{C}}_t^\top
\frac{d(\boldsymbol{\mu}_{t|t-1} - \boldsymbol{\omega})}{ d (\vect \boldsymbol{K})^\top} 
\bigg)^\top \otimes \boldsymbol{I}_N \bigg], 
\end{align*}
\endgroup
and finally, we conclude the derivations with
\begingroup
\allowdisplaybreaks
\begin{align*}
\widehat{\boldsymbol{DE}}_t^\prime
=
\frac{\partial [d(\boldsymbol{\mu}_{t|t-1} - \boldsymbol{\omega})]}{\partial (\vect \boldsymbol{\Phi})
d (\vect \boldsymbol{K})^\top} 
=&
\bigg[ \bigg(
\boldsymbol{\mathcal{C}}_t^\top
\frac{d(\boldsymbol{\mu}_{t|t-1} - \boldsymbol{\omega})}{ d (\vect \boldsymbol{\Phi})^\top} 
\bigg)^\top \otimes \boldsymbol{I}_N \bigg].
\end{align*}
\endgroup
We therefore have obtained a new set of recursions composed by
\begingroup
\allowdisplaybreaks
\begin{align*}
\frac{d^2(\boldsymbol{\mu}_{t+1|t} - \boldsymbol{\omega})}{d \nu^2} =& 
\boldsymbol{X}_t
\frac{d^2(\boldsymbol{\mu}_{t|t-1} - \boldsymbol{\omega})}{d \nu^2} \\
&+ \boldsymbol{K} 
\bigg( \frac{d(\boldsymbol{\mu}_{t|t-1} - \boldsymbol{\omega})}{ d \nu} \bigg)^\top
\boldsymbol{\mathcal{C}}^\prime_t 
\bigg( \frac{d(\boldsymbol{\mu}_{t|t-1} - \boldsymbol{\omega})}{ d \nu} \bigg) 
+ \boldsymbol{K} \boldsymbol{a}^\prime_t, \nonumber \\
\frac{d^2(\boldsymbol{\mu}_{t+1|t} - \boldsymbol{\omega})}{d (\vecth (\boldsymbol{\Omega}))d
 (\vecth (\boldsymbol{\Omega}))^\top} =&
\boldsymbol{X}_t
\frac{d^2(\boldsymbol{\mu}_{t|t-1} - \boldsymbol{\omega})}{d (\vecth (\boldsymbol{\Omega})) 
d (\vecth (\boldsymbol{\Omega}))^\top} \\
&+ \boldsymbol{K} 
\bigg( \frac{d(\boldsymbol{\mu}_{t|t-1} - \boldsymbol{\omega})}{ d (\vecth (\boldsymbol{\Omega}))^\top} \bigg)^\top
\boldsymbol{\mathcal{C}}^\prime_t 
\bigg( \frac{d(\boldsymbol{\mu}_{t|t-1} - \boldsymbol{\omega})}{ d (\vecth (\boldsymbol{\Omega}))^\top} \bigg) 
+ \boldsymbol{K} \boldsymbol{B}^\prime_t, 
\end{align*}
\endgroup
\begingroup
\allowdisplaybreaks
\begin{align*}
\frac{d^2(\boldsymbol{\mu}_{t+1|t} - \boldsymbol{\omega})}{d (\vecth (\boldsymbol{\Omega}))d \nu} =&
\boldsymbol{X}_t
\frac{d^2(\boldsymbol{\mu}_{t|t-1} - \boldsymbol{\omega})}{d (\vecth (\boldsymbol{\Omega})) 
d \nu} \\
&+ \boldsymbol{K} 
\bigg( \frac{d(\boldsymbol{\mu}_{t|t-1} - \boldsymbol{\omega})}{ d (\vecth (\boldsymbol{\Omega}))^\top} \bigg)^\top
\boldsymbol{\mathcal{C}}^\prime_t 
\bigg( \frac{d(\boldsymbol{\mu}_{t|t-1} - \boldsymbol{\omega})}{ d \nu} \bigg) 
+ \boldsymbol{K} \widehat{\boldsymbol{aB}}^\prime_t, \nonumber
\end{align*}
\endgroup
which continue with
\begingroup
\allowdisplaybreaks
\begin{align*}
\frac{d^2(\boldsymbol{\mu}_{t+1|t} - \boldsymbol{\omega})}{d (\vect \boldsymbol{\Phi})
d (\vect \boldsymbol{\Phi})^\top} =&
\boldsymbol{X}_t
\frac{d^2(\boldsymbol{\mu}_{t|t-1} - \boldsymbol{\omega})}{d (\vect \boldsymbol{\Phi}) 
d (\vect \boldsymbol{\Phi})^\top}\\ 
&+ \boldsymbol{K} 
\bigg( \frac{d(\boldsymbol{\mu}_{t|t-1} - \boldsymbol{\omega})}{ d (\vect \boldsymbol{\Phi})^\top} \bigg)^\top
\boldsymbol{\mathcal{C}}^\prime_t 
\bigg( \frac{d(\boldsymbol{\mu}_{t|t-1} - \boldsymbol{\omega})}{ d (\vect \boldsymbol{\Phi})^\top} \bigg) 
+ \boldsymbol{D}^\prime_t,\nonumber \\
\frac{d^2(\boldsymbol{\mu}_{t+1|t} - \boldsymbol{\omega})}{d (\vect \boldsymbol{K})
d (\vect \boldsymbol{K})^\top} =&
\boldsymbol{X}_t
\frac{d^2(\boldsymbol{\mu}_{t|t-1} - \boldsymbol{\omega})}{d (\vect \boldsymbol{K}) 
d (\vect \boldsymbol{K})^\top} \\
&+ \boldsymbol{K} 
\bigg( \frac{d(\boldsymbol{\mu}_{t|t-1} - \boldsymbol{\omega})}{ d (\vect \boldsymbol{K})^\top} \bigg)^\top
\boldsymbol{\mathcal{C}}^\prime_t 
\bigg( \frac{d(\boldsymbol{\mu}_{t|t-1} - \boldsymbol{\omega})}{ d (\vect \boldsymbol{K})^\top} \bigg) 
+ \boldsymbol{E}^\prime_t, \\
\frac{d^2(\boldsymbol{\mu}_{t+1|t} - \boldsymbol{\omega})}{d (\vect \boldsymbol{\Phi})
d (\vect \boldsymbol{K})^\top} =&
\boldsymbol{X}_t
\frac{d^2(\boldsymbol{\mu}_{t|t-1} - \boldsymbol{\omega})}{d (\vect \boldsymbol{\Phi}) 
d (\vect \boldsymbol{K})^\top}\\ 
&+ \boldsymbol{K} 
\bigg( \frac{d(\boldsymbol{\mu}_{t|t-1} - \boldsymbol{\omega})}{ d (\vect \boldsymbol{\Phi})^\top} \bigg)^\top
\boldsymbol{\mathcal{C}}^\prime_t 
\bigg( \frac{d(\boldsymbol{\mu}_{t|t-1} - \boldsymbol{\omega})}{ d (\vect \boldsymbol{K})^\top} \bigg) 
+ \widehat{\boldsymbol{DE}}_t^\prime, \nonumber
\end{align*}
\endgroup
and conclude with
\begingroup
\allowdisplaybreaks
\begin{align*}
\frac{d^2(\boldsymbol{\mu}_{t+1|t} - \boldsymbol{\omega})}{d (\nu)
d (\vect \boldsymbol{\Phi})^\top} =&
\boldsymbol{X}_t \frac{d^2(\boldsymbol{\mu}_{t|t-1} - \boldsymbol{\omega})}{d (\nu)
d (\vect \boldsymbol{\Phi})^\top} \\ 
&+ \boldsymbol{K} 
\bigg( \frac{d(\boldsymbol{\mu}_{t|t-1} - \boldsymbol{\omega})}{ d \nu} \bigg)^\top
\boldsymbol{\mathcal{C}}^\prime_t 
\bigg( \frac{d(\boldsymbol{\mu}_{t|t-1} - \boldsymbol{\omega})}{ d (\vect \boldsymbol{\Phi})^\top} \bigg),
\nonumber \\
\frac{d^2(\boldsymbol{\mu}_{t+1|t} - \boldsymbol{\omega})}{d (\nu) d (\vect \boldsymbol{K})^\top} =&
\boldsymbol{X}_t \frac{d^2(\boldsymbol{\mu}_{t|t-1} - \boldsymbol{\omega})}{d (\nu)
d (\vect \boldsymbol{K})^\top} \\ 
&+ \boldsymbol{K} 
\bigg( \frac{d(\boldsymbol{\mu}_{t|t-1} - \boldsymbol{\omega})}{ d \nu} \bigg)^\top
\boldsymbol{\mathcal{C}}^\prime_t 
\bigg( \frac{d(\boldsymbol{\mu}_{t|t-1} - \boldsymbol{\omega})}{ d (\vect \boldsymbol{K})^\top} \bigg), \\
\frac{d^2(\boldsymbol{\mu}_{t+1|t} - \boldsymbol{\omega})}{d (\nu)
d (\vect \boldsymbol{\Phi})^\top} =&
\boldsymbol{X}_t \frac{d^2(\boldsymbol{\mu}_{t|t-1} - \boldsymbol{\omega})}{d (\vecth(\boldsymbol{\Omega}))
d (\vect \boldsymbol{\Phi})^\top} \\ 
&+ \boldsymbol{K} 
\bigg( \frac{d(\boldsymbol{\mu}_{t|t-1} - \boldsymbol{\omega})}{d (\vecth(\boldsymbol{\Omega}))^\top} \bigg)^\top
\boldsymbol{\mathcal{C}}^\prime_t 
\bigg( \frac{d(\boldsymbol{\mu}_{t|t-1} - \boldsymbol{\omega})}{ d (\vect \boldsymbol{\Phi})^\top} \bigg), 
\nonumber \\
\frac{d^2(\boldsymbol{\mu}_{t+1|t} - \boldsymbol{\omega})}{d (\vecth(\boldsymbol{\Omega}))
 d (\vect \boldsymbol{K})^\top} =&
\boldsymbol{X}_t \frac{d^2(\boldsymbol{\mu}_{t|t-1} - \boldsymbol{\omega})}{d (\vecth(\boldsymbol{\Omega}))
d (\vect \boldsymbol{K})^\top} \\ 
&+ \boldsymbol{K} 
\bigg( \frac{d(\boldsymbol{\mu}_{t|t-1} - \boldsymbol{\omega})}{ d (\vecth(\boldsymbol{\Omega}))^\top} \bigg)^\top
\boldsymbol{\mathcal{C}}^\prime_t 
\bigg( \frac{d(\boldsymbol{\mu}_{t|t-1} - \boldsymbol{\omega})}{ d (\vect \boldsymbol{K})^\top} \bigg).
\nonumber
\end{align*} 
\endgroup
The construction of the Hessian can now be completed by deriving the remaining second-order partial derivatives of the second differential in \eqref{secdiffloglik}.

By virtue of this representation, one can show that
\begingroup
\allowdisplaybreaks
\begin{align*}
\alpha^\prime_t =
\frac{\partial^2 \ell_t (\boldsymbol{\theta})}
{\partial \nu^2} 
=
\frac{1}{2} \bigg[ \frac{1}{2} \psi^{\prime} \bigg( \frac{\nu + N}{2} \bigg) 
&- \frac{1}{2} \psi^{\prime} \bigg( \frac{\nu}{2} \bigg) + \frac{N}{\nu^2} 
- \frac{N}{\nu^2} b_t 
- \frac{\nu + N}{\nu^2} b_t (1-b_t) + \frac{1}{\nu} b_t \bigg],
\\
\boldsymbol{\beta}^\prime_t =
\frac{\partial^2 \ell_t (\boldsymbol{\theta})}
{\partial (\vecth(\boldsymbol{\Omega})) \partial (\vecth(\boldsymbol{\Omega}))^\top} 
=&
\bigg[ \frac{\nu + N}{2\nu^2} (1- b_t)^2
\boldsymbol{\mathcal{D}}_N^\top
(\boldsymbol{\Omega}^{-1/2} \otimes \boldsymbol{\Omega}^{-1/2})
(\boldsymbol{\epsilon}_t  \boldsymbol{\epsilon}_t^\top \otimes
 \boldsymbol{\epsilon}_t  \boldsymbol{\epsilon}_t^\top) \\
& \times 
(\boldsymbol{\Omega}^{-1/2} \otimes \boldsymbol{\Omega}^{-1/2})
\boldsymbol{\mathcal{D}}_N
\bigg] \nonumber\\
&-
\bigg[ \frac{\nu + N}{\nu} (1- b_t) 
\boldsymbol{\mathcal{D}}_N^\top 
(\boldsymbol{\Omega}^{-1} \otimes \boldsymbol{\Omega}^{-1/2}
\boldsymbol{\epsilon}_t \boldsymbol{\epsilon}_t^\top \boldsymbol{\Omega}^{-1/2})  
\boldsymbol{\mathcal{D}}_N \bigg] \nonumber\\
&+
\bigg[\frac{1}{2}
 \boldsymbol{\mathcal{D}}_N^\top
(\boldsymbol{\Omega}^{-1} \otimes \boldsymbol{\Omega}^{-1}) 
\boldsymbol{\mathcal{D}}_N \bigg],
\end{align*}
\begin{align*}
\boldsymbol{\varsigma}^\prime_t =
\frac{\partial^2 \ell_t (\boldsymbol{\theta})} 
{\partial \boldsymbol{\mu}_{t|t-1} \partial \boldsymbol{\mu}_{t|t-1}^\top} 
=&
\bigg[ \frac{\nu + N}{\nu^2}  2 (1- b_t)^2 
\boldsymbol{\Omega}^{-1/2} \boldsymbol{\epsilon}_t
\boldsymbol{\epsilon}_t^\top \boldsymbol{\Omega}^{-1/2}
\bigg] \\ 
&- 
\bigg[ \frac{\nu + N}{\nu}  (1- b_t) 
\boldsymbol{\Omega}^{-1} \bigg],
\end{align*}
\endgroup
and 
\begingroup
\allowdisplaybreaks
\begin{align*}
\widehat{\alpha \boldsymbol{\beta}}^\prime_t =
\frac{\partial^2 \ell_t (\boldsymbol{\theta})}
{\partial (\vecth(\boldsymbol{\Omega})) \partial \nu} 
=&
\frac{1}{2} \boldsymbol{\mathcal{D}}_N^\top
(\boldsymbol{\Omega}^{-1/2} \otimes \boldsymbol{\Omega}^{-1/2}) 
(\boldsymbol{\epsilon}_t \otimes \boldsymbol{\epsilon}_t )\nonumber\\
&\times\bigg[ 
  \frac{\nu + N}{\nu^2} b_t (1 - b_t) 
  - \frac{N}{\nu^2} (1 - b_t)
  \bigg],
\\
\widehat{\alpha \boldsymbol{\varsigma}}^\prime_t =
\frac{\partial^2 \ell_t (\boldsymbol{\theta})}
{\partial \boldsymbol{\mu}_{t|t-1} \partial \nu} 
=&
\boldsymbol{\Omega}^{1/2} \boldsymbol{\epsilon}_t
\bigg[ 
  \frac{\nu + N}{\nu^2} b_t (1 - b_t) 
  - \frac{N}{\nu^2} (1 - b_t)
\bigg],
\\
\widehat{\boldsymbol{\beta} \boldsymbol{\varsigma}}^\prime_t =
\frac{\partial^2 \ell_t (\boldsymbol{\theta})}
{\partial \boldsymbol{\mu}_{t|t-1} \partial (\vecth(\boldsymbol{\Omega}))^\top} 
=&
\bigg[ \frac{\nu + N}{\nu^2} (1- b_t)^2
(\boldsymbol{\Omega}^{-1/2} \boldsymbol{\epsilon}_t 
 \boldsymbol{\epsilon}_t^\top  \boldsymbol{\Omega}^{-1/2}
\otimes \boldsymbol{\epsilon}_t^\top \boldsymbol{\Omega}^{-1/2} ) \boldsymbol{\mathcal{D}}_N \bigg]
 \nonumber\\
&+ 
\bigg[ \frac{\nu + N}{\nu} 2 (1- b_t) 
(\boldsymbol{\epsilon}_t^\top \boldsymbol{\Omega}^{-1/2} \otimes 
\boldsymbol{\Omega}^{-1}) \boldsymbol{\mathcal{D}}_N \bigg].
\end{align*}
\endgroup
which completes the construction of the Hessian matrix.

\subsection{The Conditional Information Matrix}
\label{conditional_information_matrix}

Taking the conditional expectation of the negative Hessian matrix yields the  conditional information matrix needed for the Fisher's scoring method. Likewise to the score and the Hessian, we start the discussion by taking advantage  from the differentials of the $\log$-likelihood function.

\begingroup
\allowdisplaybreaks
\begin{align*}
\mathbb{E}_{t-1}[\mathrm{d}^2 \ell_t (\boldsymbol{\theta})] =& 
\bigg[ \frac{1}{4} \psi^{\prime} \bigg( \frac{\nu + N}{2} \bigg) 
- \frac{1}{4} \psi^{\prime} \bigg( \frac{\nu}{2} \bigg) 
+ \frac{N(\nu + N + 4)}{2\nu(\nu +N)(\nu+N+2)}\bigg] (\mathrm{d}^2 \nu) \nonumber\\
&+
\bigg[
 \frac{1}{2(\nu + N + 2)} 
 (\mathrm{d} \vect \boldsymbol{\Omega})^\top 
 (\vect \boldsymbol{\Omega}^{-1}) (\vect \boldsymbol{\Omega}^{-1})^\top
 (\mathrm{d} \vect \boldsymbol{\Omega}) \bigg] \nonumber\\
&-
\bigg[
\frac{\nu + N}{2(\nu + N + 2)} 
(\mathrm{d} \vect \boldsymbol{\Omega})^\top 
(\boldsymbol{\Omega}^{-1} \otimes \boldsymbol{\Omega}^{-1})
(\mathrm{d} \vect \boldsymbol{\Omega}) \bigg] \nonumber\\
&+
\bigg[
\frac{1}{(\nu + N) (\nu + N + 2)} (\mathrm{d} \vect \boldsymbol{\Omega})^\top 
(\vect \boldsymbol{\Omega}^{-1}) (\mathrm{d}\nu) \bigg] \nonumber\\
&-
\bigg[
\frac{\nu + N}{\nu + N + 2} (\mathrm{d} \boldsymbol{\mu}_{t|t-1})^\top
\boldsymbol{\Omega}^{-1} (\mathrm{d} \boldsymbol{\mu}_{t|t-1}) \bigg].
\end{align*}
\endgroup
The calculations of this matrix require for the first set $\boldsymbol{\xi} = (\boldsymbol{\omega}^\top, (\vecth(\boldsymbol{\Omega}))^\top, \nu)^\top$,
\begin{equation*}
\boldsymbol{\mathcal{I}}^{(\boldsymbol{\xi})}_t(\boldsymbol{\theta}) = 
-\mathbb{E}_{t-1} \bigg[
\frac{d^2 \ell_t(\boldsymbol{\theta})}{d \boldsymbol{\xi}d \boldsymbol{\xi}^\top} \bigg]
=
\boldsymbol{\mathcal{I}}^{(\boldsymbol{\xi})}(\boldsymbol{\theta})
+ 
\bigg( \frac{d (\boldsymbol{\mu}_{t|t-1} - \boldsymbol{\omega})}{d \boldsymbol{\xi}^\top} \bigg)^\top
\boldsymbol{\mathcal{I}}^{(\boldsymbol{\mu})}(\boldsymbol{\theta})
\bigg( \frac{d (\boldsymbol{\mu}_{t|t-1} - \boldsymbol{\omega})}{d \boldsymbol{\xi}^\top} \bigg),
\end{equation*}
for the second vector $\boldsymbol{\psi} = ((\vect \boldsymbol{\Phi})^\top, (\vect \boldsymbol{K})^\top)^\top$, 
\begingroup
\allowdisplaybreaks
\begin{equation*}
\boldsymbol{\mathcal{I}}^{(\boldsymbol{\psi})}_t(\boldsymbol{\theta}) = 
-\mathbb{E}_{t-1} \bigg[
\frac{d^2 \ell_t(\boldsymbol{\theta})}{d \boldsymbol{\psi}d \boldsymbol{\psi}^\top} \bigg] 
=
\bigg( \frac{d (\boldsymbol{\mu}_{t|t-1} - \boldsymbol{\omega})}{d \boldsymbol{\psi}^\top} \bigg)^\top
\boldsymbol{\mathcal{I}}^{(\boldsymbol{\mu})}(\boldsymbol{\theta})
\bigg( \frac{d (\boldsymbol{\mu}_{t|t-1} - \boldsymbol{\omega})}{d \boldsymbol{\psi}^\top} \bigg),
\end{equation*}
\endgroup
and in conclusion, the negative conditional expected value of the cross-second derivatives are
\begin{align*}
\boldsymbol{\mathcal{I}}^{(\boldsymbol{\xi, \psi})}_t(\boldsymbol{\theta}) = 
-\mathbb{E}_{t-1} \bigg[
\frac{d^2 \ell_t(\boldsymbol{\theta})}{d \boldsymbol{\xi}d \boldsymbol{\psi}^\top} \bigg]
=
\bigg( \frac{d (\boldsymbol{\mu}_{t|t-1} - \boldsymbol{\omega})}{d \boldsymbol{\xi}^\top} \bigg)^\top
\boldsymbol{\mathcal{I}}^{(\boldsymbol{\mu})}(\boldsymbol{\theta})
\bigg( \frac{d (\boldsymbol{\mu}_{t|t-1} - \boldsymbol{\omega})}{d \boldsymbol{\psi}^\top} \bigg).
\end{align*}
Now, by equation \eqref{sysMSREs} the calculations boils down to the static terms of the matrix. Specifically, 
\begin{align*}
\boldsymbol{\mathcal{I}}^{(\boldsymbol{\mu})}(\boldsymbol{\theta})
=
- \mathbb{E}_{t-1} \bigg[ 
\frac{\partial^2 \ell_t(\boldsymbol{\theta})}{
\partial \boldsymbol{\mu}_{t|t-1} \partial \boldsymbol{\mu}_{t|t-1}^\top} \bigg]
=
\frac{\nu + N}{\nu + N + 2} \boldsymbol{\Omega}^{-1},
\end{align*}
while the terms of the static matrix $\boldsymbol{\mathcal{I}}^{(\boldsymbol{\xi})}(\boldsymbol{\theta})$ are
\begin{align*}
\boldsymbol{\mathcal{I}}^{(\nu)} (\boldsymbol{\theta}) 
= 
- \mathbb{E}_{t-1} \bigg[ 
\frac{\partial^2 \ell_t(\boldsymbol{\theta})}{
\partial \nu^2 } \bigg]
=
\frac{1}{4} \bigg[ \psi^{\prime} \bigg( \frac{\nu}{2} \bigg) 
&-  \psi^{\prime} \bigg( \frac{\nu + N}{2} \bigg) 
- \frac{2 N(\nu + N + 4)}{\nu(\nu +N)(\nu+N+2)}\bigg],
\end{align*}
\begingroup
\allowdisplaybreaks
\begin{align*}
\boldsymbol{\mathcal{I}}^{(\mathrm{v}(\boldsymbol{\Omega}))} (\boldsymbol{\theta}) 
=
-\mathbb{E}_{t-1} \bigg[& 
\frac{\partial^2 \ell_t (\boldsymbol{\theta})}
{\partial (\vecth(\boldsymbol{\Omega})) \partial (\vecth(\boldsymbol{\Omega}))^\top}
\bigg]
= 
\frac{\nu + N}{2(\nu + N + 2)}
\boldsymbol{\mathcal{D}}_N^\top 
(\boldsymbol{\Omega}^{-1} \otimes \boldsymbol{\Omega}^{-1})
\boldsymbol{\mathcal{D}}_N \\
&\tab-  \frac{1}{2(\nu + N + 2)}
\boldsymbol{\mathcal{D}}_N^\top 
 (\vecth (\boldsymbol{\Omega}^{-1})) (\vecth (\boldsymbol{\Omega}^{-1}))^\top
 \boldsymbol{\mathcal{D}}_N,
\end{align*}
\endgroup
and lastly the cross terms
\begin{align*}
\boldsymbol{\mathcal{I}}^{(\mathrm{v}(\boldsymbol{\Omega}), \nu)} (\boldsymbol{\theta}) 
=
- \mathbb{E}_{t-1} \bigg[ 
\frac{\partial^2 \ell_t (\boldsymbol{\theta})}
{\partial (\vecth(\boldsymbol{\Omega})) \partial \nu}
\bigg]
=
- \frac{1}{(\nu + N) (\nu + N + 2)} 
\boldsymbol{\mathcal{D}}_{N}^\top (\vecth (\boldsymbol{\Omega}^{-1})).
\end{align*}
With these last derivations, we have completed the derivations for the Fisher's scoring method in the multivariate DCS-\emph{t} set up.

\subsection{Third differentials}
This section derives the third differential of the conditional $\log$-likelihood with respect to the dynamic location, auxiliary to the proof of the asymptotic normality of the MLE, see Lemma \ref{STARTED_likelihood_second_diff_effect}.
By differentiating equation \eqref{secdiffloglik} with respect $\boldsymbol{\mu}_{t|t-1}$ one obtains
\begin{align}
\label{thirdiffloglik}
\mathrm{d}^3_{\boldsymbol{\mu}_{t|t-1}} \ell_t ( \boldsymbol{\theta}) =&
\bigg[ 8 \frac{\nu + N}{\nu^3} (1- b_t)^3 (\mathrm{d} \boldsymbol{\mu}_{t|t-1})^\top 
\boldsymbol{\Omega}^{-1/2} \boldsymbol{\epsilon}_{t} 
(\mathrm{d} \boldsymbol{\mu}_{t|t-1})^\top 
\boldsymbol{\Omega}^{-1/2} \boldsymbol{\epsilon}_{t} \boldsymbol{\epsilon}_{t}^\top 
(\mathrm{d} \boldsymbol{\mu}_{t|t-1}) \bigg] \nonumber\\
&+
\bigg[ 2 \frac{\nu + N}{\nu^2} (1- b_t)^2 
(\mathrm{d} \boldsymbol{\mu}_{t|t-1})^\top 
[ \boldsymbol{\Omega}^{-1/2} \boldsymbol{\epsilon}_{t} \otimes \boldsymbol{I}_N + 
\boldsymbol{I}_N \otimes  \boldsymbol{\epsilon}_{t} \boldsymbol{\Omega}^{-1/2} ]
(\mathrm{d} \boldsymbol{\mu}_{t|t-1})^2  \bigg] \nonumber\\
&-
 \bigg[ 2 \frac{\nu + N}{\nu^2} (1- b_t)^2 
(\mathrm{d} \boldsymbol{\mu}_{t|t-1})^\top 
\boldsymbol{\Omega}^{-1/2} \boldsymbol{\epsilon}_{t}
(\mathrm{d} \boldsymbol{\mu}_{t|t-1})^\top 
\boldsymbol{\Omega}^{-1}
(\mathrm{d} \boldsymbol{\mu}_{t|t-1}) \bigg] \nonumber\\
&-
\bigg[ 2 \frac{\nu + N}{\nu^2} (1- b_t)^2 
(\mathrm{d} \boldsymbol{\mu}_{t|t-1})^\top 
\boldsymbol{\Omega}^{-1/2} \boldsymbol{\epsilon}_{t}
(\mathrm{d}^2 \boldsymbol{\mu}_{t|t-1})
\boldsymbol{\Omega}^{-1/2} \boldsymbol{\epsilon}_{t} \bigg] \nonumber\\
&-
\bigg[\frac{\nu + N}{\nu} (1- b_t) (\mathrm{d}^2 \boldsymbol{\mu}_{t|t-1})^\top 
\boldsymbol{\Omega}^{-1/2} (\mathrm{d} \boldsymbol{\mu}_{t|t-1}) \bigg] \nonumber\\
&-
\bigg[\frac{\nu + N}{\nu} (1- b_t) (\mathrm{d}^3 \boldsymbol{\mu}_{t|t-1})^\top 
\boldsymbol{\Omega}^{-1/2} \boldsymbol{\epsilon}_{t} \bigg].
\end{align}

\newpage

\section{Lemmata}
\label{appendix_lemmata}
This Appendix contains the proofs of the auxiliary lemmata used to establish consistency and asymptotic normality of the MLE of Section \ref{Maximum_Likelihood_Estimation}.

\subsection*{Lemmata for the Proof of Consistency}
\label{Lemmata for the Proof of Consistency}

\subsubsection*{Proof of Lemma \ref{Identifiability}}
Consider the $t$-th contribution to the $\log$-likelihood, $\ell_t(\boldsymbol{\theta} )$. We have that
\begin{align*}
\mathbb{E}
\bigg[&
\sup_{\boldsymbol{\theta} \in \boldsymbol{\Theta}}   
| \ell_t(\boldsymbol{\theta} ) |
\bigg] \\
&\leq
\sup_{\boldsymbol{\theta} \in \boldsymbol{\Theta}}  
\bigg| \ln \Gamma\bigg( \frac{\nu + N}{2} \bigg) \bigg|
+ \sup_{\boldsymbol{\theta} \in \boldsymbol{\Theta}}  
\bigg| \ln \Gamma\bigg( \frac{\nu }{2} \bigg) \bigg|
+ \sup_{\boldsymbol{\theta} \in \boldsymbol{\Theta}} \bigg| \frac{N}{2} \ln(\pi \nu)\bigg| 
+ \sup_{\boldsymbol{\theta} \in \boldsymbol{\Theta}}  
\bigg| \, \frac{1}{2} \ln |\boldsymbol{\Omega}|\,  \bigg| \\
&\tab +
\frac{\nu +N}{2}
 \mathbb{E}\bigg[
\sup_{\boldsymbol{\theta} \in \boldsymbol{\Theta}} 
\bigg|
\ln \bigg(
1 + \frac{(\boldsymbol{y}_t - \boldsymbol{\mu}_{t|t-1})^\top
 \boldsymbol{\Omega}^{-1} (\boldsymbol{y}_t - \boldsymbol{\mu}_{t|t-1})}{\nu}
\bigg)
\bigg|
 \bigg] < \infty,
\end{align*}
since the compactness of the parameter space $\boldsymbol{\Theta}$ with $0 < \nu < \infty$ ensures that the first three terms are finite, there exist $\Omega_{-} >0$ and $\Omega_{+} < \infty$ such that $\Omega_{-} < | \boldsymbol{\Omega} | < \Omega_{+}$ and moreover, the logarithmic moment in the last term exists as a consequence of Lemmata \ref{SE_Dynamic_Location}, \ref{bounded_moments} and \ref{INV_Dynamic_location} with $m>0$. In particular, we can show that
\begin{align*}
\mathbb{E}\bigg[
\sup_{\boldsymbol{\theta} \in \boldsymbol{\Theta}}   
\bigg| (\boldsymbol{y}_t - \boldsymbol{\mu}_{t|t-1})^\top
 \boldsymbol{\Omega}^{-1} (\boldsymbol{y}_t - \boldsymbol{\mu}_{t|t-1})/\nu \bigg|^m
 \bigg]
 < \infty,
\end{align*}
is always satisfied for some $m>0$ and with $ \nu > 0$, implying the existence of the required logarithmic moment.

Clearly, the result obtained above also implies that $\mathbb{E}
\big[ 
| \ell_t(\boldsymbol{\theta}_0 )|
\big] < \infty$, and then, we can turn to the last statement.

To prove the uniqueness and identifiability of $\boldsymbol{\theta}_0$ it is sufficient to consider the sequence $\{ \ell_t(\boldsymbol{\theta}) - \ell_t(\boldsymbol{\theta}_0)\}_{t \in \mathbb{Z}}$ under the assumption that $(\nu, \vecth \boldsymbol{\Omega} )^\top = ( \nu_0, \vecth \boldsymbol{\Omega}_0)^\top$. We prove the argument by contradiction. 

By denoting with $\boldsymbol{\mu}_{t|t-1}(\boldsymbol{\theta})$ and $\boldsymbol{\mu}_{t|t-1}(\boldsymbol{\theta}_0)$ the dynamic location vector evaluated at $\boldsymbol{\theta}$ and the true parameter vector $\boldsymbol{\theta}_0$ respectively, and as $\boldsymbol{v}_{t}(\boldsymbol{\theta}) = \boldsymbol{y}_t - \boldsymbol{\mu}_{t|t-1}(\boldsymbol{\theta})$ and $\boldsymbol{v}_{t}(\boldsymbol{\theta}_0) = \boldsymbol{y}_t - \boldsymbol{\mu}_{t|t-1}(\boldsymbol{\theta}_0)$ the difference becomes 
\begin{align*}
&\ell_t(\boldsymbol{\theta}) - \ell_t(\boldsymbol{\theta}_0) \\
&\propto
\ln \Big[ 1 + (\boldsymbol{v}_{t}(\boldsymbol{\theta}))^\top 
\boldsymbol{\Omega}_0^{-1}(\boldsymbol{v}_{t}(\boldsymbol{\theta})) / \nu_0
\Big] -
\ln \Big[ 1 + (\boldsymbol{v}_{t}(\boldsymbol{\theta}_0))^\top 
\boldsymbol{\Omega}_0^{-1}(\boldsymbol{v}_{t}(\boldsymbol{\theta}_0)) / \nu_0 
\Big]
\nonumber\\
&=
\ln\Big(\Big[
1+(\boldsymbol{v}_{t}(\boldsymbol{\theta}))^\top 
\boldsymbol{\Omega}_0^{-1}(\boldsymbol{v}_{t}(\boldsymbol{\theta})) / \nu_0
\Big]
\Big/
\Big[
 1 + (\boldsymbol{v}_{t}(\boldsymbol{\theta}_0))^\top 
\boldsymbol{\Omega}_0^{-1}(\boldsymbol{v}_{t}(\boldsymbol{\theta}_0)) / \nu_0
\Big]
\Big),
\end{align*}
where the latter equation holds if and only if $\boldsymbol{\mu}_{t|t-1}(\boldsymbol{\theta}) = \boldsymbol{\mu}_{t|t-1}(\boldsymbol{\theta}_0)$ almost surely since $\boldsymbol{\Omega}_0$ is symmetric positive definite and $0 < \nu < \infty $. Under maintained assumptions, it is clear that $\{\boldsymbol{\mu}_{t|t-1}(\boldsymbol{\theta})\}_{t\in\mathbb{Z}}$ and $\{\boldsymbol{\mu}_{t|t-1}(\boldsymbol{\theta}_0)\}_{t\in\mathbb{Z}}$ are stationary and ergodic sequences, which implies that the same holds true for the sequence by $\{(\boldsymbol{\mu}_{t+1|t}(\boldsymbol{\theta}) - \boldsymbol{\mu}_{t+1|t}(\boldsymbol{\theta}_0))\}_{t\in\mathbb{Z}}$. Thus, it is possible to write the difference recursion as
\begin{align*}
(\boldsymbol{\mu}_{t+1|t}&(\boldsymbol{\theta}) - \boldsymbol{\mu}_{t+1|t}(\boldsymbol{\theta}_0))
\\
&=
(\boldsymbol{\omega} - \boldsymbol{\omega}_{0})
+
(\boldsymbol{\Phi} - \boldsymbol{\Phi}_{0}) \boldsymbol{\omega}_{0}
+
(\boldsymbol{\Phi} - \boldsymbol{\Phi}_{0}) \boldsymbol{\mu}_{t|t-1}(\boldsymbol{\theta}_0)
+
(\boldsymbol{K} - \boldsymbol{K}_{0})
\boldsymbol{u}_t,
\end{align*}
and the relation above entails the fact that if $\boldsymbol{\mu}_{t|t-1}(\boldsymbol{\theta}) = \boldsymbol{\mu}_{t|t-1}(\boldsymbol{\theta}_0) $  $\forall t$ almost surely, then
\begin{align*}
(\boldsymbol{\omega} - \boldsymbol{\omega}_{0})
+
(\boldsymbol{\Phi} - \boldsymbol{\Phi}_{0})
\boldsymbol{\omega}_{0}
=&
(\boldsymbol{\Phi} - \boldsymbol{\Phi}_{0}) \boldsymbol{\mu}_{t|t-1}(\boldsymbol{\theta}_0)
+
(\boldsymbol{K} - \boldsymbol{K}_{0})
\boldsymbol{u}_t,
\end{align*}
almost surely. Nonetheless, as $\det \boldsymbol{K} \neq 0$, the whole multivariate system of equations is stochastic, and one cannot find a nontrivial solution of the system that will cancel out the driving force $\boldsymbol{u}_t$ of the dynamic location vector. As a result, the only available option reduces to the equivalence between all the parameters, that is $\boldsymbol{\omega} = \boldsymbol{\omega}_{0}$, $\boldsymbol{\Phi} = \boldsymbol{\Phi}_{0}$ and $\boldsymbol{K} = \boldsymbol{K}_0$.

Therefore, we have shown that $\mathbb{E}[\ell_t(\boldsymbol{\theta})] < \mathbb{E}[\ell_t(\boldsymbol{\theta}_0)]$ for every $\boldsymbol{\theta} \neq \boldsymbol{\theta}_0$.
$\square$

\subsubsection*{Proof of Lemma \ref{ULLN_likelihood}}

We apply a mean-value expansion of the $\log$-likelihood around $\hat{\boldsymbol{\mu}}_{t|t-1}^\star$ which is on the chord between the started filtered location $\hat{\boldsymbol{\mu}}_{t|t-1}$ and ${\boldsymbol{\mu}}_{t|t-1} $. We take the supremum over the compact parameter space and see that
\begin{align*}
\sup_{\boldsymbol{\theta} \in \boldsymbol{\Theta}} 
| \widehat{\mathcal{L}}_T(\boldsymbol{\theta}) -  {\mathcal{L}}_T(\boldsymbol{\theta}) |
&\leq
 \sup_{\boldsymbol{\theta} \in \boldsymbol{\Theta}} 
 \bigg\|
\frac{\partial \widehat{\mathcal{L}}_T(\boldsymbol{\theta})}
{\partial \hat{\boldsymbol{\mu}}_{t|t-1}^{\star\top}}
\bigg\| 
\sup_{\boldsymbol{\theta} \in \boldsymbol{\Theta}} \|
\hat{\boldsymbol{\mu}}_{t|t-1}
-
{\boldsymbol{\mu}}_{t|t-1} 
\|,
\end{align*}
where by direct calculation and the triangle inequality we get
\begin{align*}
\sup_{\boldsymbol{\theta} \in \boldsymbol{\Theta}} \bigg\|&
\frac{\partial \widehat{\mathcal{L}}_T(\boldsymbol{\theta})}
{\partial \hat{\boldsymbol{\mu}}_{t|t-1}^{\star\top}}
\bigg\| \\
&\leq
\frac{1}{T} \sum_{t=1}^{T} 
\sup_{\boldsymbol{\theta} \in \boldsymbol{\Theta}} \bigg\|
\boldsymbol{\Omega}^{-1} 
\frac{\nu + N}{\nu } 
\frac{(\boldsymbol{y}_t - \hat{\boldsymbol{\mu}}_{t|t-1}^{\star})}
{1 + (\boldsymbol{y}_t - \hat{\boldsymbol{\mu}}_{t|t-1}^{\star})^\top 
\boldsymbol{\Omega}^{-1} 
(\boldsymbol{y}_t - \hat{\boldsymbol{\mu}}_{t|t-1}^{\star})/\nu} 
 \bigg\| \\ 
&\leq
{c}_{\boldsymbol{\Omega}} 
\bigg(
\max_{\boldsymbol{\theta} \in \boldsymbol{\Theta}}
\frac{\nu + N}{\nu } 
\bigg)
\frac{1}{T} \sum_{t=1}^{T} 
\sup_{\boldsymbol{\theta} \in \boldsymbol{\Theta}} 
\|\boldsymbol{y}_t - \hat{\boldsymbol{\mu}}_{t|t-1}^{\star} \| 
\\
&\tab\times
\sup_{\boldsymbol{\theta} \in \boldsymbol{\Theta}} 
\bigg| \,\,\,
\big[
1 + (\boldsymbol{y}_t - \hat{\boldsymbol{\mu}}_{t|t-1}^{\star})^\top 
\boldsymbol{\Omega}^{-1} 
(\boldsymbol{y}_t - \hat{\boldsymbol{\mu}}_{t|t-1}^{\star})/\nu
\big]^{-1}
\bigg|.
\end{align*}
Note that the compactness of the parameter space imposed by condition \ref{as3} is crucial here. Moreover, if we treat the dynamic location vector as a fixed parameter with value $\hat{\boldsymbol{\mu}}_{t|t-1}^{\star}$ and let $\boldsymbol{y}_t \rightarrow \infty$ the entire term in the right hand side of the latter inequality will vanish. Hence, we obtain that
$
\sup_{\boldsymbol{\theta} \in \boldsymbol{\Theta}} \big\|
\frac{\partial \widehat{\mathcal{L}}_T(\boldsymbol{\theta})}
{\partial \hat{\boldsymbol{\mu}}_{t|t-1}^{\star\top}}
\big\| 
=
O_p(1),
$
which is enough to ensure the existence of $\log$-moments. Furthermore, conditions \ref{as1} and \ref{as2} are needed in order to keep the data stationary and ergodic and the filter invertible, respectively. Thus, we can apply Lemma \ref{INV_Dynamic_location} and obtain
$
\sup_{\boldsymbol{\theta} \in \boldsymbol{\Theta}} \|
\hat{\boldsymbol{\mu}}_{t|t-1}
-
{\boldsymbol{\mu}}_{t|t-1}
\| \xrightarrow[]{\text{e.a.s.}} 0.
$
In conclusion, by Lemma 2.1 in \cite{Straumann_Mikosh2006} the claimed almost sure convergence holds. 

Now, for the second result, we note that Theorem A.2.2 of \cite{White1994}, i.e. the Uniform Law of Large Numbers in its version for stationary and ergodic processes, applies straightforwardly to our case since: $(1)$ the parameter space is compact, $(2)$ the empirical likelihood function $\mathcal{L}_T(\boldsymbol{\theta} )$ defined in \eqref{Stationary_Eergodic_likelihood_equation} is continuous in $\boldsymbol{\theta}$  $\forall$ $\boldsymbol{y}_t$ and $\forall$ $\boldsymbol{\theta} \in \boldsymbol{\Theta}$ is measurable in $\boldsymbol{y}_t$, which is stationary and ergodic, and $(3)$ by Lemma \ref{Identifiability} we obtain the moment bound $\mathbb{E}\big[ \sup_{\boldsymbol{\theta} \in \boldsymbol{\Theta}}| \ell_t(\boldsymbol{\theta} )|\big] < \infty$ which ensure the dominance condition.

Thus, all the conditions of Theoerm A.2.2 in \cite{White1994} are met and the proof is complete. $\square$

\subsection*{Lemmata for the Proof of Asymptotic Normality}
\label{Lemmata for the Proof of Asymptotic Normality}

\subsubsection*{Proof of Lemma \ref{CLT_first_differential_Likelihood}}
The first derivatives of the $\log$-likelihood contribution at time $t$ with respect to the true parameter vector $\boldsymbol{\theta}_0$ can be retrieved from the differential in equation \eqref{diffloglik}.

We have 
\begin{align*}
\mathbb{E}_{t-1}[\mathrm{d}&\ell_t(\boldsymbol{\theta}_0)]\\
 =&
 \frac{1}{2} \bigg[ \psi \bigg( \frac{\nu_0 + N}{2} \bigg) - \psi \bigg( \frac{\nu_0}{2}\bigg) - \frac{N}{\nu_0} + \frac{\nu_0 + N }{\nu_0} \, 
 \mathbb{E}_{t-1}[b_t] - \mathbb{E}_{t-1}[\ln w_t] \bigg] (\mathrm{d} \nu_0) \nonumber\\
&+ \frac{1}{2} 
(\mathrm{d} \vecth (\boldsymbol{\Omega}_0))^\top \boldsymbol{\mathcal{D}}_{N}^\top
(\boldsymbol{\Omega}_0^{-1/2} \otimes \boldsymbol{\Omega}_0^{-1/2})
\bigg[ \frac{\nu_0 + N}{ \nu_0 } 
\mathbb{E}_{t-1}[ 
(\boldsymbol{\epsilon}_t \otimes  \boldsymbol{\epsilon}_t)/w_t]
- \vect \boldsymbol{I}_N \bigg]\nonumber\\
&+ \frac{\nu_0 + N}{\nu }(\mathrm{d} \tilde{\boldsymbol{\mu}}_{t|t-1})^\top
\boldsymbol{\Omega}_0^{-1} 
\mathbb{E}_{t-1}[(\boldsymbol{y}_t - \tilde{\boldsymbol{\mu}}_{t|t-1})/w_t],
\end{align*}
since the derivatives obtained from $\mathrm{d} \tilde{\boldsymbol{\mu}}_{t|t-1}$ are $\mathcal{F}_{t-1}$-measurables.

Then, one has 
\begin{align*}
\mathbb{E}_{t-1}[ b_t ] &= \frac{N}{\nu + N}, \\
\mathbb{E}_{t-1} [\ln (1/w_t) ]  &= \mathbb{E}_{t-1} [\ln (1 - b_t) ] = 
\psi \bigg( \frac{\nu}{2}\bigg) - \psi \bigg( \frac{\nu + N}{2} \bigg), \\
\mathbb{E}_{t-1}
[(\boldsymbol{\epsilon}_t \otimes  \boldsymbol{\epsilon}_t)
/w_t]
&= 
\nu
\mathbb{E}_{t-1}[(\mathbf{z}_t \otimes  \mathbf{z}_t)] 
 \mathbb{E}_{t-1}[b_t] 
= \frac{\nu}{\nu + N} \vect \boldsymbol{I}_N, \\
\mathbb{E}_{t-1}[(\boldsymbol{y}_t - \tilde{\boldsymbol{\mu}}_{t|t-1})/w_t]
&= \sqrt{\nu} \mathbb{E}_{t-1}[\sqrt{b_t (1-b_t) }]
\, \boldsymbol{\Omega}^{1/2} \mathbb{E}_{t-1}[ \mathbf{z}_t] = \boldsymbol{0},
\end{align*}
where the first and the second equality follow from the properties of the beta distribution, see equation \eqref{b_t}. The third and the fourth equalities are obtained based on the stochastic representation of the model given in equation \eqref{stochastic_rep2}. Thus, by substitutions, we obtain the martingale difference property. 

The second claim follows  by Lemmata \ref{uniformly_boundedness_ut}, \ref{SE_Dynamic_Location}, \ref{INV_Dynamic_location}, \ref{bounded_moments}, \ref{Identifiability} and by an application of the continuous mapping theorem to ${d}\ell_t(\boldsymbol{\theta}_0)$.

With the support of the Cram\'{e}r-Wold device (see \cite{VanDerVaart1998} pag. 16) the CLT for martingales of \cite{Billingsley1961} directly applies to the linear combination
$
\sqrt{T} \mathcal{L}^{\prime}_T(\boldsymbol{\theta}_0)
=
\sqrt{T} 
\frac{1}{T} \sum_{t=1}^T
\frac{d \ell_t(\boldsymbol{\theta}_0)}{d \boldsymbol{\theta}_0}
\xRightarrow[]{} \mathcal{N}(\boldsymbol{0}, \boldsymbol{V}).
$ $\square$

\subsubsection*{Proof of Lemma \ref{STARTED_diff_likelihood_effect}}

The claimed convergence in probability can be proved based on the invertibility of the location filter, see  Lemma \ref{INV_Dynamic_location}, and its derivatives, see Lemma \ref{INV_first_differential_Dynamic_location}. Invertibility  also ensures that the perturbed first differential of the dynamic location evaluated at $\boldsymbol{\theta}=\boldsymbol{\theta}_0$ will converge to the unique stationary ergodic solution,
\begin{align}
\label{eas_filters}
&\sup_{\boldsymbol{\theta} \in \boldsymbol{\Theta}}
\| 
\hat{\boldsymbol{\mu}}_{t|t-1} 
-
\tilde{\boldsymbol{\mu}}_{t|t-1} 
\| 
\xrightarrow[]{\text{e.a.s.}} 0
\,\,\,\,\,
 \text{and} \nonumber
\\
&\sup_{\boldsymbol{\theta} \in \boldsymbol{\Theta}}
\| 
{d}\hat{\boldsymbol{\mu}}_{t|t-1} 
-
{d}\tilde{\boldsymbol{\mu}}_{t|t-1} 
\| 
\xrightarrow[]{\text{e.a.s.}} 0
\,\,\,\,\,  \text{as} \,\,\,\,\, t \rightarrow \infty.
\end{align}
Hence, we can rely on a multivariate mean value expansion about all the elements of the vectors $\hat{\boldsymbol{\mu}}_{t|t-1}^{\star}$ and ${d}\hat{\boldsymbol{\mu}}_{t|t-1}^{\star}$, which lie on the chords between $(\hat{\boldsymbol{\mu}}_{t|t-1}, \tilde{\boldsymbol{\mu}}_{t|t-1})$ and $({d}\hat{\boldsymbol{\mu}}_{t|t-1}, {d}\tilde{\boldsymbol{\mu}}_{t|t-1})$ respectively, yielding
\begin{align*}
\sqrt{T}
\|& 
\widehat{\mathcal{L}}^\prime_T(\boldsymbol{\theta}_0) 
-
{\mathcal{L}}^\prime_T(\boldsymbol{\theta}_0) 
\| \nonumber\\
&\leq
\sqrt{T}
\begin{Vmatrix}
 \frac{ \partial (\widehat{\mathcal{L}}^\prime_T(\boldsymbol{\theta}_0) )}
 {\partial \hat{\boldsymbol{\mu}}_{t|t-1}^{\star\top}}\\
\frac{  \partial (\widehat{\mathcal{L}}^\prime_T(\boldsymbol{\theta}_0) )}
 {\partial ({d}\hat{\boldsymbol{\mu}}_{t|t-1}^{\star\top})}\\
\end{Vmatrix}
\begin{Vmatrix}
(\hat{\boldsymbol{\mu}}_{t|t-1}
-
\tilde{\boldsymbol{\mu}}_{t|t-1}  )\\
({d}\hat{\boldsymbol{\mu}}_{t|t-1} 
-
{d}\tilde{\boldsymbol{\mu}}_{t|t-1})
\end{Vmatrix}.
\end{align*}
The first term on the right hand of the inequality is uniformly bounded. Exponentially fast almost sure convergence of the second term in the right hand side is obtained by Lemma \ref{INV_first_differential_Dynamic_location}. 

By means of analogous arguments as in Lemma \ref{ULLN_likelihood} we can show that
\begin{align*}
\bigg\|
\frac{ \partial (\widehat{\mathcal{L}}^\prime_T(\boldsymbol{\theta}_0) )}
 {\partial \hat{\boldsymbol{\mu}}_{t|t-1}^{\star\top}}
\bigg\| 
=
O_p(1),
\,\,\,\,\, \text{and} \,\,\,\,\,
\bigg\|
\frac{  \partial (\widehat{\mathcal{L}}^\prime_T(\boldsymbol{\theta}_0) )}
 {\partial ({d}\hat{\boldsymbol{\mu}}_{t|t-1}^{\star\top})}
\bigg\| 
=
O_p(1).
\end{align*}
Moreover, the results obtained in \eqref{eas_filters} imply that for $t$ large enough
\begin{align*}
\max\{ \| 
\hat{\boldsymbol{\mu}}_{t|t-1} 
-
\tilde{\boldsymbol{\mu}}_{t|t-1} 
\| ,
\| 
{d}\hat{\boldsymbol{\mu}}_{t|t-1} 
-
{d}\tilde{\boldsymbol{\mu}}_{t|t-1} 
\| 
 \} <1.
\end{align*}
By using the Chebyshev and the $c_m$ inequalities we then have that for $\varepsilon>0$ and some $m>0$
\begin{align*}
\mathbb{P} \Big(
\sqrt{T}
\| 
\widehat{\mathcal{L}}^\prime_T(\boldsymbol{\theta}_0) 
-
{\mathcal{L}}^\prime_T(\boldsymbol{\theta}_0) 
\|
> \varepsilon
\Big)
\leq&
\frac{\sqrt{T}}{\varepsilon^m}
\mathbb{E}
[
\| 
\widehat{\mathcal{L}}^\prime_T(\boldsymbol{\theta}_0) 
-
{\mathcal{L}}^\prime_T(\boldsymbol{\theta}_0) 
\|^m
]
\\
\leq&
\frac{1}{T^{m/2}\varepsilon^m}
\sum_{t=1}^T
\mathbb{E}
\bigg[
\bigg\| 
\frac{d \widehat{\ell}_t(\boldsymbol{\theta}_0)}{d \boldsymbol{\theta}_0} 
-
\frac{d \ell_t(\boldsymbol{\theta}_0)}{d \boldsymbol{\theta}_0}
\bigg\|^m
\bigg]
\\
\leq&
\frac{1}{T^{m/2}\varepsilon^m}
O_p(t\varrho^t),
\end{align*}
which is $O_p(T^{-m/2})$ and this implies the claimed convergence in probability. $\square$

\subsubsection*{Proof of Lemma \ref{STARTED_likelihood_second_diff_effect}}
The second derivatives of the likelihood are nonlinear functions of the filtered location vector and its first end second derivatives. Hence, the mean value theorem is applied for each dynamic equation. As a result,
\begin{align*}
\sup_{\boldsymbol{\theta} \in \boldsymbol{\Theta}}
\|
\widehat{\mathcal{L}}^{\prime\prime}_T(\boldsymbol{\theta} )
-
\mathcal{L}^{\prime\prime}_T(\boldsymbol{\theta} )
\|
\leq
\sup_{\boldsymbol{\theta} \in \boldsymbol{\Theta}}
\begin{Vmatrix}
\frac{\partial \widehat{\mathcal{L}}^{\prime\prime}_T(\boldsymbol{\theta})}
{\partial \hat{\boldsymbol{\mu}}_{t|t-1}^{\star\top}}\\
\frac{\partial \widehat{\mathcal{L}}^{\prime\prime}_T(\boldsymbol{\theta})}
{\partial ({{d}\hat{\boldsymbol{\mu}}}_{t|t-1}^{\star\top})}\\
\frac{\partial \widehat{\mathcal{L}}^{\prime\prime}_T(\boldsymbol{\theta})}
{\partial ({{d}^2\hat{\boldsymbol{\mu}}}_{t|t-1}^{\star\top})}
\end{Vmatrix}
\sup_{\boldsymbol{\theta} \in \boldsymbol{\Theta}}
\begin{Vmatrix}
(\hat{\boldsymbol{\mu}}_{t|t-1} 
-
{\boldsymbol{\mu}}_{t|t-1})\\
({d}\hat{\boldsymbol{\mu}}_{t|t-1} 
-
{d}{\boldsymbol{\mu}}_{t|t-1})\\
({d}^2\hat{\boldsymbol{\mu}}_{t|t-1} 
-
{d}^2{\boldsymbol{\mu}}_{t|t-1})
\end{Vmatrix}.
\end{align*} 
Thus, the proof follows by the same arguments of the proof of Lemma \ref{ULLN_likelihood}, i.e. by obtaining the uniformly boundedness of the first term and the exponentially fast convergence of the second term in the right hand side respectively. Note that  the last component of the first term in the right hand side involves a third order differential, which can be found in \eqref{thirdiffloglik} and is uniformly bounded. Subsequent applications of Lemma 2.1 of \cite{Straumann_Mikosh2006} yield the desired result. $\square$

\subsubsection*{Proof of Lemma \ref{ULLN_likelihood_second_diff}}
Note that $\mathcal{L}^{\prime\prime}_T(\boldsymbol{\theta} )$ is a continuous function of $\{ \boldsymbol{y}_t \}_{t \in \mathbb{Z}}$ and therefore stationary and ergodic. The Lemma follows straightforwardly from Lemma \ref{Properties_second_diff_likelihood} and The Uniform Law of Large Numbers for ergodic stationary processes, see Theorem A.2.2 in \cite{White1994} and Lemma \ref{ULLN_likelihood}. $\square$

\subsubsection*{Proof of Lemma \ref{Properties_second_diff_likelihood}}
This Lemma is a multivariate extension of the Theorem 5 of \cite{Harvey2013}. Thus, we only discuss the relevant arguments.

The complete equation of the second differential is more subtle than the first, thus we leave it in \eqref{secdiffloglik}. We prove the arguments by considering equation \eqref{chainrule_hessian}, namely
\begin{align*}
\frac{d^2 \ell_t(\boldsymbol{\theta})}{d \boldsymbol{\theta}d \boldsymbol{\theta}^\top}
 =&
\frac{\partial^2 \ell_t(\boldsymbol{\theta})}{\partial \boldsymbol{\theta} \partial \boldsymbol{\theta}^\top}
+ 
\bigg(
\frac{d (\boldsymbol{\mu}_{t|t-1} - \boldsymbol{\omega})}{d \boldsymbol{\theta}^\top}
\bigg)^\top
\frac{\partial^2 \ell_t(\boldsymbol{\theta})}{\partial \boldsymbol{\mu}_{t|t-1}
 \partial \boldsymbol{\mu}_{t|t-1}^\top}
\bigg(
\frac{d (\boldsymbol{\mu}_{t|t-1} - \boldsymbol{\omega})}{d \boldsymbol{\theta}^\top}
\bigg)\\
&+
\frac{\partial \ell_t(\boldsymbol{\theta})}{\partial \boldsymbol{\mu}_{t|t-1}^\top}
\frac{d^2 (\boldsymbol{\mu}_{t|t-1} - \boldsymbol{\omega})}{d \boldsymbol{\theta}d \boldsymbol{\theta}^\top}.
\end{align*}
By taking the expectation, we get a finite and static term in the first summand on the right hand side, while by the independence and the martingale difference sequence properties of the score vector, the last term becomes null. Thus, we can focus our attention on the middle term. Define
\begin{align*}
\boldsymbol{\mathcal{I}}^{(\boldsymbol{\mu}_{t|t-1})}(\boldsymbol{\theta}) =&
-\mathbb{E}\bigg[
\bigg( \frac{d (\boldsymbol{\mu}_{t|t-1} - \boldsymbol{\omega})}{d \boldsymbol{\theta}^\top} \bigg)^\top
\frac{\partial^2 \ell_t(\boldsymbol{\theta})}{\partial \boldsymbol{\mu}_{t|t-1}
 \partial \boldsymbol{\mu}_{t|t-1}^\top}
\bigg( \frac{d (\boldsymbol{\mu}_{t|t-1} - \boldsymbol{\omega})}{d \boldsymbol{\theta}^\top} \bigg) \bigg].
\end{align*}
Note that, by independence, we can express the vectorized counterpart as
\begin{align*}
\vect \boldsymbol{\mathcal{I}}^{(\boldsymbol{\mu}_{t|t-1})}(\boldsymbol{\theta}) =&
\mathbb{E}\bigg[
\bigg( \frac{d (\boldsymbol{\mu}_{t|t-1} - \boldsymbol{\omega})}{d \boldsymbol{\theta}^\top} \bigg)
\otimes
\bigg( \frac{d (\boldsymbol{\mu}_{t|t-1} - \boldsymbol{\omega})}{d \boldsymbol{\theta}^\top} \bigg) 
\bigg]^\top
\vect \boldsymbol{\mathcal{I}}^{(\boldsymbol{\mu})} (\boldsymbol{\theta}).
\end{align*}
By Lemmata \ref{INV_Dynamic_location} and \ref{INV_first_differential_Dynamic_location}, the dynamic location filter and its differentials are invertible and achieve their own unique stationary ergodic solution with an unbounded number of finite moments.

 Thus, we obtain the desired result by repeated applications of the law of iterated expectation  to the following equality
\begin{align*}
\mathbb{E}_{t-1}\bigg[&
\bigg( \frac{d (\boldsymbol{\mu}_{t+1} - \boldsymbol{\omega})}{d \boldsymbol{\theta}^\top} \bigg)
\otimes
\bigg( \frac{d (\boldsymbol{\mu}_{t+1} - \boldsymbol{\omega})}{d \boldsymbol{\theta}^\top} \bigg) 
\bigg]^\top \nonumber \\
=&
\mathbb{E}_{t-1}\bigg[
\bigg( 
\boldsymbol{X}_t
\frac{d (\boldsymbol{\mu}_{t|t-1} - \boldsymbol{\omega})}{d \boldsymbol{\theta}^\top} 
+ \frac{d \boldsymbol{R}_{t} }{d \boldsymbol{\theta}^\top}
\bigg)
\otimes
\bigg( \boldsymbol{X}_t 
\frac{d (\boldsymbol{\mu}_{t|t-1} - \boldsymbol{\omega})}{d \boldsymbol{\theta}^\top}
+ \frac{d \boldsymbol{R}_{t} }{d \boldsymbol{\theta}^\top}
\bigg) 
\bigg]^\top \nonumber\\
=&
\bigg(
\frac{d (\boldsymbol{\mu}_{t|t-1} - \boldsymbol{\omega})}{d \boldsymbol{\theta}^\top} 
\otimes
\frac{d (\boldsymbol{\mu}_{t|t-1} - \boldsymbol{\omega})}{d \boldsymbol{\theta}^\top}
\bigg)^\top
\mathbb{E}_{t-1} \bigg[
\bigg( \boldsymbol{X}_t \otimes  \boldsymbol{X}_t \bigg)
\bigg]^\top  \nonumber\\
&+ 
\mathbb{E}_{t-1}\bigg[
\bigg( 
\boldsymbol{X}_t
\frac{d (\boldsymbol{\mu}_{t|t-1} - \boldsymbol{\omega})}{d \boldsymbol{\theta}^\top} 
\otimes 
\frac{d \boldsymbol{R}_{t} }{d \boldsymbol{\theta}^\top}
\bigg) 
\bigg]^\top
+ 
\mathbb{E}_{t-1}\bigg[
\bigg( 
\frac{d \boldsymbol{R}_{t} }{d \boldsymbol{\theta}^\top}
\otimes 
\boldsymbol{X}_t
\frac{d (\boldsymbol{\mu}_{t|t-1} - \boldsymbol{\omega})}{d \boldsymbol{\theta}^\top} 
\bigg)
\bigg]^\top\\
&+
\mathbb{E}_{t-1}\bigg[
\bigg( 
\frac{d \boldsymbol{R}_{t} }{d \boldsymbol{\theta}^\top}
\otimes 
\frac{d \boldsymbol{R}_{t} }{d \boldsymbol{\theta}^\top}
\bigg) \bigg]^\top.
\end{align*}
Note that the contraction conditions \ref{as2} and \ref{as5} are more than enough to ensure the stability of the recursions, while Lemma \ref{SE_First_differentials_dynamic_location} ensures the existence of the required moments. $\square$

\subsection*{Auxiliary Lemmata}
\label{Auxiliary Lemmata}

\begin{lemma}
\label{SE_First_differentials_dynamic_location}
Consider the stochastic difference equation
\begin{align*}
{d} (\boldsymbol{\mu}_{t+1|t} - \boldsymbol{\omega})
=
\boldsymbol{X}_t {d}(\boldsymbol{\mu}_{t|t-1} - \boldsymbol{\omega})
+
\boldsymbol{R}_t,
\end{align*}
where $\boldsymbol{X}_t$ and $\boldsymbol{R}_t$ are defined in \eqref{Xt} and \eqref{Rt}, respectively.
 
Assume that conditions \ref{as1}, \ref{as2} and \ref{as3} in Assumption \ref{consistency_theorem_assumptions} are satisfied. Then, there exist a unique sequence $\{ d (\tilde{\boldsymbol{\mu}}_{t+1|t} - \boldsymbol{\omega}) \}_{t \in \mathbb{Z}}$ which is stationary and ergodic. A causal stationary solution exists and can be expressed as
\begin{align*}
d (\tilde{\boldsymbol{\mu}}_{t+1|t} - \boldsymbol{\omega})
=
\sum_{j=0}^{\infty}\bigg( 
 \prod_{k=1}^j 
\boldsymbol{X}_{t-k}
\bigg) \boldsymbol{R}_{t-j}.
\end{align*}
Furthermore, 
$\mathbb{E}[ \| d(\tilde{\boldsymbol{\mu}}_{t|t-1} - \boldsymbol{\omega})\|^m ] < \infty$ for every $m > 0$.
\end{lemma}
\begin{proof}
The proof follows the arguments of the proof of Lemma \ref{SE_Dynamic_Location}, which now applies by rewriting $\boldsymbol{X}_{t}$ and all the components of $\boldsymbol{R}_{t} $ in terms of the innovations and independently of $\tilde{\boldsymbol{\mu}}_{t|t-1}$ so that a stationary ergodic sequence $\{(\boldsymbol{X}_t , \boldsymbol{R}_t) \}_{t\in \mathbb{Z}}$ can be generated.

It follows from Lemma \ref{SE_Dynamic_Location} that the first condition is used in order to keep the multivariate system stable and the matrices $\boldsymbol{X}_t$ random, while the contraction condition \ref{as2} for linear stochastic difference equations gives us the sufficient condition which ensures that $d(\tilde{\boldsymbol{\mu}}_{t+1|t} - \boldsymbol{\omega})$ is the unique stationary and ergodic solution, see \cite{Bougerol1993}.

Moreover, the H{\"o}lder and Minkowsky inequalities imply that
\begin{align*}
\mathbb{E}
\bigg[ 
\| d(\tilde{\boldsymbol{\mu}}_{t+1|t} - \boldsymbol{\omega})\|^m 
\bigg] 
\leq
\bigg\{
\sum_{j=0}^{\infty}
\mathbb{E}
\bigg[ 
 \bigg\|
\prod_{k=0}^{j}  \boldsymbol{X}_{t-k} \bigg\|^{m} \bigg]^{1/m}
\, \mathbb{E}
\bigg[ 
 \| \boldsymbol{R}_{t-j} \|^{m}
\bigg]^{1/m}
\bigg\}^m.
\end{align*}
In addition, from equation \eqref{alternvecCt} we note that when $\boldsymbol{\theta} = \boldsymbol{\theta}_0$ 
\begin{align*}
\mathbb{E} \bigg[&
 \| \boldsymbol{X}_t \|^m
\bigg]\\
&\leq
 \| \boldsymbol{\Phi} \|^m
+
\mathbb{E} \bigg[
 \| \boldsymbol{K} \boldsymbol{\mathcal{C}}_t \|^m
 \bigg] \\
&\leq
\bar{\rho}^m +
c_{\boldsymbol{K}}
\mathbb{E} \bigg[ b_t^{m/2} (1- b_t)^{m/2}
\bigg]
\mathbb{E} \big[ 
\| (\mathbf{z}_t \otimes \mathbf{z}_t) \|^m  \big] 
+
c_{\boldsymbol{K}}
N^{m/2}  
\mathbb{E}\big[ 
(1- b_t)^{m/2}
\big]\\
&=
\bar{\rho}^m +
c_{\boldsymbol{K}}
\mathbb{E} \big[ \| \mathbf{z}_t  \|^{2m} \big]
\frac{B\big( \frac{N+m}{2} , \frac{\nu+m}{2} \big)}
{B\big( \frac{N}{2} , \frac{\nu}{2}\big)}
+
c_{\boldsymbol{K}}
N^{m/2}
\frac{B\big( \frac{N}{2} , \frac{\nu+m}{2} \big)}
{B\big( \frac{N}{2} , \frac{\nu}{2}\big)} \\
&=
\bar{\rho}^m +
\frac{ c_{\boldsymbol{K}} }{N^{m}}
\frac{B\big( \frac{N+m}{2} , \frac{\nu+m}{2} \big)}
{B\big( \frac{N}{2} , \frac{\nu_0}{2}\big)}
+
c_{\boldsymbol{K}}
N^{m/2}
\frac{B\big( \frac{N}{2} , \frac{\nu+m}{2} \big)}
{B\big( \frac{N}{2} , \frac{\nu}{2}\big)}
 < \infty,
\end{align*}
by Lemma \ref{uniformly_boundedness_ut}. Note that the condition \ref{as1} is needed in order to keep the matrix $\boldsymbol{X}_t$ random and identifiable.

It remains to prove the moment bounds of $\boldsymbol{R}_t$ for every $m > 0$. We have,
\begin{align*}
\mathbb{E}\bigg[
\|
\boldsymbol{a}_t \|^m \bigg]
&=
\mathbb{E}\bigg[
 b_t^{3m/2} (1- b_t)^{m/2}/ \nu^{m/2}
\bigg]
\mathbb{E}\bigg[ 
\| \boldsymbol{\Omega}^{1/2} \mathbf{z}_t\|^m \bigg] \\
&\leq
\frac{{c}_{\boldsymbol{\Omega}}}{N^{m/2}}
\frac{B\big( \frac{N + 3m}{2} , \frac{\nu+m}{2} \big)}
{B\big( \frac{N}{2} , \frac{\nu_0}{2}\big)} < \infty,
\end{align*}
In addition,  
\begin{align*}
\mathbb{E}\bigg[&
\|
\vect \boldsymbol{B}_t \|^m \bigg]\\
=&
\mathbb{E}\bigg[
  \nu^{m/2} b_t^{3m/2} (1- b_t)^{m/2}
\bigg]
\mathbb{E}\bigg[
\bigg\|
(\boldsymbol{\Omega}^{-1/2} \mathbf{z}_t \otimes 
\boldsymbol{\Omega}^{-1/2} \mathbf{z}_t \otimes 
\boldsymbol{\Omega}^{1/2}\mathbf{z}_t )
\bigg\|^m \bigg] \\
\leq& 
{c}_{\boldsymbol{\Omega}}
\mathbb{E} \big[ \| \mathbf{z}_t  \|^{3m} \big]
\frac{B\big( \frac{N + 3m}{2} , \frac{\nu+m}{2} \big)}
{B\big( \frac{N}{2} , \frac{\nu}{2}\big)} 
=
\frac{{c}_{\boldsymbol{\Omega}}}{N^{3m/2}}
\frac{B\big( \frac{N + 3m}{2} , \frac{\nu+m}{2} \big)}
{B\big( \frac{N}{2} , \frac{\nu}{2}\big)} 
< \infty,
\end{align*}
and
\begin{align*}
\mathbb{E}\bigg[
\|
\boldsymbol{D}_t \|^m \bigg]
=&
\mathbb{E}\bigg[
\big\|
 \big[(\tilde{\boldsymbol{\mu}}_{t|t-1} - \boldsymbol{\omega})^\top \otimes \boldsymbol{I}_N \big]
\big\|^m \bigg] \\
\leq&
\bigg\{
\sqrt{N}
\bar{c}
\sum_{j=0}^\infty
\bar{\rho}^j 
\bigg(
\mathbb{E} \Big[ 
\| 
\boldsymbol{u}_{t-j} 
\|^{m}
 \Big]
\bigg)^{1/m} \bigg\}^{m} 
 < \infty,
\end{align*}
by Lemma \ref{SE_Dynamic_Location}, and finally, by Lemma \ref{uniformly_boundedness_ut}, 
\begin{align*}
\mathbb{E}\bigg[
\|
\boldsymbol{E}_t \|^m \bigg]
&=
\mathbb{E}\bigg[
\big\|
\big[(\boldsymbol{u}_{t})^\top \otimes \boldsymbol{I}_N \big]
\big\|^m \bigg] 
\leq
N^{m/2}
\mathbb{E}\bigg[
\|
\boldsymbol{u}_{t}
\|^m \bigg] \\
&\leq
{c}_{\boldsymbol{\Omega}}
 \nu^{m/2}
\frac{B\big( \frac{N+m}{2} , \frac{\nu+m}{2} \big)}
{B\big( \frac{N}{2} , \frac{\nu}{2} \big)}.
 < \infty.
\end{align*}

\end{proof}

\begin{lemma}
\label{SE_Second_differentials_dynamic_location}
Consider the stochastic difference equation
\begin{equation*}
{d}^2(\boldsymbol{\mu}_{t+1|t} - \boldsymbol{\omega} )
=
\boldsymbol{X}_t {d}^2 ( \boldsymbol{\mu}_{t|t-1} - \boldsymbol{\omega} )
+
\boldsymbol{K} 
{d}( \boldsymbol{\mu}_{t|t-1} - \boldsymbol{\omega} )^\top
\boldsymbol{\mathcal{C}}^\prime_t 
{d}( \boldsymbol{\mu}_{t|t-1} - \boldsymbol{\omega} ) 
+
\boldsymbol{Q}_t, 
\end{equation*}
where $\boldsymbol{X}_t$, $\boldsymbol{Q}_t$ and $\boldsymbol{\mathcal{C}}^\prime_t$ are defined in \eqref{Xt}, \eqref{Qt} and \eqref{partialscores_C_prime_t}, respectively.
 
Assume that conditions \ref{as1} and \ref{as2} in Assumption \ref{consistency_theorem_assumptions} are satisfied. Then, there exist a unique sequence $\{ d^2 (\tilde{\boldsymbol{\mu}}_{t+1|t} - \boldsymbol{\omega}) \}_{t \in \mathbb{Z}}$ which is stationary and ergodic. A causal stationary solution exists and can be expressed as
\begin{align*}
d^2 (\tilde{\boldsymbol{\mu}}_{t+1|t} - \boldsymbol{\omega})
=&
\sum_{j=0}^{\infty} 
\bigg\{ \bigg( 
\prod_{k=1}^j 
\boldsymbol{X}_{t-k}
\bigg) \\
&\times \bigg[
\boldsymbol{K} 
d( \tilde{\boldsymbol{\mu}}_{t-j|t-j-1} - \boldsymbol{\omega} )^\top
\boldsymbol{\mathcal{C}}^\prime_{t-j} 
d( \tilde{\boldsymbol{\mu}}_{t-j|t-j-1} - \boldsymbol{\omega} ) 
+ \boldsymbol{Q}_{t-j} 
\bigg]\bigg\}
\end{align*}
Furthermore, 
$\mathbb{E}[ \|d^2(\tilde{\boldsymbol{\mu}}_{t|t-1} - \boldsymbol{\omega})\|^m ] < \infty$ for every $m > 0$.
\end{lemma}
\begin{proof}
From Lemma \ref{SE_Dynamic_Location}, the first two conditions ensure the existence of a unique stationary and ergodic sequence $\{ d^2 (\tilde{\boldsymbol{\mu}}_{t+1|t} - \boldsymbol{\omega}) \}_{t \in \mathbb{Z}}$.

Moreover, by the H{\"o}lder and Minkowsky inequalities along with the independence between each component, imply that
\begin{align*}
\mathbb{E}
\bigg[ 
\|& d^2(\tilde{\boldsymbol{\mu}}_{t+1|t} - \boldsymbol{\omega})\|^m 
\bigg] 
\leq
\bigg\{
\sum_{j=0}^{\infty}
\mathbb{E}
\bigg[ 
 \bigg\|
\prod_{k=0}^{j}  \boldsymbol{X}_{t-k} \bigg\|^{2m} \bigg]^{1/2m} \nonumber\\
&\times
\bigg( 
c_{\boldsymbol{K}}
\mathbb{E}
\bigg[ 
 \|
d( \tilde{\boldsymbol{\mu}}_{t-j|t-j-1} - \boldsymbol{\omega} )
\|^{4m} \bigg]^{1/4m}
\mathbb{E}
\bigg[ 
  \|
\boldsymbol{\mathcal{C}}^\prime_{t-j} 
\|^{2m} \bigg]^{1/2m}   \\
&+
\mathbb{E}
\bigg[ 
 \|
\boldsymbol{Q}_{t-j} 
\|^{m} \bigg]^{1/m} 
\bigg)
\bigg\}^m,
\end{align*}
from which we can see that, by  Lemma \ref{SE_First_differentials_dynamic_location}, the first two terms are uniformly bounded and the third is the second derivative of the driving force with respect to the dynamic location vector.

In the same spirit of Lemma \ref{SE_First_differentials_dynamic_location}, let us consider equation \eqref{partialscores_C_prime_t}. Then, when $\boldsymbol{\theta} = \boldsymbol{\theta}_0$, the $c_m$-inequality establishes that
\begin{align*}
\mathbb{E}
\bigg[
  \|
\boldsymbol{\mathcal{C}}^\prime_{t}
&\|^{m} \bigg] 
\leq
\mathbb{E}
\bigg[
[8(1-b_t)^3/\nu^2 ]^m 
\Big\| \Big\{ \big[
\boldsymbol{I}_N   \otimes 
\boldsymbol{v}_t \boldsymbol{v}_t^\top
  \big] \vect \boldsymbol{\Omega}^{-1} 
\Big\} \Big\|^m
\Big\| \big[ 
\boldsymbol{v}_t ^\top \boldsymbol{\Omega}^{-1}
\big] \Big\|^m \bigg]  \nonumber\\
&+
\mathbb{E}
\bigg[
[2(1-b_t)^2/ \nu ]^m  
\Big\| \Big\{ \big[
\boldsymbol{\Omega}^{-1}  \otimes \boldsymbol{I}_N \big]
\big[ \boldsymbol{v}_t \otimes \boldsymbol{I}_N 
+ \boldsymbol{I}_N  \otimes \boldsymbol{v}_t
\big] \Big\} \Big\|^m \bigg] \nonumber\\
&+
\mathbb{E}
\bigg[
[2(1-b_t)^2/\nu ]^m
\Big\| \Big\{  \big[ 
\boldsymbol{\Omega}^{-1}  \otimes \boldsymbol{I}_N \big]
 \big[  \boldsymbol{v}_t \otimes \boldsymbol{I}_N 
 \big] \Big\} \Big\|^m \bigg] \\
&\leq
{C}_4 \mathbb{E}
\big[ \| \mathbf{z}_t \|^{3m} \big]
+
{C}_3 \mathbb{E}
\big[ \| \mathbf{z}_t \|^{2m} \big]
+
{C}_3 \mathbb{E}
\big[ \| \mathbf{z}_t \|^{2m} \big]
< \infty.
\end{align*}
Straightforward calculations show that analogous results hold for each component of $\boldsymbol{Q}_{t}$, so that $\boldsymbol{Q}_{t}$ is uniformly bounded.
\end{proof}

\begin{lemma}
\label{INV_first_differential_Dynamic_location}
Let the conditions of Lemmata \ref{SE_Dynamic_Location}, \ref{SE_First_differentials_dynamic_location} and \ref{SE_Second_differentials_dynamic_location} hold true. Consider further the filtering equation \eqref{score_driven_filter} under the condition of Lemma \ref{INV_Dynamic_location}. Then, for any initialization of the filter $ \hat{\boldsymbol{\mu}}_{1|0} $ and its first derivatives in $ d \hat{\boldsymbol{\mu}}_{1|0} $, the perturbed first  and second derivative sequences of the dynamic location filter, i.e. $\{ d(\hat{\boldsymbol{\mu}}_{t|t-1} - \boldsymbol{\omega}) \}_{t \in \mathbb{N}}$ and $\{ d^2(\hat{\boldsymbol{\mu}}_{t|t-1} - \boldsymbol{\omega}) \}_{t \in \mathbb{N}}$, converge exponentially fast almost surely to the unique stationary ergodic solution $\{ d({\boldsymbol{\mu}}_{t|t-1} - \boldsymbol{\omega}) \}_{t \in \mathbb{Z}}$ and $\{ d^2({\boldsymbol{\mu}}_{t|t-1} - \boldsymbol{\omega}) \}_{t \in \mathbb{Z}}$.

Furthermore, for any $m>0$
\begin{align*}
&\mathbb{E}[\sup_{\boldsymbol{\theta}\in \boldsymbol{\Theta}}
 \|d(\hat{\boldsymbol{\mu}}_{t|t-1} - \boldsymbol{\omega})\|^m ] < \infty
\,\,\,\,\, \text{and} \,\,\,\,\,
\mathbb{E}[\sup_{\boldsymbol{\theta}\in \boldsymbol{\Theta}}
 \|d^2(\hat{\boldsymbol{\mu}}_{t|t-1} - \boldsymbol{\omega})\|^m ] < \infty,\\
&\mathbb{E}[\sup_{\boldsymbol{\theta}\in \boldsymbol{\Theta}}
 \|d(\tilde{\boldsymbol{\mu}}_{t|t-1} - \boldsymbol{\omega})\|^m ] < \infty
\,\,\,\,\,\text{and} \,\,\,\,\,
 \mathbb{E}[\sup_{\boldsymbol{\theta}\in \boldsymbol{\Theta}}
 \|d^2(\tilde{\boldsymbol{\mu}}_{t|t-1} - \boldsymbol{\omega})\|^m ] < \infty.
\end{align*}
\end{lemma}
\begin{proof}
We provide a detailed discussion for the first case, that is the convergence of the perturbed first derivatives, since the proof for the convergence of the perturbed second derivatives follows the same line.

The proof of this Lemma builds upon the arguments of Theorem 2.10 in \cite{Straumann_Mikosh2006} for perturbed SRE. In particular, the perturbed SREs corresponds to the derivatives in
$
d (\hat{\boldsymbol{\mu}}_{t+1|t} - \boldsymbol{\omega})
=
\widehat{\boldsymbol{X}}_t d(\hat{\boldsymbol{\mu}}_{t|t-1} - \boldsymbol{\omega})
+
\widehat{\boldsymbol{R}}_t,
$
which are nonlinear functions of the initialized filtered sequence $\{(\hat{\boldsymbol{\mu}}_{t|t-1} - \boldsymbol{\omega}) \}_{t \in \mathbb{N}}$. The relevant contraction condition \eqref{inv_contraction} of Lemma \ref{INV_Dynamic_location} holds and the required convergence of the recurrence equation is obtained if 
\begin{align}
\label{required_condition_pert_SRE}
\sup_{\boldsymbol{\theta}\in \boldsymbol{\Theta}}
\| \widehat{\boldsymbol{X}}_t - \widetilde{\boldsymbol{X}}_t \| \xrightarrow[]{\text{e.a.s.}} 0 
\,\,\,\,\,  \text{and} \,\,\,\,\,
\sup_{\boldsymbol{\theta}\in \boldsymbol{\Theta}}
 \| \widehat{\boldsymbol{R}}_t - \widetilde{\boldsymbol{R}}_t \| \xrightarrow[]{\text{e.a.s.}} 0
 \,\,\,\,\, \text{as} \,\,\,\,\, t \rightarrow \infty.
\end{align}
In order to verify these conditions, we use the mean value theorem, giving
\begin{align}
\label{pert_SRE_Xt}
\sup_{\boldsymbol{\theta}\in \boldsymbol{\Theta}}
\| \widehat{\boldsymbol{X}}_t - \widetilde{\boldsymbol{X}}_t \|
\leq
\sup_{\boldsymbol{\theta}\in \boldsymbol{\Theta}}
\|
\boldsymbol{\mathcal{C}}^\prime_{t}
\|
\,
\sup_{\boldsymbol{\theta}\in \boldsymbol{\Theta}}
\|
\hat{\boldsymbol{\mu}}_{t|t-1} 
-
\tilde{\boldsymbol{\mu}}_{t|t-1}
\|,
\end{align}
and
\begin{align*}
\sup_{\boldsymbol{\theta}\in \boldsymbol{\Theta}}
\| \widehat{\boldsymbol{R}}_t - \widetilde{\boldsymbol{R}}_t \|
\leq
\sup_{\boldsymbol{\theta}\in \boldsymbol{\Theta}}
\begin{Vmatrix}
\boldsymbol{\mathcal{C}}^\prime_{t}\\
\widehat{\boldsymbol{B\mathcal{C}}}^\prime_t\\
\widehat{\boldsymbol{a\mathcal{C}}}^\prime_t
\end{Vmatrix}
\,
\sup_{\boldsymbol{\theta}\in \boldsymbol{\Theta}}
\|
\hat{\boldsymbol{\mu}}_{t|t-1} 
-
\tilde{\boldsymbol{\mu}}_{t|t-1}
\|,
\end{align*}
where the expression for $\boldsymbol{\mathcal{C}}^\prime_{t}$, $\widehat{\boldsymbol{B\mathcal{C}}}^\prime_t$ and $\widehat{\boldsymbol{B\mathcal{C}}}^\prime_t$ can be found in \eqref{partialscores_C_prime_t}, \eqref{partialscores_aC_prime_t} and \eqref{partialscores_BC_prime_t} respectively. We can combine the results obtained in Lemma \ref{SE_Second_differentials_dynamic_location} together with the almost sure exponentially fast convergence \eqref{eas_dynamic_location} in Lemma \ref{INV_Dynamic_location}, in order to achieve the required convergences in \eqref{required_condition_pert_SRE}. As in Lemma \ref{SE_First_differentials_dynamic_location} we can show by direct calculations that the property of uniformly boundedness applies to each these derivatives, since they are continuous functions of $w_t$ in equation \eqref{u_t}.

We obtain that
\begin{align*}
\sup_{\boldsymbol{\theta}\in \boldsymbol{\Theta}}
\|
\boldsymbol{\mathcal{C}}^\prime_{t}
\|
=
O_p(1),
\,\,\,\,\, 
\sup_{\boldsymbol{\theta}\in \boldsymbol{\Theta}}
\begin{Vmatrix}
\boldsymbol{\mathcal{C}}^\prime_{t}\\
\widehat{\boldsymbol{B\mathcal{C}}}^\prime_t\\
\widehat{\boldsymbol{a\mathcal{C}}}^\prime_t
\end{Vmatrix}
=
O_p(1)
\end{align*}
and
\begin{align*}
\sup_{\boldsymbol{\theta}\in \boldsymbol{\Theta}}
\|
\hat{\boldsymbol{\mu}}_{t|t-1}
-
\tilde{\boldsymbol{\mu}}_{t|t-1} 
\|
=
o_{e.a.s.}(1)
\,\,\,\,\, \text{as} \,\,\,\,\, t \rightarrow \infty.
\end{align*}
Thus, repeated applications of Lemma 2.1 in \cite{Straumann_Mikosh2006} ensure the required convergence in \eqref{pert_SRE_Xt}.

Summarising, we have
\begin{align*}
\sup_{\boldsymbol{\theta}\in \boldsymbol{\Theta}}
\| 
d(\hat{\boldsymbol{\mu}}_{t|t-1} - \boldsymbol{\omega}) 
-
d(\tilde{\boldsymbol{\mu}}_{t|t-1} - \boldsymbol{\omega}) 
\| 
\xrightarrow[]{\text{e.a.s.}} 0
\tab  \text{as} \tab t \rightarrow \infty.
\end{align*}
Since the sequence $\{ d^2(\hat{\boldsymbol{\mu}}_{t|t-1} - \boldsymbol{\omega}) \}_{t \in \mathbb{N}}$ is a nonlinear function of both the perturbed recurrence $\{ d(\hat{\boldsymbol{\mu}}_{t|t-1} - \boldsymbol{\omega}) \}_{t \in \mathbb{N}}$ and the filter $\{( \hat{\boldsymbol{\mu}}_{t|t-1} - \boldsymbol{\omega}) \}_{t \in \mathbb{N}}$ the same arguments apply sequentially, yielding
\begin{align*}
\sup_{\boldsymbol{\theta}\in \boldsymbol{\Theta}}
\| 
{d}^2(\hat{\boldsymbol{\mu}}_{t|t-1} - \boldsymbol{\omega}) 
-
{d}^2(\tilde{\boldsymbol{\mu}}_{t|t-1} - \boldsymbol{\omega}) 
\| 
\xrightarrow[]{\text{e.a.s.}} 0
\tab  \text{as} \tab t \rightarrow \infty.
\end{align*}
The second claim for the moment bounds follows by the continuous mapping theorem, since the derivatives are nonlinear continuous functions of $\tilde{\boldsymbol{\mu}}_{t|t-1}$, which has unbounded moments, see Lemma \ref{INV_Dynamic_location}.
\end{proof}


\bibliography{winnower_template}
\bibliographystyle{chicago}